# Coherent methods in the X-ray sciences

**Keith A Nugent**

**ARC Centre of Excellence for Coherent X-ray Science**

**School of Physics**

**The University of Melbourne, Vic., 3010**

**AUSTRALIA**

| | |
|---|---|
| Correspondence: | Prof Keith A Nugent |
| | School of Physics |
| | The University of Melbourne, Vic., 3010 |
| | Australia |
| | Phone: +613 8344 5446 |
| | Email: keithan@unimelb.edu.au |

# Abstract

X-ray sources are developing rapidly and their coherent output is growing extremely rapidly. The increased coherent flux from modern X-ray sources is being matched with an associated rapid development in experimental methods. This article reviews the literature describing the ideas that utilise the increased brilliance from modern X-ray sources. It explores how ideas in coherent X-ray science are leading to developments in other areas, and *vice versa*. The article describes measurements of coherence properties and uses this discussion as a base from which to describe partially-coherent diffraction and X-ray phase contrast imaging, with its applications in materials science, engineering and medicine. Coherent diffraction imaging methods are reviewed along with associated experiments in materials science. Proposals for experiments to be performed with the new X-ray free-electron-lasers are briefly discussed. The literature on X-ray photon correlation spectroscopy is described and the features it has in common with other coherent X-ray methods are identified. Many of the ideas used in the coherent X-ray literature have their origins in the optical and electron communities and these connections are explored. A review of the areas in which ideas from coherent X-ray methods are contributing to methods for the neutron, electron and optical communities is presented.

# Table of contents

# 1   Introduction

Highly coherent sources of energetic X-rays are developing with breathtaking rapidity. The advent of the X-ray free electron laser will see the coherent output of the brightest X-rays sources increase by something like ten orders of magnitude. Transitions of this scale inevitably reveal fascinating new insights about nature and the next few years is certain to be an exciting time in all fields of science for which X-rays are a powerful and sensitive probe.

However, while there is no doubt that the availability of X-ray laser sources will be transformational, this is simply the latest stage of a development that has seen the coherent output of more conventional sources, such as synchrotrons, increase by an order of magnitude every two years since the seventies [1]. As a result, the use of the coherent properties of X-rays is emerging as a critically important aspect of X-ray science, a trend that is simply being accelerated by the development of laser sources.

One could argue that, as evidenced by the influence of the laser on the optical sciences, X-ray laser sources will revolutionise all aspects of X-ray science. To an extent this will be true, but there are likely to be some key differences. One key difference is that ionisation by the X-ray photons will significantly impact the degree to which the interaction of the probe with the object must be considered. Another current limit is that the scale and cost of the sources themselves will limit the applicability of X-ray laser ideas in fields such as medicine and industrial processing. However, from the scientific perspective, a more important difference will be that the ideas already generated and established by the optics community will be able to be rapidly adopted by the X-ray science community; one can anticipate a rapid convergence of the two areas, a merging of ideas and concepts, and a rapid interchange of experimental methods.

Similarly, the community that uses electrons as a probe has also been developing highly coherent sources of its own, in the form of, for example, the field emission gun [2, 3]. And others are emerging [4]. While there are key points of difference revolving around the fermionic nature of the electrons, many of the ideas created by this community will see application in the X-ray science community. As will be demonstrated by this review, there are already very strong indications that this is happening. Moreover, ideas being developed for coherent X-rays are also feeding back into the electron, neutron and optical sciences.

The key theme of this review is to explore some areas in which the fields of coherent electron, optical and X-ray science are converging. This convergence is happening over a broad range of scientific areas that cannot be reviewed in a single article. Developments include new approaches to the study of materials properties, ultrafast X-ray science and so on. In this review, then, the scope will be limited to the methods by which structures may be probed via the scattering or diffraction of X-ray fields. In particular, this review will largely but not exclusively concentrate on the science that can be explored by the conceptually straightforward experiment shown schematically in Figure 1.

In this broad scheme, a highly, but not necessarily completely, coherent beam of X-rays is incident on an object and the scattered or diffracted light is detected some distance downstream. This is, in a sense, the simplest possible experiment; it involves only the observation of the photons scattered or diffracted by a object. The review is largely concerned with the experimental systems that conform to this broad schematic, and for which coherence effects are important.

But note that even this configuration is too broad as it includes the case where the object is crystalline: the enormous field of crystallography. This field is far too large to be covered by this review and so considerations of cases in which the object can be treated as an infinite extended periodic object will be explicitly excluded. However the study of nanocrystals, for which the shape of the crystal has measurable consequences, will be considered.

Currently X-ray sources available for experiments are not themselves entirely coherent spatially or temporally. While third-generation sources produce radiation with a relatively high degree of spatial coherence, they fall well short of the essentially perfect coherence of a modern optical laser. X-ray free electron laser sources are expected to provide excellent spatial coherence but they will not provide perfect temporal coherence. Furthermore, there will be a great deal of shot-to-shot variation in the output of these sources and so, for data accumulated over many shots, the variation will very likely need to be treated as an ensemble average which will, for all practical purposes, emulate a partially coherent data-set. For the foreseeable future, then, the role of partial coherence will be an important part of X-ray science. As such, this review will consider the experimental methodologies in the context of partially coherent X-ray fields.

This review will look at the capacity to recover structural information from the scattered photons and will do so on spatial scales ranging from relatively large scale imaging at the micron scale down to coherent diffractive imaging at the nanoscale.

The review begins in section 2 with a discussion of diffraction by partially coherent X-rays. The unifying theme here is the exploration of the forms of information that can be extracted from measurements of the diffraction of highly coherent X-ray fields and so a clear exposition of the essential elements of that process is needed. In section 3, methods are explored by which one can be sure of the degree to which the field is coherent. Laser science can safely assume that the incident field is, for all practical

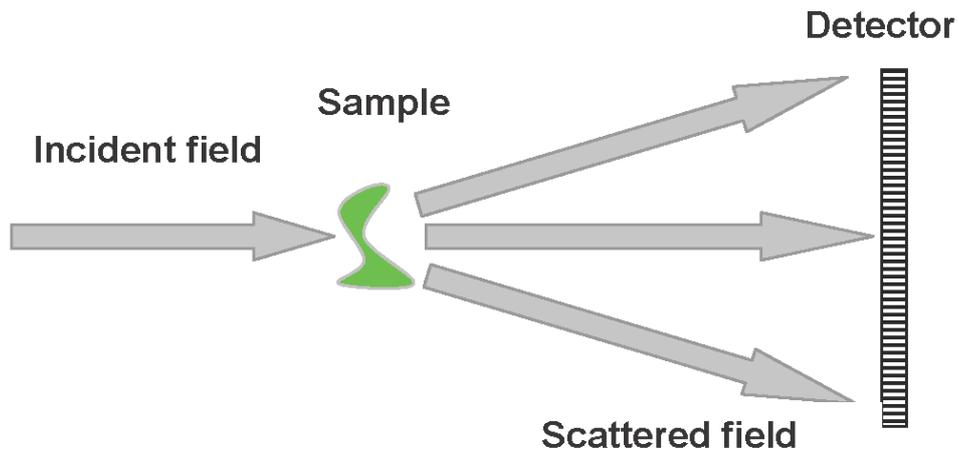

*Figure 1*: Conceptual layout of the experimental system that is the topic of this review. The incident field is assumed to have a degree of partial coherence and information about the sample is extracted from the measured intensity distribution

considerations, perfectly coherent. X-ray sources have not yet achieved this level of development and it is, in general, rather unsafe to assume that the incident field has a perfect degree of either spatial or temporal coherence. This review explores the approaches and results of the attempts to fully characterise the coherence of the light emerging from a modern X-ray source.

The development of X-ray facilities has been largely driven by the quest to create ever brighter sources. Increased brightness is entirely equivalent to increased coherent output. The mid-nineties saw the emergence of the first large third-generation sources, sources in which magnetic structures such as undulators and wigglers are inserted into the facility to reduce the emittance and so enhance the coherent output. Section 2 reviews some of the fundamental concepts and definitions in the study of partial coherence. Section 3 then considers progress on the methods by which partial coherence can be measured. Coherent effects immediately began to emerge with the development of third-generation sources. Amongst the first effect to be observed was the phenomenon of X-ray phase contrast, and it was immediately recognised that this effect could form the basis of an important new experimental technique. Section 4 reviews the methods of phase contrast imaging and its applications. The methods of phase contrast imaging can be made quantitative and section 5 reviews the literature describing the development and implementation of approaches to quantitative phase imaging. The appropriate algorithms and methodologies are described.

The ideas that have been developed for X-ray sciences have a complex interplay with developments in optical, electron and neutron sciences and a failure to include a discussion of this interplay would lead to an incomplete and possibly misleading picture of the sources of the ideas and the areas in which they can have the most impact. In this review, then, while the prime theme will be developments in X-ray sciences, the discussion will take us into a consideration of developments beyond this area, including science with electrons, neutrons and visible light. These methods and developments will be discussed in section 5.6.

Section 6 explores the development of very high resolution imaging using coherent diffraction, known as coherent diffractive imaging (CDI). This is an important emerging area both in the context of the X-ray free electron laser sources and with synchrotron sources. The fundamental ideas are outlined and applications to date are described. CDI is currently seen as having huge promise and applications of the method to important scientific problems are now emerging. However it is early days and CDI is an area that will continue its current rapid development.

Finally, section 7 reviews the application of coherent X-rays to the study of materials using X-ray photon correlation spectroscopy and the adaptation of techniques developed in electron microscopy such as fluctuation microscopy. These methods have clear origins in the optical and electron communities, and coherent X-rays offer a range highly complementary analytical approaches.

In section 8 the field is briefly summarised and some thoughts are presented on future directions in the field.

## 2     **Fundamental Concepts**

### 2.1    **Some basic ideas, definitions & terminology**

In this section some of the basic ideas of optical coherence theory will be briefly reviewed and concepts that underpin much of this review will be presented. The reader is also recommended to consult the article by Sutton [5].

X-ray sources that are emerging over the next ten years will largely be accelerator based sources, such as synchrotron sources [6], energy recovery linac sources [7] and X-ray free electron lasers [8] based on the self-amplified spontaneous emission process [9]. Other sources are also emerging, including the creation of high harmonic light from intense laser pulses [10] and the use of inverse Compton scattering of visible photons off electron beams [11]. In general, the X-rays emerge as a coherent or partially coherent beam with low divergence characteristics. These sources are therefore such that the photons, even after scattering, are still propagating at a small angle with respect to the direction of the beam; the beam-like quality of the light allows most theoretical treatments to adopt the paraxial approximation, which assumes $\sin\theta \approx \theta$, where $\theta$ is the angle subtended between the direction of energy propagation and the axis of the beam. The same conditions also permit the use of the scalar formulation of diffraction theory allowing polarisation effects in the scattering to be neglected.

Consider an electromagnetic field with a time and space-varying electric field, $E(\boldsymbol{\rho},t)$, written as a complex function with amplitude and phase and where $\boldsymbol{\rho}$ denotes position in three-dimensional space and *t* is time. One can then write the first order correlation function for this field using the so-called mutual coherence function (MCF),

$$\Gamma(\boldsymbol{\rho}_1,\boldsymbol{\rho}_2,\tau) = \langle E(\boldsymbol{\rho}_1,t) E^*(\boldsymbol{\rho}_2,t+\tau) \rangle, \qquad (1)$$

where this is treated as an *ensemble* average over the realisations of the field. It is, of course, also possible to write higher-order correlation functions. To date, X-ray sources of relevance to this review have an essentially thermal, Gaussian, character and so the first-order mutual coherence function (eq1) completely defines the field [12]. It may, from time to time, be more efficacious to access experimentally measurements of higher-order correlations and such techniques will be explored later in the review.

An important concept for this review is the degree of coherence, which is essentially the normalised mutual coherence function [13]

$$\gamma^{(1)}(\boldsymbol{\rho}_1,\boldsymbol{\rho}_2,\tau) = \frac{\Gamma(\boldsymbol{\rho}_1,\boldsymbol{\rho}_2,\tau)}{\sqrt{\Gamma(\boldsymbol{\rho}_1,\boldsymbol{\rho}_1,0)\Gamma(\boldsymbol{\rho}_2,\boldsymbol{\rho}_2,0)}}. \qquad (2)$$

This is clearly a second order correlation function in terms of the fields, but is often termed the first order degree of coherence. It can also be conveniently written in the form

$$\gamma^{(1)}(\boldsymbol{\rho}_1,\boldsymbol{\rho}_2,\tau) = \frac{\langle E(\boldsymbol{\rho}_1,t) E^*(\boldsymbol{\rho}_2,t+\tau) \rangle}{\sqrt{\langle I(\boldsymbol{\rho}_1,t)\rangle \langle I(\boldsymbol{\rho}_2,t)\rangle}}. \qquad (3)$$

Textbook treatments introduce the ideas of coherence through the consideration of Young's two-pinhole experiments, where the coherence function describes the location and contrast of the interference fringes. In this interpretation, $\boldsymbol{\rho}_1$ and $\boldsymbol{\rho}_2$ denote the positions of the pinholes and $\tau$ describes the time delay between the arrival of the light from the two pinholes at the detector. A full exploration of the coherence function requires that the two pinholes each explore a two-dimensional surface, leading to an extremely demanding experiment requiring a four-dimensional data set.

As will be seen, there are a number of approaches to the measurement of phase and coherence that do not depend on interference. However the most conceptually simple of these do employ an observation of the contrast of interference fringes, implying high demands on mechanical and optical stability. For this reason, Hanbury-Brown and Twiss proposed a coherence measurement based on intensity correlations [14] that substantially eases the experimental requirements and which uses the fourth-order field correlations, or second order intensity correlations in the form

$$\gamma^{(2)}(\boldsymbol{\rho}_1,\boldsymbol{\rho}_2,\tau) = \frac{\langle I(\boldsymbol{\rho}_1,t)I(\boldsymbol{\rho}_2,t+\tau)\rangle}{\langle I(\boldsymbol{\rho}_1,t)\rangle\langle I(\boldsymbol{\rho}_2,t)\rangle}. \qquad (4)$$

This paper will be dealing entirely with thermal light in which the fluctuations in the electric fields have a Gaussian distribution. In this limit, the correlation functions at a given point are related by a simple expression,

$$\gamma^{(2)}(\boldsymbol{\rho},\boldsymbol{\rho},\tau) = 1 + \left|\gamma^{(1)}(\boldsymbol{\rho},\boldsymbol{\rho},\tau)\right|^2, \qquad (5)$$

known as the Siegert relation, enabling a connection to be drawn between the two correlation functions and allowing coherence measurements to be obtained using intensity correlation measurements. The applications of these ideas will be of most significance in the method of photon correlation spectroscopy, reviewed in section 7, but also apply to the measurement of the coherence properties of the X-ray field (section 3).

Naturally, all experimental systems ultimately rely on a measurement of the intensity distribution over a detector and, in the language of optical coherence theory, the intensity distribution is the self-correlation of the field:

$$I(\boldsymbol{\rho}) = \Gamma(\boldsymbol{\rho}, \boldsymbol{\rho}, 0). \tag{6}$$

This is ultimately the quantity that is to be compared with experiment.

To an experimental readership, it is perhaps more intuitive to consider the temporal correlations in the field in terms of its optical frequencies, in which case the cross-spectral density function may be defined

$$W(\boldsymbol{\rho}_1, \boldsymbol{\rho}_2, \omega) = \int \Gamma(\boldsymbol{\rho}_1, \boldsymbol{\rho}_2, \tau) \exp[i\omega\tau] d\tau. \tag{7}$$

The subject matter of this review is concerned primarily with relatively narrow bandwidth electromagnetic fields, a limit described via the so-called quasi-monochromatic approximation

$$\Gamma(\boldsymbol{\rho}_1, \boldsymbol{\rho}_2, \tau) \approx J(\boldsymbol{\rho}_1, \boldsymbol{\rho}_2) \exp[-i\omega_0 \tau], \tag{8}$$

where $\omega_0$ is the central angular frequency of the distribution and $J(\boldsymbol{\rho}_1, \boldsymbol{\rho}_2)$ is referred to as the mutual optical intensity (MOI). This approximation assumes that the electric fields in the wave are well approximated by a harmonic variation in time. In this case, one may write

$$W(\boldsymbol{\rho}_1, \boldsymbol{\rho}_2, \omega) \approx J(\boldsymbol{\rho}_1, \boldsymbol{\rho}_2) \delta(\omega - \omega_0). \tag{9}$$

That is, to sufficient precision, the field can be considered as consisting of a single optical frequency. Alternatively, one can regard this as meaning that deviation from perfect temporal coherence is small on the relevant spatial scales in the experiment, an assumption that is generally true for the experimental arrangements discussed here. Indeed, in much of this field, the experiments are designed so as to ensure that this is the case.

It is perhaps now apparent that a fully general description of the coherence properties of a field can be quite complex. In practice, the concept of a coherence length is often used. The spatial coherence length is the distance over which correlations in the field are reduced to some pre-determined level and really has only a strict meaning when applied to a known distribution of correlations. If it is assumed that the complex degree of coherence, eq2, for a quasi-monochromatic source has the form

$$\gamma^{(1)}\left(r_1 - r_2\right) = \exp\left[-\frac{|r_1 - r_2|^2}{\ell_c^2}\right] \quad (10)$$

then we use this as the implicit definition of the coherence length, $\ell_c$, a definition adopted for the remainder of this review. Note that the correlations are unchanged if the position variables are interchanged and so, in terms of this correlation function, the coherence length characterises the *separation* of the points at which the correlations have dropped to a value of $e^{-1}$. The definition provided by eq10 is convenient, but differs from the definition used in the visible optics regime, $\ell_c^{opt}$ in which it is defined as the point at which the correlations drop to a value of 0.88, consistent with the dip between two incoherent point sources that are just resolved by the Rayleigh criterion. The relationship between the two is therefore $\ell_c^{opt} = \ell_c \sqrt{-\ln(0.88)} \approx 0.36 \ell_c$. The definition adopted for visible optics is therefore, in a sense, rather more strict. A further common definition is to use the half-width at half maximum, $\ell_c^{HW}$. In this case, $\ell_c^{HW} = \ell_c\sqrt{\ln(2)} \approx 0.83\ell_c$, which is the point separation at which the correlation falls to half of its maximum value. The book by Attwood [15] is a good resource on this matter, as well as many other aspects of X-ray optics.

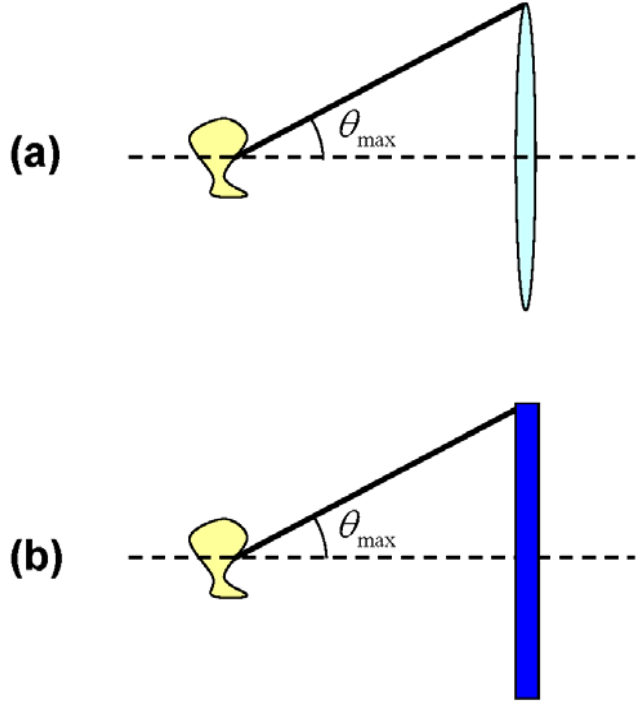

*Figure 2: (a) The numerical aperture for a lens based imaging system is defined as $NA = \sin\theta_{max}$, where $\theta_{max}$ is defined by the maximum angle at which light is collected, and it is assumed that the refractive imdex of the medium surrounding the lens does not deviate significantly from unity. For lensless imaging (b) one can sensibly define the numerical aperture as the maximum angle at which light can be detected or, more commonly, the maximum light at which light is detected and reliably assigned a phase. Note that, in the text, $s_{max} = \sin\theta_{max}$.*

In this review, we will have rather less concern with the concept of longitudinal coherence but it is nonetheless worth adopting a definition that is consistent with the definition for spatial coherence. Let us suppose that we have a wavefield that has a Gaussian distribution of power over optical frequency

$$S(\omega) = \exp\left[-\frac{\ln(2)}{4}\frac{(\omega-\omega_0)^2}{\Delta\omega^2}\right], \quad (11)$$

so that in this case $\Delta\omega$ is the full-width at half maximum of the frequency distribution. The temporal coherence length may be obtained by taking the Fourier transform of this, eq7, so that the temporal coherence function has the form

$$\gamma(\tau) = \exp[-i\omega_0\tau]\exp\left[-\frac{\Delta\omega^2}{\ln(2)}\tau^2\right]. \quad (12)$$

The corresponding longitudinal coherence length, then, is given by

$$\ell_c^{long} = \frac{\sqrt{\ln(2)}}{\Delta\omega} = \frac{\sqrt{\ln(2)}}{2\pi}\frac{\lambda^2}{\Delta\lambda}, \quad (13)$$

where we have also introduced an expression in terms of the wavelength distribution. As will become clear through this review, these definitions are at best broad measures of the degree of coherence in an X-ray field, but they are nonetheless useful quantities.

A second vital concept is the definition of resolution in coherent X-ray imaging. The question of the achievable resolution is the subject of considerable debate surrounding the reliability at which the maximum spatial frequency is reconstructed, a topic that will be considered in more detail in section 6. In optical imaging the definition most commonly adopted is the Rayleigh criterion, a criterion that is only strictly valid for incoherently illuminated objects. The Rayleigh criterion is defined in terms of the numerical aperture of the imaging system (see figure 2(a)) and does not consider the degree to which the object scatters the light. In lensless imaging, a technique to which the present review devotes considerable space (see section 6), the numerical aperture is defined by the detector (see figure 2(b)), but the consensus is that the simple fact of having a large detector does not warrant the claim of high resolution; rather one considers the resolution via the highest scattered angle at which the photons are both detected and reliably assigned a phase. This is, in turn, rather unsatisfactory as it prevents the assignment of a resolution to an object that is, in fact, genuinely relatively featureless, a problem that no longer concerns the microscopy community. However the concepts of contrast and resolution are rarely independent and the X-ray field is still seeking an agreed methodology by which resolution may be assigned to an image. We here consider the definition of resolution by applying the Rayleigh criterion for coherent imaging in the case where the effective numerical aperture of the system is defined by the maximum scattering angle at which the intensity is measured and reliably phased. We therefore also here bypass the inadequacy of the applicability of this definition to low contrast objects.

Let us suppose that we measure the diffraction pattern of some scattering object with a complex distribution $T(\mathbf{r})$. We suppose that far-field diffraction pattern of this object is measured and the phase recovered using some approach so as to enable the complex field, $\sigma(\mathbf{s})$, defined by

$$\sigma(\mathbf{s}) = \int_{-\infty}^{+\infty} T(\mathbf{r}) \exp[-ik\mathbf{s} \bullet \mathbf{r}] d\mathbf{r}, \quad (14)$$

to be measured out to some maximum scattered direction, $\mathbf{s}_{max}$ and $k = 2\pi/\lambda$. The reconstructed field can be described by

$$\sigma_{rec}(\mathbf{s}) = \sigma(\mathbf{s}) \Pi\left(\frac{s_x}{2s_{max}}\right) \Pi\left(\frac{s_y}{2s_{max}}\right) \quad (15)$$

where

$$\Pi(x) = \begin{cases} 1 & x \leq 0.5 \\ 0 & x > 0.5 \end{cases}, \qquad (16)$$

and we have implicitly assumed that the maximum frequency is measured out to a region in frequency spaced described by a square, a result that is consistent with the measurement of data to a resolution limited by the size of a square detector with an ideal modulation transfer function. Using the convolution theorem, it follows that

$$S_{rec}(x,y) = S(x,y) \otimes \left\{ \frac{\sin[k s_{max} x]}{k s_{max} x} \frac{\sin[k s_{max} y]}{k s_{max} y} \right\}. \qquad (17)$$

where $\otimes$ denotes the convolution operation. The Rayleigh criterion suggests that two points are to be considered resolved if the maximum of one point lies on the first zero of the point spread function of the neighbouring point. Eq17 suggests that the separation, $\Delta_{res}$, consistent with this criterion is given by

$$\Delta = \frac{1}{2} \frac{\lambda}{s_{max}}. \qquad (18)$$

In terms of diffraction angles, $s_{max} = \sin \theta_{max}$, where $\theta_{max}$ is the maximum angle to which scattering is observed as measured from the optical axis (figure 2), so we obtain $\Delta = \frac{\lambda}{2 \sin \theta_{max}}$ (for a circular aperture this becomes the more familiar expression $\Delta = 1.22 \left( \frac{\lambda}{2 \sin \theta_{max}} \right)$). Note that this adaptation of the Rayleigh criterion for coherent imaging is precisely the sampling ratio for the discrete Fourier transform and implies that the resolution from a fully illuminated array of pixel is given by the size of the pixel in the object plane, not by twice the pixel size as is sometimes assumed.

Note also that coherent imaging methods return the amplitude of the coherent field. This means that, unlike in optical and electron microscopy, one has a well defined amplitude spread function, a well-defined coherent spatial resolution and therefore one avoids the non-linearities arising from the conversion from amplitude to intensity that occur in microscopy with partially coherent illumination.

## 2.2 Partially coherent diffraction

The paraxial approximation has been adopted and so propagation over free space is described by the Fresnel diffraction integral. In this case a coordinate system may be used in which the three-dimensional position vector, $\boldsymbol{\rho}$, is explicitly written in terms of a two-dimensional vector, $\boldsymbol{r}$, and position along the chosen optical axis, $z$. That is, $\boldsymbol{\rho} = (\boldsymbol{r}, z)$.

Consider the electric field component of an electromagnetic field incident on a thin two-dimensional object with complex transmission function $T(\boldsymbol{r})$ and placed perpendicular to the incident beam. The electric field emerging from the object has the form $E(\boldsymbol{r})T(\boldsymbol{r})$ so, using eq1, the MOI of the emerging field has the form

$$J_{out}(\boldsymbol{r}_1, \boldsymbol{r}_2) = J_{in}(\boldsymbol{r}_1, \boldsymbol{r}_2) T(\boldsymbol{r}_1) T^*(\boldsymbol{r}_2) \tag{19}$$

This expression will be re-visited for three-dimensional objects shortly.

Propagation is viewed as the transformation of the field from a plane at $z_1$ to another plane at $z_2$. For simplicity, let us write $Z = z_2 - z_1$. This transformation is described by the expression

$$E(\boldsymbol{r}, z_2) = -i \frac{k}{2\pi} \frac{1}{Z} \int E(\boldsymbol{r}', z_1) \exp\left[\frac{ik}{2Z}|\boldsymbol{r}' - \boldsymbol{r}|^2\right] d\boldsymbol{r}', \tag{20}$$

where $k = 2\pi/\lambda$. Thus, the propagation function for the MOI can be written

$$J(\mathbf{r}_1,\mathbf{r}_2,z_2) = \frac{k^2}{4\pi^2 Z^2}\int J(\mathbf{r}'_1,\mathbf{r}'_2,z_1)\exp\left[\frac{ik}{2Z}\{|\mathbf{r}'_1-\mathbf{r}_1|^2-|\mathbf{r}'_2-\mathbf{r}_2|^2\}\right]d\mathbf{r}'_1 d\mathbf{r}'_2. \quad (21)$$

Using eq19 and eq21 MOI for the partially coherent diffraction by a thin two dimensional object is described by

$$J(\mathbf{r}_1,\mathbf{r}_2,Z) = \frac{k^2}{4\pi^2 Z^2}\int J_{inc}(\mathbf{r}'_1,\mathbf{r}'_2,0)T(\mathbf{r}'_1)T^*(\mathbf{r}'_2)\exp\left[\frac{ik}{2Z}\{|\mathbf{r}'_1-\mathbf{r}_1|^2-|\mathbf{r}'_2-\mathbf{r}_2|^2\}\right]d\mathbf{r}'_1 d\mathbf{r}'_2. \quad (22)$$

Importantly the measured intensity at a distance Z downstream from the scattering object is described by

$$I(\mathbf{r},Z) = \frac{k^2}{4\pi^2 Z^2}\int J_{inc}(\mathbf{r}'_1,\mathbf{r}'_2,0)T(\mathbf{r}'_1)T^*(\mathbf{r}'_2)\exp\left[\frac{ik}{2Z}\{r'^2_1-r'^2_2-2(\mathbf{r}'_1-\mathbf{r}'_2)\bullet\mathbf{r}\}\right]d\mathbf{r}'_1 d\mathbf{r}'_2. \quad (23)$$

A couple of other aspects of diffraction physics are needed before the complete context of partially-coherent diffraction may be formed. First, many experiments are conducted in the far-zone of the diffracted field, which is to say that all field curvature is considered to be negligible. The conditions for the far-zone approximation are well covered in many texts and so will not be reviewed here in any detail. The essence of this approximation is to write $\boldsymbol{\rho}_1 = \rho_1\mathbf{s}_1$ $\boldsymbol{\rho}_2 = \rho_2\mathbf{s}_2$, where $\rho_1 = |\boldsymbol{\rho}_1|$ $\rho_2 = |\boldsymbol{\rho}_2|$ and $\mathbf{s}_1$ and $\mathbf{s}_2$ are unit vectors pointing from the origin to the points at which the field is observed. If the observation point is sufficiently far away then propagation of the MOI may be written in the form

$$J_\infty(r_1\mathbf{s}_1,r_2\mathbf{s}_2) = \frac{1}{\rho_1\rho_2}\exp[ik(\rho_2-\rho_1)]\int J(\mathbf{r}'_1,\mathbf{r}'_2)\exp[-ik(\mathbf{s}_1\bullet\mathbf{r}'_1-\mathbf{s}_2\bullet\mathbf{r}'_2)]d\mathbf{r}'_1 d\mathbf{r}'_2. \quad (24)$$

For simplicity, this can be written

$$J_\infty(r_1\mathbf{s}_1,r_2\mathbf{s}_2) = \frac{1}{\rho_1\rho_2}\exp[ik(\rho_2-\rho_1)]L(\mathbf{s}_1,\mathbf{s}_2), \quad (25)$$

with the obvious definition for $L(s_1, s_2)$, known as the radiant cross-intensity [13]. This expression describes the partially coherent field in the far-zone as a four-dimensional Fourier transform of the MOI modulated by a spherical wave with a radius of curvature equal to the object-to-detector distance.

The final piece of theoretical groundwork that is needed is the Born approximation. There is an ongoing debate in the field at the moment about the precise role that coherent X-ray science can have, given the very advanced development of electron science, and the potential applications of electron microscopy and electron diffraction. There is no doubt, however, that electrons and X-rays will continue to be valuable and complementary probes and the matter of the benefits of each for high-resolution imaging will be clarified over the coming years. A key difference is the manner by which the probe interacts with the object, the electron seeing a Coulomb interaction and the photon an electromagnetic interaction. It is immediately clear that the electron will interact with matter far more strongly (something like four orders of magnitude) than the photon. It follows therefore that the object must see a far greater flux of energy quanta with an X-ray probe than with an electron probe for the same number of scattering events. However, X-rays have a place, as witnessed by their extensive applications in many fields, and one driver for this, as with medical imaging, is the ability to penetrate deep into an object. For the analysis of small objects, this need not be particularly advantageous, but a certain clear advantage is that the multiple scattering effects that plague and complicate electron science are much less dominant. As a result, the image analysis is far simpler and one can often confidently adopt the single scattering, or Born, approximation, or its more general and more widely applicable extension, the Rytov approximation [16, 17].

Consider a object with a three-dimensional scattering potential, $V(r,z)$. The scattered field is observed in the far-zone and it is assumed that the potential interacts only weakly with an incident field. The field is again written here in terms of the electric field and it is assumed that the interaction is so weak that it has a negligible effect on the incident field, so that the total field is described by

$$E_{tot}(r\mathbf{s}) \approx E_{inc}(r\mathbf{s}) + E_f(r\mathbf{s}), \tag{26}$$

where

$$E_f(r\mathbf{s}) = -ik^2 \frac{\exp[ik\rho]}{\rho} \int E_{inc}(r'\mathbf{s}) V(\mathbf{r}',z) \exp\left[-ik\left(\mathbf{s}\bullet\mathbf{r}' + z\sqrt{1-s^2}\right)\right] d\mathbf{r}'dz, \tag{27}$$

is the diffracted field and and $s = |\mathbf{s}|$. For small scattering angles this leads to a far-zone radiant cross-intensity for the scattered component of the radiation described by [13]

$$L(\mathbf{s}_1,\mathbf{s}_2) = \int J_{inc}(\mathbf{r}_1,\mathbf{r}_2) S_3(\mathbf{r}_1,\mathbf{r}_2) \exp\left[-ik(\mathbf{s}_2\bullet\mathbf{r}_2 - \mathbf{s}_1\bullet\mathbf{r}_1)\right] d\mathbf{r}_1 d\mathbf{r}_2 \tag{28}$$

where the scattering potential is defined as

$$S_3(\mathbf{r}_1,\mathbf{r}_2) = \int V(\mathbf{r}_1,z) V^*(\mathbf{r}_2,z) \exp\left[-i\frac{1}{2}kz(r_1^2 - r_2^2)\right] dz. \tag{29}$$

Finally, the far-zone intensity distribution is conveniently described as a function of angle via

$$I_f(\mathbf{s}) = \int J_{inc}(\mathbf{r}_1,\mathbf{r}_2) S_3(\mathbf{r}_1,\mathbf{r}_2) \exp\left[-ik\mathbf{s}\bullet(\mathbf{r}_2 - \mathbf{r}_1)\right] d\mathbf{r}_1 d\mathbf{r}_2 \tag{30}$$

Eq30 is a central expression for much of this review. A large part of coherent X-ray science is the extraction of structural information from a measurement of the scattered intensity described by eq30.

## 2.3 The projection approximation

The three-dimensional characteristics of the scattering object are encapsulated in eq30, and, in particular, in the complex exponential therein. The criterion that the three-dimensionality can be ignored is that the curvature in that exponential may be ignored; ie. $\exp\left[-i\frac{1}{2}kz\left(s_1^2 - s_2^2\right)\right] \approx 1$. The z dimension characterises thickness through the object and is assumed to have a maximum range of $\left[-\frac{T}{2}, +\frac{T}{2}\right]$, where T is the maximum thickness. The scattered photons are also scattered out to a maximum measured angle, limited either by the angular distribution of the scattering or the spatial extent of the detector. Let us again call the magnitude of the maximum scattering angle $s_{max}$. The scattering potential is Taylor expanded so to explore where the exponential in the integrand begins to deviate from unity and is therefore written in the form $\exp\left[-i\frac{1}{2}kz\left(s_1^2 - s_2^2\right)\right] \approx 1 - i\frac{1}{2}kz\left(s_1^2 - s_2^2\right)$. The requirement that the exponential in this integral not deviate significantly from unity, which is to say the effects of the three-dimensionality of the object can be neglected, is that $\frac{1}{2}kz\left(s_1^2 - s_2^2\right) \ll 1$ everywhere. On re-writing, it can be seen that this requires that $T \ll \frac{2}{ks_{max}^2}$. Much of the discussion in this paper concerns resolution and, if the Rayleigh criterion $\delta_{res} = \lambda/2s_{max}$ (eq18) is adopted, then the criterion for a "thin" object is that

$$T \ll \frac{4}{\pi}\frac{\delta_{res}^2}{\lambda} \qquad (31)$$

If the object obeys this condition then

$$S_3(r_1,r_2) \to \int V(r_1,z)V^*(r_2,z)dz, \qquad (32)$$

in which case the properties of the three-dimensional diffracting structure can be treated via a simple integral along the optical axis – a projection through the object. Note that when $V(r,z) \to T(r)\delta(z)$ one recovers the limit $S_3(r_1,r_2) \to T(r_1)T^*(r_2)$, as required. Hence, this condition is known as the projection approximation and is implicit in many treatments.

## 2.4 The Weak Object Approximation

There is one further approximation that is frequently adopted and which needs to be described. Imagine a coherent field incident on a three-dimensional object. In general, the field leaving the object – the exit surface wave – will have a well-defined amplitude and phase distribution. In the case of a weakly interacting object, for which the Born approximation holds, then this relationship is encapsulated in eq30. If both the Born- and the projection-approximations are adopted then the analysis can be simplified considerably, and in a manner that it often adopted in the literature. The complex transmission function is written in the form

$$T(r) = |T(r)|\exp[i\phi(r)] = \exp[-\eta(r)+i\phi(r)], \qquad (33)$$

where $\eta(r)$ describes the absorption of the object via $\eta(r) = -\ln[|T(r)|]$, and $\phi(r)$ is the phase shift imparted by the object.. The interactions are assumed to be sufficiently weak that both terms may be Taylor expanded in the form

$$T(r) \approx (1-\eta(r))(1+i\phi(r)) = 1-\eta(r)+i\phi(r), \qquad (34)$$

in which case

$$S_3(r_1,r_2) = 1-(\eta(r_1)+\eta(r_2))+i(\phi(r_1)-i\phi(r_2)). \qquad (35)$$

This is a simple linear form that connects the physical scattering object to the scattered wave in a form that is amenable to analytic study. However one must always be cautious as to whether the rather stringent assumptions that have been made are applicable to the problem at hand.

The pieces are now in place to enable to present a unified description of experiments that conform to the scheme shown in figure 1 for partially coherent illumination.

## 2.5 The Wigner Function

A number of studies of coherence effects in X-ray physics, and other aspects of science touched on in this review, have been based around the Wigner function [18]. An appreciation of much of the literature benefits from an understanding of this powerful theoretical tool.

The properties of the Wigner function have been extensively explored [19-21] in the context of partially coherent optics and these papers are valuable sources for its relevant mathematical and physical properties. As will be outlined in this section, the Wigner function has a very geometric interpretation which makes its application to short-wavelength optics particularly powerful [22-28], in part due to wave-effects being relatively minor. An alternative, but largely equivalent, formalism is based in the so-called ambiguity function [29], which is simply the two-dimensional Fourier transform of the Wigner function, and a number of studies [30-32] have used this form instead.

In the context of quantum mechanics, the Wigner function is regarded as a quasi-probability distribution that simultaneously provides the distribution of the light field in terms of position and momentum. As with quantum mechanics, it is not possible to measure the position and momentum of a light field simultaneously and so the Wigner

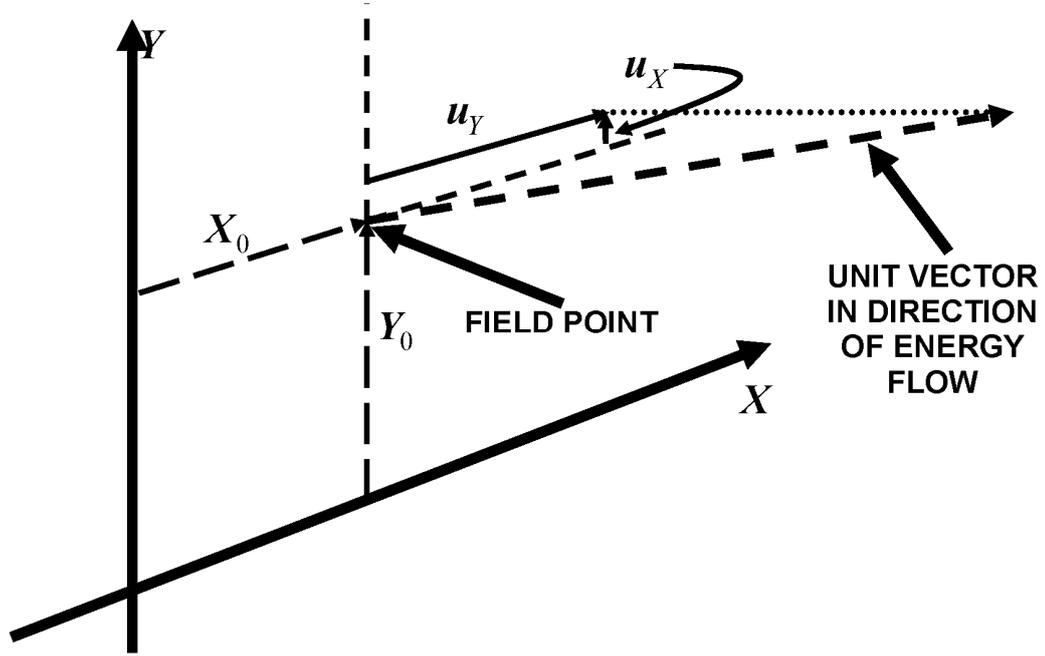

*Figure 3:* Schematic outlining the geometrical interpretation of the Wigner function for partially coherent wave propagation. The field at point $(X_0, Y_0)$ has a distribution of propagation directions, where the variables $(u_X, u_Y)$ indicate the projections of the unit vector in the direction of energy flow on to the plane

function cannot have all the properties required of a true probability distribution; in particular, it can assume negative values. Its value here, however, lies in its description of a light field in terms of its phase-space density, a language that naturally lends itself to the description of partially coherent fields.

The Wigner function of a quasi-monochromatic field can be written in terms of the mutual optical intensity in the form

$$B(r,u) = \int J\left(r+\frac{x}{2}, r-\frac{x}{2}\right) \exp[-ik u \bullet x] dx, \qquad (36)$$

where $r$ denotes position and $u$ is an angular variable defined via the k-vector by $k = ku$, where $k = 2\pi/\lambda$. In the context of a quantum mechanical interpretation, we can see that the momentum of the photon is given by $p = \hbar k$ so that the angular variable $u$ is related to the distribution of photon momentum via $p = k\hbar u$. If, as is done throughout this review, the paraxial approximation is adopted, and the standard formulae for the propagation of the partially coherent field from one plane to another are used, then one quickly obtains the following simple expressions for the transport of the Wigner function [19-21]

$$B_z(r,u) = B_0(r - zu, u), \tag{37}$$

where $B_z(r,u)$ is the Wigner function for the field over the plane located at a position z along the optical axis. Moreover, the intensity at a given plane is obtained using the simple integral:

$$I_z(r) = \int B_z(r,u) du. \tag{38}$$

where $I_z(r)$ is the intensity distribution over the plane located at a position z along the optical axis. These two expressions give rise to a simple physical picture for the propagation of light through free space. If the variable $u$ is interpreted as describing the unit vector of propagation of the light (see figure 3) then eq37 simply describes the geometric propagation of energy in straight lines. Similarly, eq38 informs us that the energy deposited at a point in space – the intensity – is simply the sum of energy striking that point over all the possible incident directions.

One must be wary of taking this interpretation too far as the non-positivity of the Wigner function would imply the presence of negative probability. However, used judiciously, it is a powerful way of thinking through the consequences of partial coherence in a given experimental system.

The Wigner function formulation describes the effects of diffraction through the manner in which light passing through a complex transmitting aperture is described. When the projection approximation is obeyed, a object can be considered thin with complex transmission $T(r)$, then the Wigner function of the complex transmission function

$$G_T(r,u) = \int T\left(r + \frac{x}{2}\right) T^*\left(r - \frac{x}{2}\right) \exp[-iku \bullet x] dx \qquad (39)$$

can introduced to describe the diffraction, so that the field leaving the object is described by

$$B_{out}(r,u) = \int B_{in}(r,u') G_T(r,u-u') du', \qquad (40)$$

which is a convolution over the variable $u$. This is a convolution over the angular (momentum) variable and so allows for the diffraction of light into a range of directions. As an example, a coherent plane wave incident on a small pinhole will diffract strongly into a range of directions.

A summary of the basic definitions for the theory of coherence is given in Table 1.

## 3 X-ray Coherence

### 3.1 Coherence measurement

This section is concerned with the measurement of the coherence properties of X-ray beams. For a source obeying Gaussian statistics, the coherence properties are determined by the first order mutual coherence function and so the temporal coherence properties are determined by the spectral properties of the field (see eq7) and so may be determined by a measurement of the spectral distribution. A precise measurement of the spectrum of an X-ray field has, of course, its own experimental challenges but these are beyond the scope of the present review and so will not be discussed further. Note that for the most part the concern here is with quasi-monochromatic distributions in which the effects of deviations from perfect temporal coherence are deemed to be negligible. However as the experimental techniques and sources develop, the measured scattering angles will get larger and so one might anticipate that, in the future, increasing attention will need to be paid to the issues of temporal coherence.

| Quantity | Mathematical Definition | Comments |
|---|---|---|
| Mutual Coherence Function | $\Gamma(r_1, r_2, \tau) = \langle E(r_1, t) E^*(r_2, t+\tau) \rangle$ | Fundamental and general starting point for a description of coherence |
| Degree of coherence | $\gamma^{(1)}(r_1, r_2, \tau) = \dfrac{\Gamma(r_1, r_2, \tau)}{\sqrt{\Gamma(r_1, r_1, 0)\Gamma(r_2, r_2, 0)}}$ | Useful for characterizing the coherence properties. |
| Cross Spectral Density Function | $W(r_1, r_2, \omega) = \int \Gamma(r_1, r_2, \tau) \exp[i\omega\tau] d\tau$ | A description of the field in terms of optical frequencies |
| Quasi-monochromatic approximation | $\Gamma(r_1, r_2, \tau) \approx \dfrac{1}{2\pi} J(r_1, r_2) \exp[i\omega_0 \tau]$ | A useful approximation where the field is essentially monochromatic |
| Mutual Optical Intensity | $J(r_1, r_2)$ | A description of a quasi-monochromatic field. |
| Radiant Cross-Intensity | $L(s_1, s_2) = \int J(r_1, r_2) \exp[-ik(s_2 \bullet r_2 - s_1 \bullet r_1)] dr_1 dr_2$ | A useful description of the field in the far-zone. |
| Far-zone intensity distribution | $I_{ff}(s) = \int J(r_1, r_2) \exp[-iks \bullet (r_2 - r_1)] dr_1 dr_2$ | The intensity in the far-zone, the experimentally significant quantity. |
| Wigner function | $B(r, u) = \int J\left(r + \dfrac{x}{2}, r - \dfrac{x}{2}\right) \exp[-iku \bullet x] dx$ | An intuitively simple formulation that offers a simple physical picture |

Table 1: Some convenient formulae for the application of partially coherent analysis to X-ray science.

The spatial coherence properties of sources are often characterised using the spatial coherence length of the field (section 2.1). The typical model experiment is to create fringes using a Young's two slit experiment and measure the visibility of the fringes as a function of slit separation [33, 34]. Typically the fringe visibility reduces as a function of slit separation and the coherence length is the slit separation for which the visibility drops to some agreed value, as discussed in section 2.1 we adopt here a value of $e^{-1}$. In the context of the Young's experiment, the observed fringes have a visibility and a phase – the MOI is therefore a complex function – and this quantity depends on the two-dimensional locations of the two pinholes over the plane of measurement. It follows that the MOI is a four-dimensional complex function; a complete measurement system must acknowledge this complexity and the quantity of information required to characterise it. Conversely, the MOI also carries a huge quantity of information if only it might be extracted.

The coherence length is a concept that reduces a complex function to a single number, a simplification that is valid for Gaussian isotropic coherence functions. The concept of the coherence length also tends to encourage a view in which points within the coherence length are fully correlated and those beyond it completely uncorrelated. While this may be a reasonable starting point, as will be seen in the section on coherent diffractive imaging, it is can also be a misleading oversimplification.

The complete mutual coherence function is a five-dimensional complex quantity and the mutual optical intensity is a four-dimensional quantity. As discussed in this section, the problem of reconstructing such a function with complete generality from experimental data is extremely difficult. As a result, the majority of coherence measurement approaches reduce the dimensionality of the inverse problem by adopting a model for the functional form of the coherence function. We here summarise some of the most important of these.

The coherent modes model [35] is described by,

$$J(r_1,r_2)=\sum_{n=1}^{N}\mu_n\psi_n(r_1)\psi_n^*(r_2), \qquad (41)$$

where the $\psi_n(r)$ are known as the coherent modes and are mutually incoherent, and the $\mu_n$ are positive real numbers that describe the occupancy in each mode. This model can offer good theoretical insight and also offers complete generality, but suffers from the drawback that the form of the coherent modes are not in general known and are difficult to recover from experimental data, though the recovery of the modes has recently been demonstrated [36].

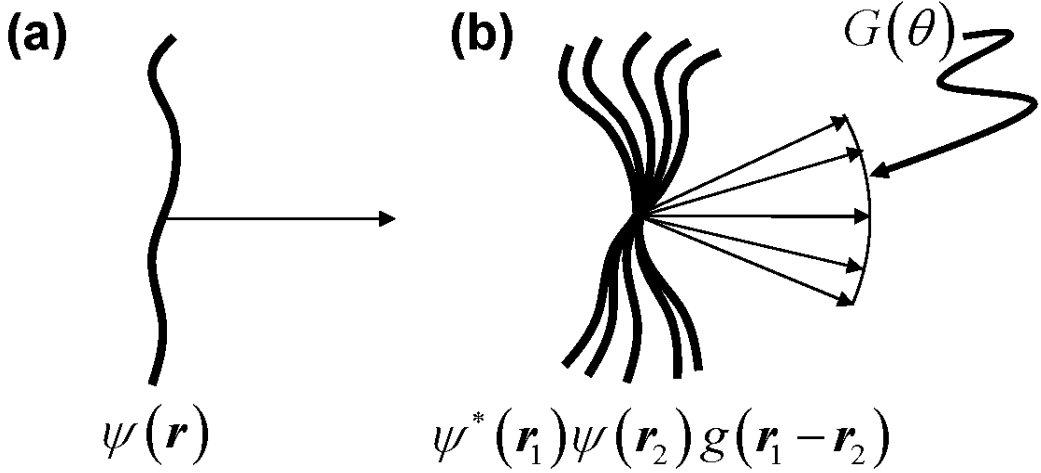

*Figure 4: Physical picture for understanding the physical meaning of the various models for coherence functions (a) A coherent field (b) A partially coherent field described by the generalised Schell model.*

The generalised Schell model [37]

$$J(r_1,r_2)=\psi(r_1)\psi^*(r_2)g(r_1-r_2) \qquad (42)$$

has been used for the examination of partially coherent X-ray diffraction [38] and is a generalisation of the model that historically carries Schell's name, the Schell model. As shown below, when the waves, $\psi(r)$, are spherical, the generalised Schell model has the form of the field described by the van Cittert-Zernike theorem [39] for an incoherent source.

| Coherence Model | Functional Form | Comments |
|---|---|---|
| Coherent Modes form | $W(r_1, r_2, \omega) = \sum_{n=1}^{N} \mu_n \Psi_n^*(r_1, \omega) \Psi_n(r_2, \omega)$ | Describes the coherence in terms of mutually incoherent fields. Theoretically useful and generally applicable, but the form of the modes is rarely known. |
| Generalised Schell Model | $J(r_1, r_2) = \psi^*(r_1)\psi(r_2)g(r_1 - r_2)$ | Useful model for calculating diffraction and for fitting of experimental data. Fields produced by incoherent sources have this form. |
| Schell Model | $J(r_1, r_2) = \sqrt{I(r_1)I(r_2)}\, g(r_1 - r_2)$ | Can be useful for fitting data, but less so than the generalized form |
| Quasi-homogeneous model | $J(r_1, r_2) = I\left(\frac{r_1 + r_2}{2}\right) g(r_1 - r_2)$ | A useful description for fields very close to an incoherent radiator. That is, for fields that are almost incoherent |
| Stationary Statistics Model | $J(r_1, r_2) = I_0 g(r_1 - r_2)$ | A simple form that is useful for analysis. Also seems to be largely consistent with the field produced by synchrotron sources. |
| | | |

*Table 2*: A summary of the most convenient functional forms of the coherence function. A simple physical interpretation of each of them is given in the text.

The Schell model,

$$J(r_1, r_2) = \sqrt{I(r_1)I(r_2)}\, g(r_1 - r_2), \tag{43}$$

has the form produced by an incoherent source in the limit of the far-zone and the statistically stationary model [39]

$$J(r_1, r_2) = I_0 g(r_1 - r_2), \tag{44}$$

is the simplest model of all and describes the limit of the Schell model in which the component fields are uniform and planar.

The quasi-homogenous model [40]

$$J(r_1, r_2) = I\left(\frac{r_1 + r_2}{2}\right) g(r_1 - r_2) \tag{45}$$

is a useful description of the field near a source that is almost completely incoherent, and so would, for example, be a good description of the field near the source within an undulator. A summary of the coherence models is shown in Table 2. Note that $g(\mathbf{0}) = 1$ in eqs43-45.

We may define the angular component of the partially coherent wave for many of these models via

$$A(\mathbf{u}) = \int g(\mathbf{x}) \exp[-ik\mathbf{u} \bullet \mathbf{x}] d\mathbf{x}, \tag{46}$$

where, as before, $\mathbf{u} = (u_x, u_y) = (\sin\theta_x, \sin\theta_y) \approx (\theta_x, \theta_y)$. The generalised Schell model consists of a series of mutually incoherent, identical waves travelling in a distribution of directions given by eq46. The Schell model has all of the component waves in generalised Schell model consisting of waves that are planar but contain amplitude variation. The statistically stationary model has all the component waves planar and uniform. The quasi-homogeneous model treats the source as a series of mutually incoherent point radiators each radiating into an angular distribution given by eq46. This physical picture for the above coherence models is outlined in Figure 4.

Suppose that we have an incoherent source described by $I(\mathbf{r})$. This has a quasi-homogeneous coherence function given by $J_0(r_1, r_2) = I\left(\frac{r_1 + r_2}{2}\right)\delta(r_1 - r_2)$, and a generalised radiance given by $B(\mathbf{r}, \mathbf{u}) = I(\mathbf{r})$. The field a distance z downstream is given by

$$B_z(\mathbf{r}, \mathbf{u}) = I(\mathbf{r} - z\mathbf{u}), \tag{47}$$

using eq37. The inverse of eq36 tells us that

$$J_z\left(\mathbf{r}+\frac{\mathbf{x}}{2}, \mathbf{r}-\frac{\mathbf{x}}{2}\right) = \left(\frac{k}{2\pi}\right)^2 \int B_z(\mathbf{r},\mathbf{u}) exp[-ik\mathbf{u}\bullet\mathbf{x}] d\mathbf{u}. \qquad (48)$$

Insertion of eq47 into eq48 gives, after some rearrangement

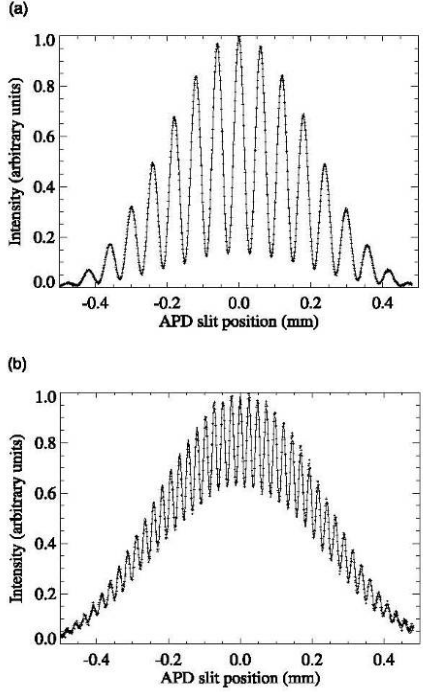

*Figure 5: Young's interference data obtained from an X-ray undulator for (a) a relatively high and (b) relatively low spatial coherence. The intensity is here plotted as a function of the position of an avalanche photo diode (APD) detector for two slit separations. The reduction in the fringe visibility for the larger slit separation (b) is obvious. The data was acquired at an X-ray energy of 2.1 keV at the 2-ID-B beamline at the Advanced Photon Source. Reprinted from Paterson et al. [33].*

$$J_z(\mathbf{r_1},\mathbf{r_2}) = \frac{1}{\lambda^2 z^2} exp\left[i\frac{k}{2z}(r_1^2 - r_2^2)\right] \int I(\mathbf{r'}) exp\left[-ik\frac{\mathbf{r'}}{z}\bullet(\mathbf{r_1}-\mathbf{r_2})\right] d\mathbf{r'}, \qquad (49)$$

which is the famous van Cittert-Zernike theorem [39]. Careful examination will also reveal that the mutual optical intensity in eq49 has the mathematical form described by the generalised Schell model, where $\psi(\mathbf{r}) = \frac{-i}{\lambda z} exp\left[ik\frac{|\mathbf{r}|^2}{2z}\right]$ is a spherical wave.

Interestingly, the physically intuitive expression, eq47, contains exactly the same physical content as eq49.

As a good rule for estimating coherence properties, consider the Gaussian intensity distribution with characteristic width D:

$$I(r) = I_0 \exp\left[-4\frac{|r|^2}{D^2}\right]. \qquad (50)$$

Substitution of eq50 into eq49 yields a coherence length given by

$$\ell_c = \frac{2}{\pi}\frac{\lambda z}{D}, \qquad (51)$$

a useful relationship relating source size, coherence length and distance of propagation. A range of approaches have been developed for the measurement of the spatial coherence. Kohn et al [41] have used a simple fitting of a known diffraction pattern with an assumed Gaussian statistically stationary MOI (eq44, where the function $g(r_1 - r_2)$ is a Gaussian) through the coherence length. Such measurements have been found to be in broad agreement with expectations based on an incoherent source within the synchrotron ring.

Other groups [33, 34] have used Young's experiments themselves, a two-beam interference experiment [42], dynamical diffraction [43] and the Talbot effect [44] (see section 4.4.2) to measure the fringe visibility as a function of fringe separation. The resulting fringe patterns (see Figure 5) are of very high quality and have yielded curves that generally conform to the Gaussian distribution assumed by Kohn et al [41]. The coherence properties of free-electron laser have also been reported recently [45] using a Young's two-slit experiment. These papers only sample one point in the four-dimensional space over which the mutual optical intensity function is defined (Figure 6) and so are rather incomplete as descriptions of the full coherence properties of the field.

A sufficiently large array of randomly distributed pinholes will have a very well-defined, sharp autocorrelation function and an image may be recovered from a diffraction pattern from such an array by cross-correlating it with the known pinhole distribution. This property of random scatterers was used by Sandy et al [46] and Abernathy et al [47] to measure coherence properties using scattering from an aerogel object. The use of a random scattering approach such as in these papers assumes that the coherence function depends only on the separation of the point in the field and is uniform in illumination –the statistically stationary model is assumed.

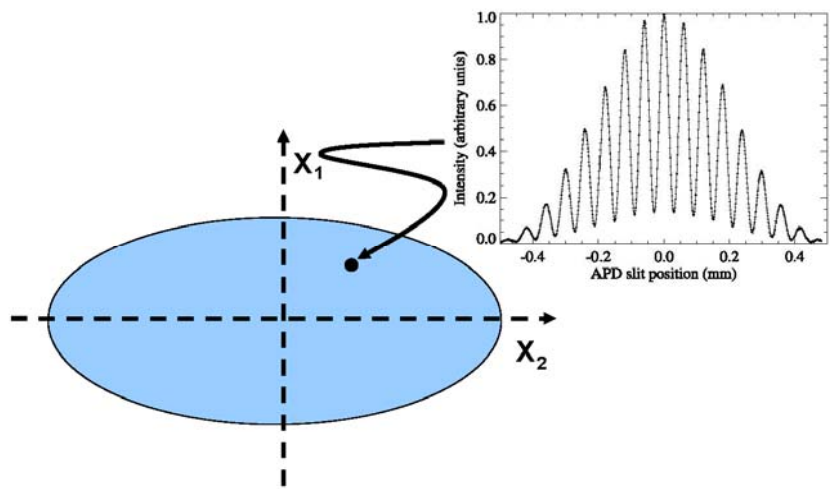

*Figure 6: Schematic indicating that the measurement of the visibility and phase of the fringes obtained from a Young's interference measurement only probe one point in a four-dimensional coherence function determined by the location of the two pinholes.*

The concept of a uniformly redundant array (URA) was developed for coded aperture imaging [48]. The URA contains a pinhole arrangement such that, on a discrete grid, all pinhole separations occur and they occur an equal number of times. The result is an array with an autocorrelation function possessing perfectly flat side lobes. From a coherence measurement perspective, the URA offers the possibility of performing many Young's experiments simultaneously. Nugent and Trebes [49] first proposed this approach to coherence measurement and used it to measure the coherence properties of a laser-pumped X-ray laser [50]. The method was adapted to measure the coherence function of an undulator source by Lin et al [51]. In this case, the coherence function could be recovered using a slightly more general model of the form The recovery of the coherence function from a URA was recovered using the generalised Schell model under the assumption that the incident component fields are spherical. The results for a synchrotron are consistent with a statistically stationary Gaussian distribution of correlations [51].

A coherence measurement that is wholly model-independent requires that a four-dimensional data set be acquired, an ideal that has yet to be achieved. One approach that enables the full function to be measured, at least in principle, is the method of phase-space tomography. This approach is based on the properties of the Wigner function outlined in section 2.5.

Consider the problem of the determination of the coherence properties of a partially coherent field at the plane $z=0$, and that one can measure the intensity of the field as it propagates through space. To perform this, the intensity is measured at a series of planes located at $z_k; k = 0,\ldots,N$. Using the methods of the Wigner function, the field at $z_j$ is described, via eq37, by

$$B(\mathbf{r},\mathbf{u},z_j) = B(\mathbf{r} - z_j\mathbf{u},\mathbf{u},0). \tag{52}$$

The intensity at this plane is, via eq38, described by

$$I(\mathbf{r},z_j) = \int_{-\infty}^{+\infty} B(\mathbf{r} - z_j\mathbf{u},\mathbf{u},0)d\mathbf{u}. \tag{53}$$

This is a projection in the tomographic sense, and a simple application of the Fourier projection theorem show that

$$\hat{I}(\boldsymbol{p}, z_j) = \mathcal{A}(\boldsymbol{p}, z_j \boldsymbol{p}, 0), \qquad (54)$$

where $\mathcal{A}$ is the two-dimensional Fourier transform of the Wigner function (ie. the ambiguity function [29]). Nugent [52] suggested that measurements over a complete range of z might allow the complete four-dimensional space of the ambiguity function to be covered and the complete coherence function therefore measured. Raymer et al [53] subsequently pointed out that a three-dimensional measurement of the intensity distribution is not, in general, enough to measure the complete coherence function - an observation that is related to the possible presence of phase vortices in the field, a subject considered in more detail in section 5.3. Raymer et al [53] proposed that the problem could be resolved through the introduction of symmetry breaking cylindrical lenses into the optical system. Three-dimensional intensity measurements are then required as a function of the orientation of a cylindrical lens. In this way, an arbitrary four-dimensional coherence function could be recovered, the cost being that a four-dimensional data set is also required; such a measurement overhead is so large that it is unlikely to be practical in the foreseeable future. A very similar approach has also been proposed in the visible optics area [54] in the explicit context of measuring the coherence function.

However the method of Nugent [52] does work for the measurement of a one-dimensional field with an associated two-dimensional coherence function. This was first demonstrated by Tran et al. [55] for the beam emerging from a slit illuminated by X-rays from an undulator source. The resulting reconstructed coherence function conformed within experimental error to a Gaussian statistically stationary form, and was consistent with an independent Young's experiment. This group went on to use the method to recover the Wigner distribution of the field diffracted by a Young's two slit experiment and, interestingly, succeeded in observing the regions of negative quasi-probability [56] produced by the effects of the interference, an effect previously only observed with quantum-mechanical fields [57].

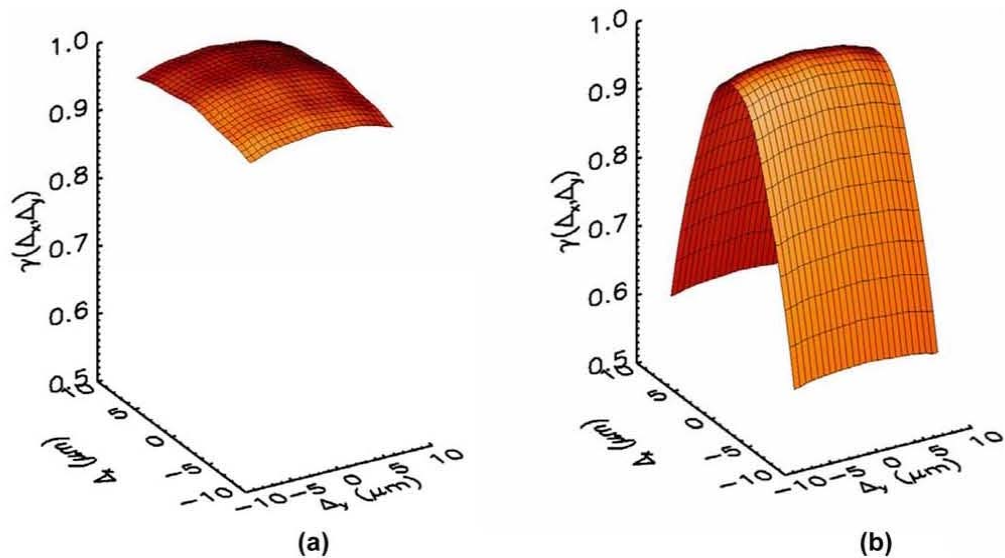

*Figure 7: Mutual optical intensity functions plotted as function of point separation for two sets of data with different spatial coherence lengths in the x (horizontal) direction. These are four-dimensional measurements that were found to be consistent with the statistically stationary form of the MOI. The data uses 2.1keV X-rays from the undulator beamline 2-ID-B at the Advanced Photon Source. Reprinted from ref [58].*

The extension to a general two-dimensional field has not been demonstrated but it can be shown that a separable field, by which is meant a field with a mutual optical intensity of the form

$$J(x_1, y_1, x_2, y_2) = J_x(x_1, x_2) J_y(y_1, y_2), \qquad (55)$$

can be recovered from the three-dimensional intensity distribution by treating it as the product of two one-dimensional fields. This analysis has been implemented for an X-ray beam emerging from a square aperture [58] and the results are shown in Figure 7. Again, the field is found to be very well described by a statistically stationary Gaussian distribution.

Tran et al [56, 59] have also made strong arguments that phase-space tomography has the potential to become a powerful imaging approach that is able to extract all of the enormous information contained in a partially coherent field, if only an experimentally tractable approach can be developed. For example, they showed how to detect and correct the action of such a lens and so correct the wavefront in software after the experiment [59]. Tran et al explore a range of related ideas [56, 59]. However, to re-iterate, such a goal requires that the complete coherence function be able to be acquired in a practicable manner, a problem that remains to be solved.

The first step to such a more flexible solution has been proposed by Rydberg et al [60] who have adapted ideas from iterative phase-retrieval methods, to be discussed in section 6.2, and suggested an efficient iterative approach that may make this possible. The algorithm proposed by this group is based on the coherent mode formulation of coherence theory [35], eq41, and iteratively seeks a coherence function that is consistent with a series of intensity measurements taken at different distances from the plane of interest. Although the simulation results look promising, this method has yet to be applied to experimental data. As a matter of principle, and on symmetry grounds (see section 5.3), phase vortices will prevent a truly unique solution.

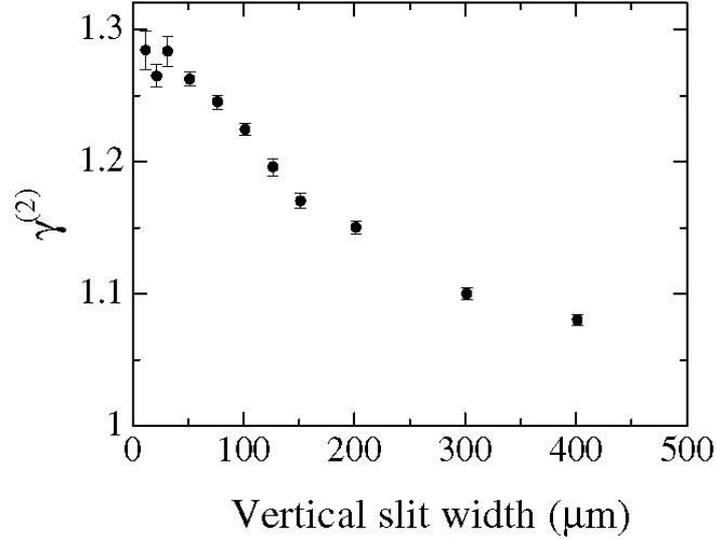

*Figure 8: Second order correlations extracted as a function of exit slit size from the 19LXU undulator beamline at the Spring8 facility using intensity interferometry methods. Note that the plot exceeds unity as required by the Siegert relationship, eq5. The data was obtained using X-rays with an energy of 14.41 keV. Reprinted from ref [64].*

Although based on the specialised assumption of separability, the paper by Tran el al [58] is the only complete measurement of X-ray spatial coherence that has yet been reported. The essential conclusion from these measurements so far is that the field emerging from the exit window of an undulator beamline is very well characterised by a Gaussian distribution obeying stationary statistics – an assumption that was made in the first and least sophisticated of the coherence measurement experiments. We will adopt this same assumption at a number of points further in this review.

Intensity correlation techniques have also been developed, relying on a time-correlation in the intensity measurements, eq4, and recovering the second order field correlations via the Siegert relation, eq5 (Figure 8). Early work was reported by Gluskin and collaborators [61-63], and also by Yabashi and colleagues [64-66]. In addition, Yabashi et al have also used intensity correlation techniques to measure the longitudinal coherence length [67]

The method of intensity correlation interferometry was initially developed for astronomy as it does not require the maintenance of highly stable optical path differences, and so allowed very long baseline interferometry. The cost is that the signal to noise ratio in the data is very low, requiring, for example, hours of data collection [65] to obtain a reliable result even on a third-generation synchrotron source. The signal to noise ratio problem has resulted in intensity interferometry being largely abandoned as a method for optical astronomy. Similar observations hold for X-ray science and so these intensity correlation measurements are interesting insofar as they demonstrate that they can be used to characterise the coherence of an X-ray source. However they have not been more successful or more accurate, and are considerably more difficult, than those that target a direct measurement of the first order correlation function.

## 3.2  Decoherence and coherence preservation

A review of the field of coherent X-ray science must deal with the concept of decoherence as it is a phenomenon that is discussed quite often in the experimental literature (see Refs [64, 65], for example).

As has been seen in the previous section, the quasi-monochromatic spatial coherence function – the mutual optical intensity – is related to the phase space density of the radiation field via a Fourier transformation (cf. eq36)). That is to say, the two physical descriptions carry identical information. The phase space density, as described by the Wigner function, is the Liouville invariant of the field and so cannot be compressed or expanded by any closed system. Strictly speaking, of course, an X-ray experiment is not a truly closed system. However external interactions with the field will be almost entirely through some time-varying contribution to the system such a vibrations, a time-varying optical element of a time-varying object (see section 7.1). In the absence of such effects, it is to be anticipated that coherence will not be degraded.

The observed coherence of the field can be viewed as the number of coherent modes that can be observed and while it is certainly possible to remove coherent modes to increase the coherence, albeit at the cost of photon flux, it is a physical impossibility to decrease the coherence of the field unless the optical system is subject to external influences. However, the field of coherent optics now places considerable importance on the requirement for "coherence preserving" optics. That is, it is observed experimentally that there is an effective loss of spatial coherence in the transport of the field from the source to the experimental chamber.

The phenomenon has been subject to some analysis and a model proposed by Robinson et al [68], and further developed by Vartanyants & Robinson [69], argues that objects in the beam can act as a secondary source of scatter which then superimposes incoherently on the scattered signal. In the case where the object is stationary (which excludes the dynamic effects used in X-ray photon correlation spectroscopy discussed in section 7), such a description violates Liouville's theorem and so cannot be complete.

Nugent et al [70] argued that every optical system is imperfect and so, when illuminated with coherent light, will produce coherent speckle. This is a familiar effect with laser optics and where it is impossible to create a surface that is so smooth that speckle is eliminated; any reflecting surface illuminated with coherent light will produce a diffraction pattern with a great deal of very finely structured speckle. In the case of X-rays, the short wavelength will ensure that the speckles are themselves very small and will in many cases be unresolved. The averaging process that arises from the unresolved speckles at the detector plane will be a *de-facto* ensemble averaging process that simulates the effects of loss of coherence. Nesterets [71] has used this model to compare theory with numerical simulation and obtained good agreement. As scattered light will contain a complex and fine-grained speckle pattern, any interference will usually be impossible to resolve and so this theoretical work leads to the conclusion that the fundamental assumption in the Robinson et al [68] and Vartanyants & Robinson [69] papers is empirically justified. Experiments in coherent X-ray optics will depend on the ongoing development of very high-quality X-ray optics that have the best possible surface quality.

## 4 X-ray phase contrast imaging

### 4.1 X-ray phase visualisation

This rapidly developing form of X-ray imaging began as a somewhat surprising observation from projection imaging using third-generation X-ray synchrotron sources. The first reported observations of phase-contrast in X-ray imaging were reported by Snigirev et al [72] and shortly thereafter by Cloetens et al [73], both at the European Synchrotron Radiation Facility (ESRF) (Figure 9). The experimenters observed the beam after it had propagated some distance downstream from a nominally smooth window and observed unexpectedly large intensity variations [74]. It was concluded that the observed intensity variations arose from thickness variations in the object resulting in refraction of the incident X-rays. This is precisely analogous to refraction of light by the atmosphere, causing stars to twinkle and degrading astronomical images.

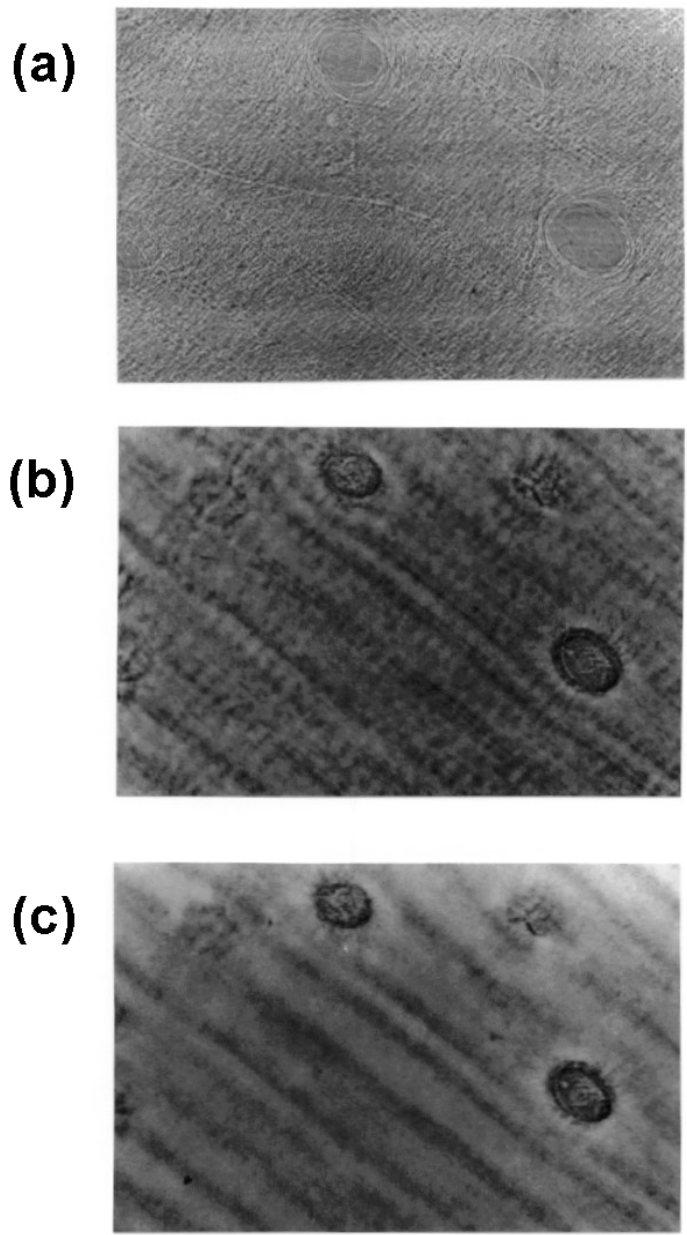

*Figure 9*: *X-ray phase contrast image of wood showing the development of image contrast on propagation. The propagation distances are (a) D = 0:05 m, (b) D = 0:5 m and (c) D = 1 m. The data were acquired from the D5 bending magnet beamline at the European Synchrotron Radiation Facility at an energy of 18.8 keV. From Cloetens et al. ref [73].*

The sensitivity to phase was not expected. That it was a surprise to the community may now, itself, be somewhat surprising. It has been anecdotally reported that phase-contrast effects can now be seen in many historical X-ray images. However these observations seem to have been universally ignored. With the development of more coherent X-ray sources, starting with the ESRF but rapidly following elsewhere, it became apparent that the effects of phase are far from negligible and are, in many cases, dominant.

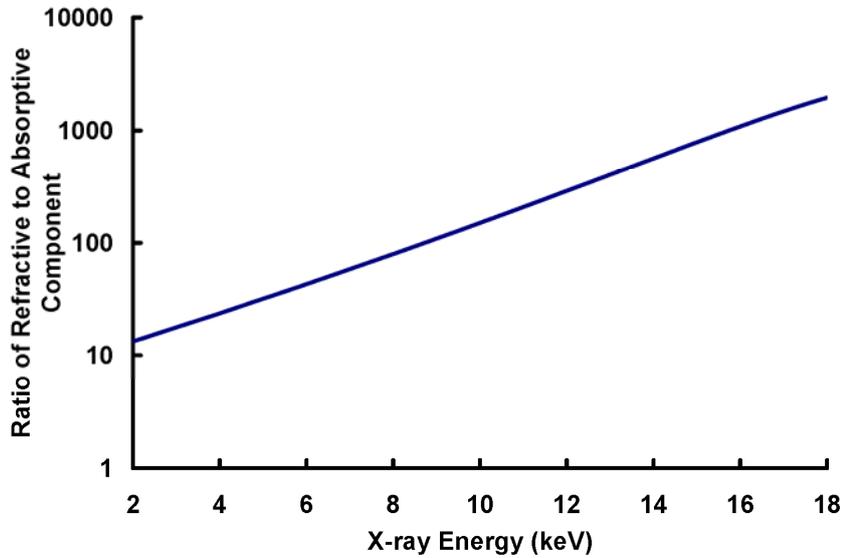

*Figure 10*: The ratio of the refractive to the absorptive parts of the refractive index of carbon as a function of X-ray energy. The plot was calculated using the website: http://henke.lbl.gov/optical_constants/

As with many developments in science, the precursors to the development are many, varied and, with the benefit of hindsight, clear. Perhaps the most important observation, now routinely repeated in the literature on X-ray phase contrast, arises from a simple examination of the interaction of X-rays with matter. The refractive index of X-rays passing through matter is generally written in the form

$$n = 1 - \delta + i\beta \qquad (56)$$

where $\delta$ and $\beta$ are wavelength dependent. Figure 10 shows a plot of the ratio of the refractive component, $\delta$, to the absorptive component, $\beta$, for carbon as a function of energy. It is obvious that the the absorptive component of the refractive index diminishes with energy must faster than does the refractive component. Exploiting the refractive contrast seems obvious. The reason that this contrast mechanism was not noted and exploited arises because the contrast due to refraction needs a reasonable degree of spatial coherence (see section 4.2) that was not available until the advent of third-generation synchrotrons. Nonetheless, shortly after the publication by Snigirev et al [72], Wilkins et al [75] showed that phase-contrast enhanced image contrast even using a conventional laboratory source. In this case, the predominant contribution to the image was still absorption but the phase contribution to the refractive index further enhanced the contrast. The paper by Wilkins et al [75] is the principle trigger for much of the recent work on laboratory-source phase contrast images and the ensuing interest in potential medical applications.

**4.2   Mathematical basis for propagation-based phase contrast**

Before reviewing the many interesting applications and results of from X-ray phase contrast imaging approach let us use the theoretical formalism in section 2.2 to understand the physical basis for the method. It will be assumed that the object obeys the projection approximation (section 2.3), a condition very well obeyed in all applications to date, and that the object interacts weakly with the object (see section 2.4), enabling the complex transmission function for the object to be written in the form given in eq34.

It is further assumed that the coherence properties of the incident field has the form of stationary statistics (section 3), and this is substituted into the expression for the intensity measured downstream from the object

$$I_z(r) = \left(\frac{k}{2\pi Z}\right)^2 \int \left\{ 1 - \left[\eta\left(r'+\frac{x}{2}\right) + \eta\left(r'-\frac{x}{2}\right)\right] + i\left[\Phi\left(r'+\frac{x}{2}\right) - \Phi\left(r'-\frac{x}{2}\right)\right]\right\} \times$$
$$g(x) \exp\left[ik\frac{(r'-r)\bullet x}{Z}\right] dr'dx \quad .(57)$$

where the functions $\eta(r)$ and $\Phi(r)$ are defined just prior to eq34, and the variable change

$$r' = \frac{1}{2}(r_1 + r_2) \quad x = r_1 - r_2 \quad (58)$$

has been introduced. This expression contains three terms, which can be written as

$$I_z(r) = I_0 + I_\eta(r) + I_\Phi(r), \quad (59)$$

where reference to eq57 will supply the appropriate definition, and the first multiplicative factor in eq57 is dropped from now on. The first term is the un-diffracted beam, something that is inevitable under the assumption of an object that interacts only weakly with the illuminating field; this component will not be further considered. The second term arises from the amplitude distribution in the wave transmitted by the object, and the third term describes the phase contribution to the image.

Eq57 leads to an expression for the amplitude contribution to the image,

$$\hat{I}_\eta(q) = -2g\left(-\frac{2\pi z}{k}q\right)\cos\left[\frac{2\pi^2}{k}z|q|^2\right]\hat{\eta}(q), \quad (60)$$

where $\hat{I}_\eta(q)$ is the Fourier transform of the amplitude component of the intensity distribution. Similarly the Fourier transform of the phase contribution to the intensity distribution is described by

$$\hat{I}_\Phi(\boldsymbol{q}) = 2g\left(-\frac{2\pi z}{k}\boldsymbol{q}\right)\sin\left[\frac{2\pi^2}{k}z|\boldsymbol{q}|^2\right]\hat{\Phi}(\boldsymbol{q}). \tag{61}$$

These expressions describe the essential properties underlying the method of phase contrast imaging.

Let us suppose that the coherence function has the Gaussian form

$$g(\boldsymbol{x}) = exp\left[-\frac{|\boldsymbol{x}|^2}{\ell_c^2}\right] \tag{62}$$

where $\ell_c$ is the spatial coherence length of the incident field; section 3.1 confirms that the assumption of Gaussian form has good experimental justification. In this case eq60 takes on the form

$$\hat{I}_\eta(\boldsymbol{q}) = 2\,exp\left[-\left(\frac{2\pi}{k}\right)^2\left(\frac{z}{\ell_c}\right)^2|\boldsymbol{q}|^2\right]\cos\left[\frac{2\pi^2}{k}z|\boldsymbol{q}|^2\right]\hat{\eta}(\boldsymbol{q}) \tag{63}$$

and the overall intensity distribution has the form

$$\hat{I}(\boldsymbol{q}) = \delta(\boldsymbol{q}) - 2\,exp\left[-\left(\frac{2\pi}{k}\right)^2\left(\frac{z}{\ell_c}\right)^2|\boldsymbol{q}|^2\right] \times \\ \left\{\cos\left[\frac{2\pi^2}{k}z|\boldsymbol{q}|^2\right]\hat{\eta}(\boldsymbol{q}) - \sin\left[\frac{2\pi^2}{k}z|\boldsymbol{q}|^2\right]\hat{\Phi}(\boldsymbol{q})\right\}. \tag{64}$$

This is essentially the expression obtained by Pogany et al [76], and is also related to the earlier work of Guigay [31, 77]. This expression captures some of the essential features of projection phase-contrast imaging.

Note that in the limit $z \to 0$ the effect of partial coherence disappears along with any contribution from the phase term. This is contact imaging which reveals a high resolution projection of the absorption distribution even when the illumination is completely incoherent; it does not reveal any phase information.

The phase sensitive term demonstrates that, for sufficient coherence, the phase contribution to the observed intensity will grow with propagation, a theoretical observation that is entirely in accord with the early observations from the ESRF [72, 73] (Figure 9). The key to the observation of phase sensitive effects, then, lies in the degree to which it is obscured by partial coherence.

Consider the imaging of an object up to a maximum spatial frequency $q_{max}$ and a small propagation distance such that

$$\frac{2\pi^2}{k} z |q_{max}|^2 \ll 1. \tag{65}$$

Then

$$\hat{I}_\eta(q) \approx \delta(q) - 2g\left(-\frac{2\pi}{k} zq\right)\left\{\hat{\eta}(q) - \frac{2\pi}{k} z|q|^2 \hat{\Phi}(q)\right\}. \tag{66}$$

An estimate of the effect of coherence on resolution might be found by asking at what frequency the phase information is maximised. That is, the phase contrast resolution limit is defined as the point at which $2\pi\lambda z|q|^2 g(-\lambda zq)$ is a maximum. Using eq65 and eq66, this frequency is given by

$$q_{max} = \frac{2\pi}{k} \frac{\ell_c}{z}. \tag{67}$$

The resolution for projection imaging without magnification will be limited by the pixel size of the detector. The maximum spatial frequency that can be measured is then the inverse of twice this pixel size. With a wavelength of 0.1nm, a propagation distance of 1m, and a detector pixel size of 10 microns, then the coherence length should be no less than 10 microns. In essence, it is this barrier that was crossed with third-generation sources and which allowed phase-contrast effects to be so strikingly revealed. The required coherence length decreases linearly as the propagation distance is reduced and so, for very short propagation distances, X-ray phase imaging is extremely forgiving of poor spatial coherence. Physically, this is because all the relevant optical path differences are very small. Limitations on the spatial resolution of the detector can be circumvented by illuminating the object with diverging beams, an approach that has been implemented with some success at the European Synchrotron Radiation Facility [78, 79]

Now consider the coherent limit $\ell_c \to \infty$ and consider a small propagation distance so that

$$\hat{I}_\eta(q) \approx \delta(q) - 2\hat{\eta}(q) - \frac{2\pi^2}{k} z |q|^2 \hat{\Phi}(q). \tag{68}$$

Inverse Fourier transformation then yields

$$I(r) = [1 - 2\eta(r)] - \frac{z}{k} \nabla^2 \Phi(r). \tag{69}$$

This describes an image of the absorption distribution with an additional contribution from the Laplacian of the phase distribution. Thus, after a small propagation distance one expects to see some additional detail arising around the edges of the object where the phase gradients are undergoing rapid variation [75].

The structure and optimisation of X-ray phase imaging, based largely around these ideas, was analysed in some detail in a useful papers by Pogany et al [76] and Zabler et al [80]. The idea of using propagation to reveal phase information has led to this method to be termed propagation-based phase imaging.

## 4.3 Applications of propagation-based phase imaging

Cloetens et al [81] used phase-contrast radiography and tomography to investigate microstructure and damage in materials. This paper also identified two regimes for phase-contrast imaging. The first they termed the "edge-detection" regime and identified it as that "where each border is imaged independently, but which does not allow the measurement of the local phase". This corresponds to the regime described by eq69 in which the phase contrast arises as the Laplacian of the phase – the phase curvature. As will be seen, the second phrase was, even at that stage, known to be overly limiting; it is possible to recover the phase from a measurement of its Laplacian, an aspect to be discussed in section 5.4. The second regime they termed the holographic regime, as that "where the image of the object is distorted, but which can give access to the phase when combining the images recorded at different distances with a suitable algorithm". The terminology is unfortunate as holography normally implies a role for a reference wave, something not necessary in this imaging modality. In reality, this regime is that described by eq64 and requires that the data be described by the Fresnel diffraction formalism.

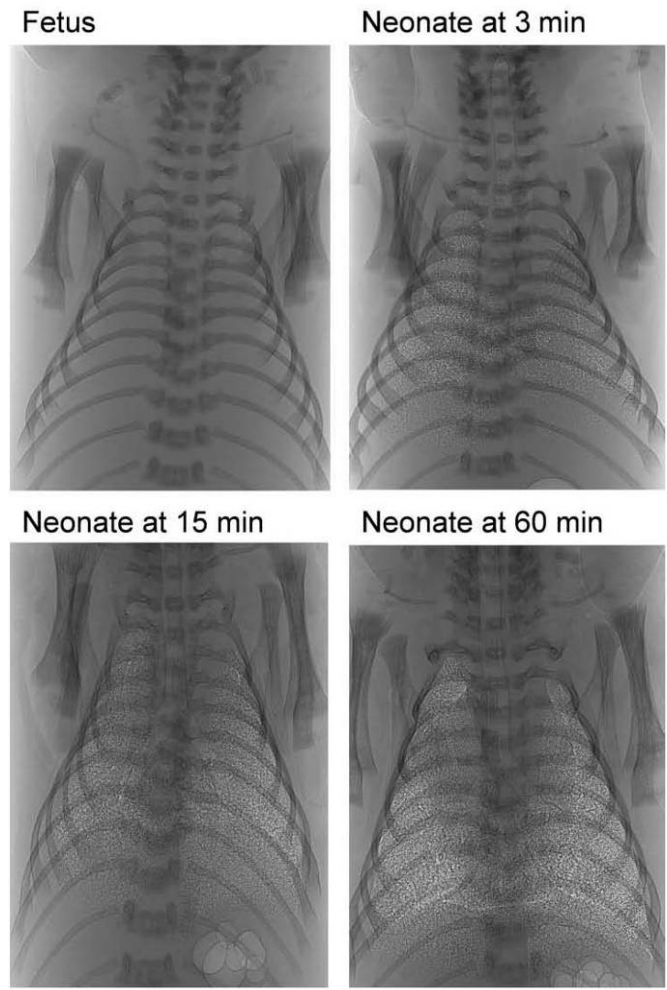

*Figure 11: Images of the flow of air into the lungs of a neonatal rabbit using X-ray phase contract imaging. The data were acquired at en energy of 25keV at the 20B2 beamline at Spring8. The panel headings indicate the age of the rabbit and the phase-contrast permits the visualisation of the air entering the animal's lungs during its first breaths. Image reproduced from ref [87].*

Much of the interest in X-ray phase contrast imaging has been stimulated by the possibility of being able to observe features that do not display sufficient absorption contrast and it is in this area that many of the applications of the ideas have developed; medical imaging has emerged as an area of particular importance. One of the first examples of the power of the method arose from the work of Spanne et al [82] who showed phase contrast of an artery specimen and discussed the potential for phase contrast in medical imaging, including the three-dimensional analysis of bone structures [83]. In 2000, publications by Arfelli et al [84, 85] demonstrated the applicability of propagation-based phase contrast to mammography. While one might hope for lower X-ray doses through the use of phase-contrast, the reality is that the physical properties of the interaction of X-rays with matter limits the degree to which this can be realised. It has been shown [85, 86] that phase-contrast can yield improvements in image quality and with a dose comparable to that of conventional mammography using a synchrotron source and with reduced dose compared to mammography using a conventional source. Some of these applications are considered further in section 4.4.

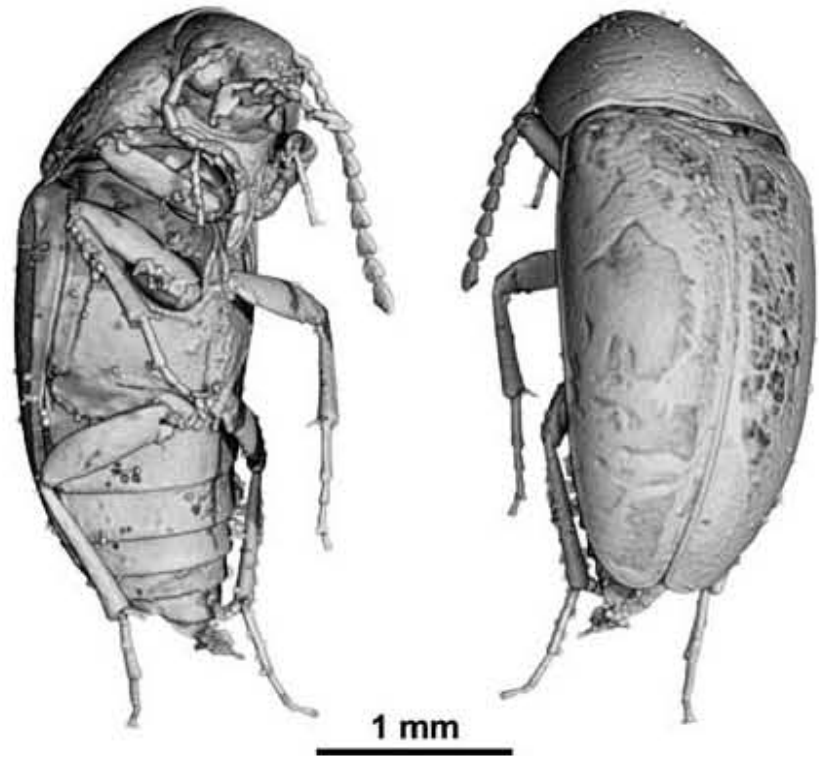

*Figure 12: X-ray phase contrast tomography is used here to non-destructively extract exquisite images of a beatle from the cretaceous era trapped in amber. The data was acquired at an energy of 20keV at the ID19 beamline at the European Synchrotron Radiation Facility. Reproduced from ref [120].*

Lewis and co-authors [87-90] have produced beautiful images of the aeration of foetal lungs, yielding insight into the processes by which fluid is cleared when an animal is born. The work has been the subject of a recent short review [91], and an example of the images acquired is shown in Figure 11.

The application of synchrotron radiation to radiology, including phase-contrast imaging, was reviewed by Meuli and colleagues [92, 93]. Other biological imaging applications have involved the imaging of fibroblasts [94], rat lung samples [95], tooth dentin [96] and soft tissue in cochlea samples [97].

Tomographic extensions have been reported by a number of workers for the investigation of materials [81], medical samples [82, 83], high-resolution imaging using zone plates [98], identification of materials via the complex refractive index [99]. Phase-contrast tomography has also been demonstrated using laboratory sources [100]. As these methods are often quantitative, they will be re-visited in section 5.4.1.9.

One of the most intriguing examples of the use of phase-contrast imaging has been in the visualisation of the physiology of living insects. This work has been pioneered by Westneat, Lee and collaborators [101-110] and X-ray phase-contrast has, as with the lung visualisation work of Lewis and collaborators, been demonstrated to be superb for visualising air-filled structures for small animals [101-107, 109-115], and also feeding structures in butterflies [116].

Another beautiful example is the application of X-ray phase contrast imaging to palaeontology, particularly to the study of plants and animals trapped in amber. In many cases the amber is opaque to visible light and so is not amenable to inspection using visible light. However the slight change in the composition between the amber and the trapped organism is sufficient to yield a measurable X-ray phase change. A object is placed in a synchrotron X-ray beam and the resulting phase-contrast image can reveal the presence of the organism within. X-ray tomographic techniques and image processing approaches may be used to extract a detailed three-dimensional image of the ancient organism [117-120], yielding morphological information that would only otherwise be available though destructive examination, (Figure 12).

X-ray phase contrast imaging was applied to non-biological materials by Lagomarsino et al [121] who observed and measured strains in buried interfaces using an X-ray waveguide optic. In 2002 a demonstration of the three-dimensional structure of paper was published [122] and they referred to the "edge contrast" imaging as "outline" imaging. In the same year the potential for performing phase-contrast tomography of integrated circuits was explored, again in "outline" imaging mode [123]. Kim et al [124] used phase-contrast radiology for the observation of crack propagation in sintered materials. X-ray phase contrast has also made contributions of significant technological importance to the understanding the properties of fuel sprays, including the visualisation of the growth of cavitation, the structure of nozzles and relating these factors to spray dynamics [125-127].

Phase-contrast imaging has been used to observe flowing substances in opaque objects, such as Teflon tubes [128], and for visualising blood flow [129] and also to characterise sail fabrics [130]. The characterisation of inertial-confinement fusion capsules using laboratory-based phase-contrast imaging has been demonstrated [131, 132].

## 4.4 Gratings, crystals and interference

### 4.4.1 *Interferometry*

Interferometer is the natural path towards phase measurement and in the X-ray region, interference generally relies on the use of crystal structures. The use of a conventional crystal-based interferometer, such as proposed by Bonse and Hart [133], is an obvious direction into which to head. A useful summary of the avenues pursued in this direction has been given by Momose [134].

Before discussing the results in some detail, we make some general comments on interferometric imaging. Interferometric data is acquired through the superposition of a known mutually coherent reference field onto the field of interest. For the sake of the argument let us write the reference field as being planar and of unit amplitude, $\psi_{ref}(r) = 1$, and we write the spatially non-uniform field to be determined in the form

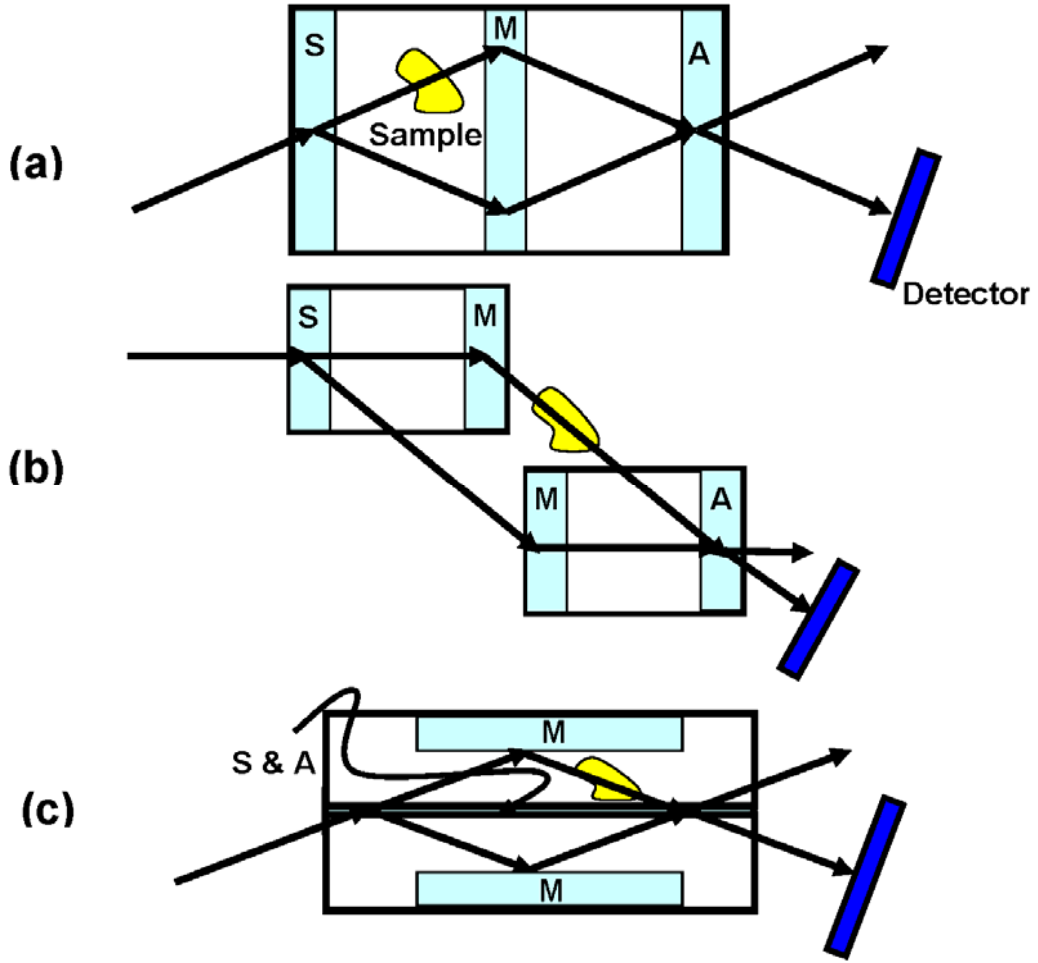

*Figure 13: Configurations used for X-ray interferometric imaging. (a) Monolithic LLL interferometer (b) interferometrr using two crystal blocks carrying two lamellae (c) BBB X-ray interferometer. S – beam splitter; M – mirror, and A – analyser. Figure and description adapted from Momose [134]*

$\sqrt{I_f(r)}\exp[i\phi_f(r)]$. We assume that the two fields are perfectly mutually coherent. The undergraduate formalism for interferometry tells us that the resulting interferogram has the form

$$I_{\text{int}}(r) = 1 + I_f(r) + 2\sqrt{I_f(r)}\cos[\phi_f(r)]. \tag{70}$$

A few moments consideration will reveal that this set of data does not yield an unambiguous reconstruction of the field – a variation in the phase can produce an identical effect on the data as a variation in the amplitude of the field. Note that most elementary treatments of interferometry ignore the possibility that the field may have a non-uniform amplitude distribution. Moreover, the phase is only determined to within some integer multiple of $2\pi$. The first problem is managed through the introduction of phase-shifting interferometry in which the phase of the reference beam is changed in a controlled and known manner so to produce a series of interferograms that can then used to solve for the amplitude and phase independently. Clearly, then, this requires the acquisition of a number of data sets. Another useful approach is to introduce a tilt to the reference beam, so it has the form $\psi_{ref}(r) = \exp[-k_{ref} \bullet r]$, and the interferogram has the form

$$I_{int}(r) = 1 + I_f(r) + 2\sqrt{I_f(r)}\cos[k_{ref} \bullet r + \phi_f(r)]. \tag{71}$$

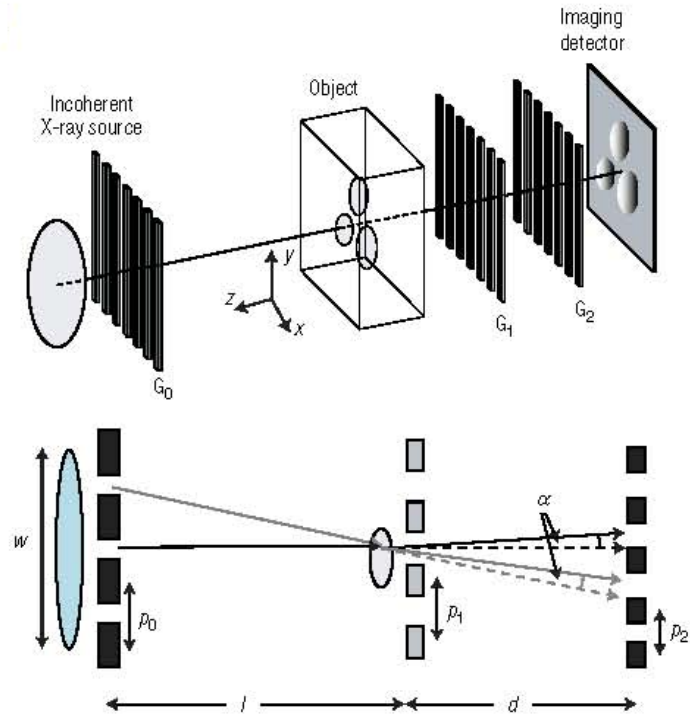

*Figure 14: Schematic of the configuration used for X-ray phase contrast imaging using Moire approaches. A source of width W illuminates a grating with pitch $p_0$, effectively breaking the source into a set of separate small sources, The X-rays pass through the object which deviates the rays by an angle $\alpha$. The rays pass through a second grating with pitch $p_1$ and the fringes produced by it are then projected onto a third grating with pitch $p_2$ to form the observed Moire fringes. Reproduced from reference [154]*

Under certain conditions, this can be solved unambiguously for the phase without the need for multiple sets of data [135]. Under certain more restrictive conditions, the amplitude may also be extracted from a single set of data [136].

The removal of the indeterminacy to within an additive integer multiple of $2\pi$ is the problem of phase-unwrapping, an area with a rich literature of its own (see [137] for a recent review) and will not be covered here.

X-ray interferometers for phase imaging have been developed in three broad forms, as indicated in figure 13. The first of these, the so-called LLL interferometer, as it relies on three Laue diffraction processes, is cut from a single crystal ingot giving considerable benefits to the stability of the alignment of the diffracting structures (lamellae), but the requirement to put the object within the interferometer limits the potential field-of-view. The second configuration using two independent crystal blocks allows a greater field of view but obviously places extremely strict requirements on the relative alignment. The third configuration [138, 139], using three Bragg diffraction processes, has some technical advantages [134] in terms of spatial resolution, but has not seen a great deal of application.

Interferometry-based X-ray phase-contrast imaging has seen application primarily in soft-tissue imaging for biomedical problems [140-143] but it was quickly realised that the primary benefits lie in the development of three-dimensional imaging approaches and this has been successfully demonstrated with the LLL interferometer [144] and a number of additional biomedical and soft-matter imaging problems have subsequently

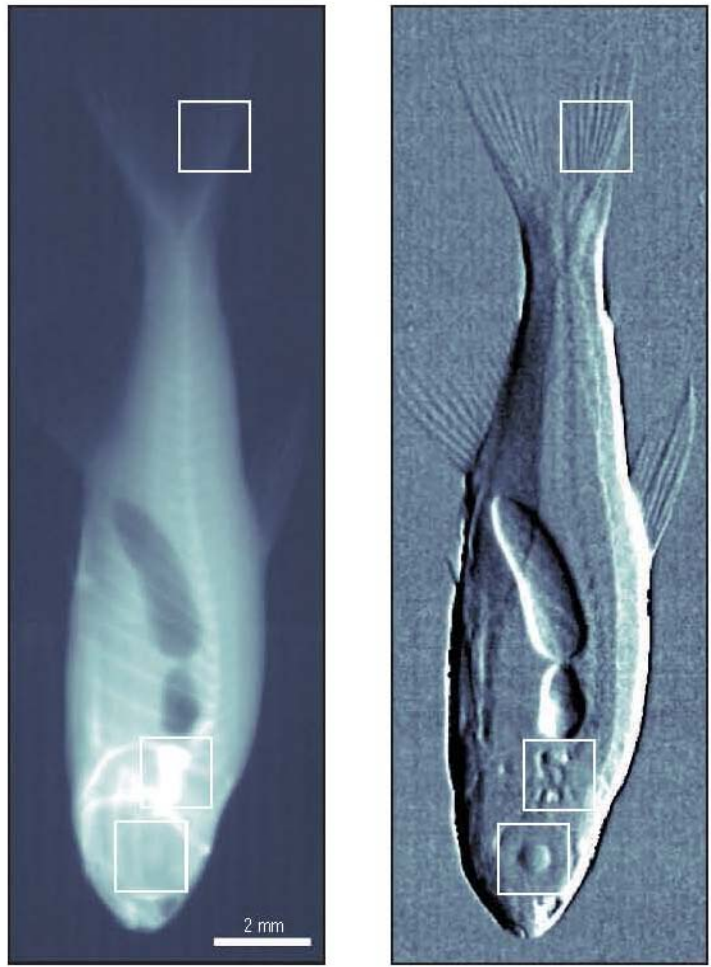

*Figure 15: Left panel shows A conventional radiograph of a small fish. The right panel shows Phase-contrast image of the fish using the More approach shown in Figure 11. Reproduced from ref [154]*

been addressed [145-149].

*4.4.2 Moiré approaches*

The operation of an LLL interferometer can also be viewed in terms of the familiar Moiré effect. The periodic crystal structure in the lamellae of an LLL interferometer are duplicated at particular planes in space through the Talbot effect [150]. (Note that in electron microscopy these image planes are known as the Fourier planes [151].) The overlapping of the Talbot image of one plane on the crystal structure of the second results in Moiré fringes, interpreted equivalently as interference fringes.

In a general sense, the interference fringes in an interferogram act as markers on the wavefront that enable its shape to be deduced. The need for coherence and stability, then, largely resides in the need to make these "marks" using interference. However one can also mark the wavefront by other methods, such as shadow-casting using a grating (see figure 14). If one indeed allows the shadow produced by one grating to project on another grating with comparable periodicity then clearly visible Moiré fringes will result. As with interference, the Moiré fringes are exquisitely sensitive to deviations in the wavefield, also in a manner strongly analogous to interference, but without the very strong requirements on coherence and stability [152]. These ideas have been used to develop very sensitive metrology tools in the visible optics regime (see [153], for example). A key difference in these non-coherent methods is that the contrast in the data arises from the phase gradient in the incident field, a theme that will re-emerge in section 5.

Moiré-based methods using the Talbot effect have more recently been developed and applied for X-ray and neutron phase-contrast methods. Pfeiffer and colleagues have used the ideas to enable imaging for X-rays [154, 155] and neutrons [156] with a particular emphasis on using the tolerance to very low levels of coherence to enable convenient imaging methods using laboratory X-ray [157] and reactor neutron [158] sources and they have extended these ideas to tomographic imaging [155, 158, 159]. A sample image from this method is shown in Figure 15.

*4.4.3 Diffraction enhanced imaging*

Much of the work is emerging in the context of the development of the broad class of techniques that are now known as diffraction enhanced imaging (DEI). This work has its origins in the work of Ingal and Beliaevskaya [160] who had been developing a form of phase contrast imaging in which crystals detect the deviation of the X-ray beam due to refraction by the object. The idea is to use diffraction from a crystal to create a highly collimated beam. This beam is then analysed by a second crystal which itself only can only accept a small range of incident angles which are then diffracted towards the detector. If an object is placed in the monochromatised beam then phase variations will introduce refraction of the incident beam out of or into the acceptance range of the analyzer crystal, producing an exquisitely sensitive X-ray phase imaging technique. Some years later, Davis et al [161] proposed a modification to this approach, and Chapman et al [162] developed yet another configuration and it is the last of these that has most promise and is now known as diffraction enhanced imaging (DEI). A significant amount of effort has been made in the direction of quantifying images [163-166] from this class of methods, but success has been very limited on experimental data. DEI is now a potentially important approach to high sensitivity medical imaging [85, 86, 162, 167]. An insightful analysis of method of this class based on Fourier methods has been published by Guigay et al [168]

# 5   Non-Interferometric Phase measurement

## 5.1   Introductory comments

The previous section explored the use of phase as a powerful visualisation tool in X-ray science. In this section a quantitative explanation of the phenomenon is developed, leading towards its use as a quantitative phase measurement technique.

The idea that the refractive effects explored in the previous section might be used to quantify phase measurements has a rich recent history in the optical sciences and so we turn to optics to provide a context for recent work in short-wavelength imaging methods.

The idea of phase is often considered from the perspective of the interference of a coherent wave. For work with X-rays, complete coherence is an unrealistic limit and so partial coherence is here an essential consideration. As a result, the concept of phase needs to be re-considered a little. At a fundamental level, phase is a property of a coherent field – it exists via the assumption that the wave has a well-defined periodic behaviour. In other words, phase, as defined through this formalism, does not exist for a partially coherent field and one must resort to partially coherent formulations that make no explicit reference to phase. In practice, however, phase imaging is not concerned with the phase of the field; its aim is to extract information about the refractive properties of the object on the wavefield, a property that is entirely independent of the incident field.

The interferometric heuristic for phase measurement has its roots in the interferometric work of Michelson [169]. Michelson showed that interferometry is a powerful tool for application to optical measurement and the direct measurement of phase can yield remarkably precise measurements. Prior to the development of the laser the application of interferometric ideas was limited. Gabor [170] recognised that interferometric encoding of phase had the potential to improve the resolution of electron microscopy through his development of holography. Zernike [171] subsequently developed his method of phase contrast microscopy to visualise phase through interference over very short path differences, with Nomarski [172] proposing a related method now known as differential interference microscopy, another method understood through the mechanism of interference. With the development of the laser, and very much longer coherence lengths, interferometry and holography became much easier to implement and their use became much more widespread. However interferometric microscopy has not been widely implemented, though some methods have been coming to the fore recently [173].

The ideas for non-interferometric phase measurement have origins in astronomical imaging and are based on the idea of visualising phase gradients and phase curvature. A major limitation on astronomical imaging has been the effect of turbulence on the light entering the atmosphere [174]. Density variations in the atmosphere introduce phase gradients which refract the light so as to aberrate wavefronts entering a telescope. This is identical to the effect used in propagation-based X-ray phase contrast imaging [72, 73]. The astronomical community also noted that the intensity contrast arises from phase curvature and that the resulting contrast can be used to recover the phase structure [175-177]. A method for sensing the phase gradient as a function of position can, via an appropriate integration, be used to deduce the phase of the wave as a function of position. This is the principle of the Hartman sensor (see [178]) a method that uses a set of pinholes, with the image of each pinhole moving by an amount proportional to the phase gradient at the pinhole, allowing the phase gradients to be mapped. The Hartman sensor (and its descendent, the Hartman-Shack sensor in which the pinholes are replaced with small lenses) is widely used in astronomy and to sense refractive errors in ophthalmology [179].

Similarly, the electron community has developed exquisite approaches to electron holography and has used them to explore aspects of the interaction of electrons with magnetic structures [180]. In parallel they have shown that a series of defocused images can be used to recover the phase distribution exiting an object in high resolution transmission electron microscopy [181]. This latter method is another example of recovering phase without the need for interferometric measurements.

The idea of using propagation to *measure* phase directly was arguably first proposed by Teague [182, 183]. In these papers he first pointed out that one could use the conservation of energy on propagation to write a differential equation for the transport of energy by an optical field, and that there was a possibility that it could be used as an approach to phase-recovery. He termed this equation "the transport of intensity equation" and this approach will be discussed in more detail in section 5.4.1.2. In a subsequent paper, Teague [183] proposed an approach to the solution of the equation using Green's functions. This approach was cumbersome, but did demonstrate feasibility. Shortly thereafter, a brief examination of the possibility of applying this approach to optical microscopy was published by Streibl et al [184], though this paper also stopped well short of demonstrating phase recovery.

In parallel, the adaptive optics community was exploring how to recover phase from optical intensity measurements so as to correct for the effects of phase distortions in the atmosphere in real time. The group let by Roddier et al were developing a technique known as curvature sensing [176, 177], and built on the observation that the entrance pupil of a telescope, if re-imaged and then defocused slightly, would display intensity contrast that was proportional the curvature of the wavefront, as described by the Laplacian of the phase distribution (see eq69), and that this signal could be used to feed back into an adaptive optic to correct the phase-error induced by the atmospheric turbulence. Roddier et al [176] also drew on transport-of-intensity equation [182]

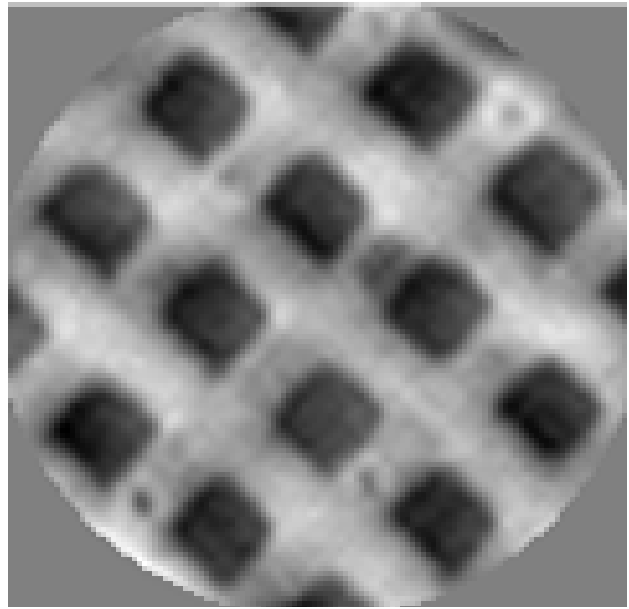

*Figure 16: A quantitative X-ray phase measurement on a thin carbon grid obtained using propagation-based X-ray phase contrast and 16 keV X-rays. The data was obtained at beamline 20A at the Photon Factory. Reprinted from ref [185].*

related ideas.

Teague's transport of intensity equation approach was re-introduced by Gureyev et al [27] with a view to performing quantitative X-ray phase measurement. In the first paper by this group [27] they showed that the solution to the transport of intensity was unique subject to certain conditions, which they connected to the presence of optical vortex structures – structures that contain a helical wave front (see section 5.3). This group was possibly also the first to recognize that non-interferometric phase recovery techniques could be applied in the field of X-ray optics, in this case primarily for the quantitative characterization of optical elements. The paper reporting X-ray phase contrast by Snigirev and colleagues appeared almost simultaneously [72]. Shortly thereafter Nugent and colleagues [185] showed that the effect reported by Snigirev et al [72] could be used to make quantitative phase measurements using the transport of intensity equation, and reported a quantitative phase image of the thin carbon foil using 16keV X-rays (see figure 16). Shortly thereafter, Cloetens et al showed that ideas from phase-recovery in transmission electron microscopy should also be applied successfully to quantitatively recover the phase distribution in an image by using a series of images at different distances [186]. They further showed that three-dimensional images can be obtained using such techniques, a method they term holotomography [186], and involves the acquisition of data over a series of planes (see section 5.4.1.7) for a range of object orientations.

## 5.2 A Poynting Vector Picture for Phase

The idea of phase is typically introduced using the idea of interference. One imagines two completely coherent waves superimposed on each other and then observes the interference pattern to deduce the phase of one of the waves with respect to the other. This picture requires the concept of interference and implies the need for coherence so as to permit the observation of the interference pattern.

Phase-contrast microscopy uses some form of interference to enable the observation of the phase shifts induced on the incident light by a transparent or semi-transparent object. In this case, the phase shift is a representation of a property of the object and so does not depend on any property, such as the coherence, of the incident field. To show interference is not necessary to understand phase and its measurement, Paganin and Nugent [187] developed a more generally applicable theoretical formalism that is now described.

The Poynting vector describes the direction and magnitude of the energy flow in the wave. The non-paraxial formulation is used here in the first instance and so the three-dimensional position vector, $\rho$, is used. In the case of coherent light with intensity $I(\rho)$, phase $\Phi(\rho)$ and wavenumber $k$, the Poynting vector is given by

$$S(\rho) = \frac{1}{k} I(\rho) \nabla \Phi(\rho). \tag{72}$$

The Poynting vector is a well-defined concept in electromagnetic propagation. The conservation of energy in the propagating field is encapsulated in the requirement that

$$\nabla \bullet S(\rho) = 0. \tag{73}$$

In the case of partially coherent light the Poynting vector will fluctuate rapidly in magnitude and direction and any measurement will only be sensitive to its average value, $\langle S(r) \rangle$, where $\langle \ \rangle$ denotes a time average over a period much longer than the coherence time. The average Poynting vector is perfectly well defined and the corresponding partially-coherent version of eq73 remains true.

Paganin & Nugent inverted the meaning of eq73 and, instead, used it to define the phase. That is, for quasi-monochromatic light the phase is defined via the expression

$$\langle S(\rho) \rangle = \frac{1}{k} I(\rho) \nabla \Phi(\rho). \tag{74}$$

Paganin and Nugent define the phase gradient, then, via the expression

$$\nabla \Phi(\rho) = k \lim_{\varepsilon \to 0^+} \frac{\langle S(\rho) \rangle}{I(\rho) + \varepsilon} \tag{75}$$

and it can be explicitly seen that this phase is also not properly defined at points of zero intensity. Furthermore, as an arbitrary vector field, the (average) Poynting vector may contain vorticity so that, in analogy with the scalar and vector potentials of electromagnetic theory, it can be written

$$\langle \mathbf{S}(\boldsymbol{\rho}) \rangle = I(\boldsymbol{\rho}) \left[ \nabla \Phi_S(\boldsymbol{\rho}) + \nabla \times \boldsymbol{\Phi}_V(\boldsymbol{\rho}) \right], \tag{76}$$

where $\Phi_S$ and $\boldsymbol{\Phi}_V$ are appropriately defined quantities that have been termed the scalar and vector phase components [187] respectively. In order to remove any ambiguity in the definitions, it is required that $\nabla \times \nabla \Phi_S(\boldsymbol{\rho}) = 0$ and $\nabla \cdot \nabla \times \boldsymbol{\Phi}_V(\boldsymbol{\rho}) = 0$. Thought of in this way, the phase can be regarded as having a vector as well as a scalar component. The scalar component is the familiar idea encountered in the undergraduate curriculum. The vector component, as with electromagnetism and fluid flow, can be associated with vorticity, or angular momentum, in the field. That electromagnetic waves can carry orbital angular momentum is now well-established, it has been shown that it is possible to transfer the angular momentum to trapped particles in the case of visible light [188], and singular optics is a well-established area of study at visible wavelengths.

One can invert eq76 to obtain explicit definitions for the two phase components, specifically

$$\phi_s(\boldsymbol{\rho}) = -\frac{1}{4\pi} \int \frac{\nabla^2 \Phi(\boldsymbol{\rho}')}{|\boldsymbol{\rho} - \boldsymbol{\rho}'|} d\boldsymbol{\rho}' \tag{77}$$

and

$$\boldsymbol{\Phi}_V(\boldsymbol{\rho}) = \frac{1}{4\pi} \int \frac{\nabla \times \nabla \Phi(\boldsymbol{\rho}')}{|\boldsymbol{\rho} - \boldsymbol{\rho}'|} d\boldsymbol{\rho}' \tag{78}$$

Careful consideration of eq78 will reveal that the vector phase is only non zero if the field contains a phase discontinuity, giving rise to the necessary presence of phase singularities in the field. It is also significant that the vector phase component is divergence-free and so cannot be observed via the simple propagation of energy (c/-eq73).

An important aspect of associating phase with energy flow is to demand that energy be conserved on propagation through free-space, eq73. If the field is written in its coherent form

$$E(\boldsymbol{\rho}) = \sqrt{I(\boldsymbol{\rho})}\exp\left[i\phi(\boldsymbol{\rho})\right] \tag{79}$$

and the paraxial approximation is re-introduced, eq79 can be cast in the form

$$E(\boldsymbol{r},z) \approx \sqrt{I(\boldsymbol{r},z)}\exp\left[i\phi(\boldsymbol{r})\right]\exp\left[ikz\right], \tag{80}$$

so that the Poynting vector at a plane z has the form

$$S(\boldsymbol{r},z) = \frac{1}{k}I(\boldsymbol{r},z)\{\nabla\phi(\boldsymbol{r}) + k\hat{z}\} \tag{81}$$

where $\hat{z}$ is the unit vector along the optical axis and the conservation of energy expression, eq73, assumes the form

$$k\frac{\partial I(\boldsymbol{r},z)}{\partial z} = -\nabla \bullet \left[I(\boldsymbol{r},z)\nabla\phi(\boldsymbol{r})\right]. \tag{82}$$

This is the transport of intensity equation. A measurement of the intensity $I(\boldsymbol{r},z)$ and its derivative along the optical axis, $\partial I(\boldsymbol{r},z)/\partial z$, will allow eq82 to be solved for the phase distribution, $\phi(\boldsymbol{r})$.

As will be seen in the next section, there are a number of approaches that will permit the extraction of quantitative phase information from intensity measurements. Many of these techniques have yet to be properly analysed in terms of their capacity to yield unique phase results; the transport of intensity equation is the one method that has been exhaustively analysed.

## 5.3 Phase vortices

Undergraduate texts generally treat phase as a smooth well-behaved continuous function. In practice the properties of phase are not this simple and the phase distribution can be discontinuous. Indeed, as has been shown by Berry and colleagues [189], one can generate a rather unified picture of the properties of waves through a consideration of the phase dislocations within them. The dislocations can be classified into edge and screw dislocations, in a precisely analogous manner to the defects observed within crystals.

Dislocations are lines or points at which the phase is not defined, and in the case of a point dislocation, the phase accumulates a multiple of $2\pi$ around the point dislocation. The intensity vanishes at the dislocation. Generically, the amplitude of an optical vortex has the form

$$\psi_{vortex}(r,\phi) = A(r,\phi)\exp[im\phi], \qquad (83)$$

where $m$ is termed the topological charge, a quantity that continuity requires be an integer. Optical vortices are characterised by their topological charge, being the number of wavelengths of phase that is accumulated in circulating the intensity zero in the vortex structure. An alternative description is that an optical vortex structure carries a quantity $m\hbar$ of orbital angular momentum.

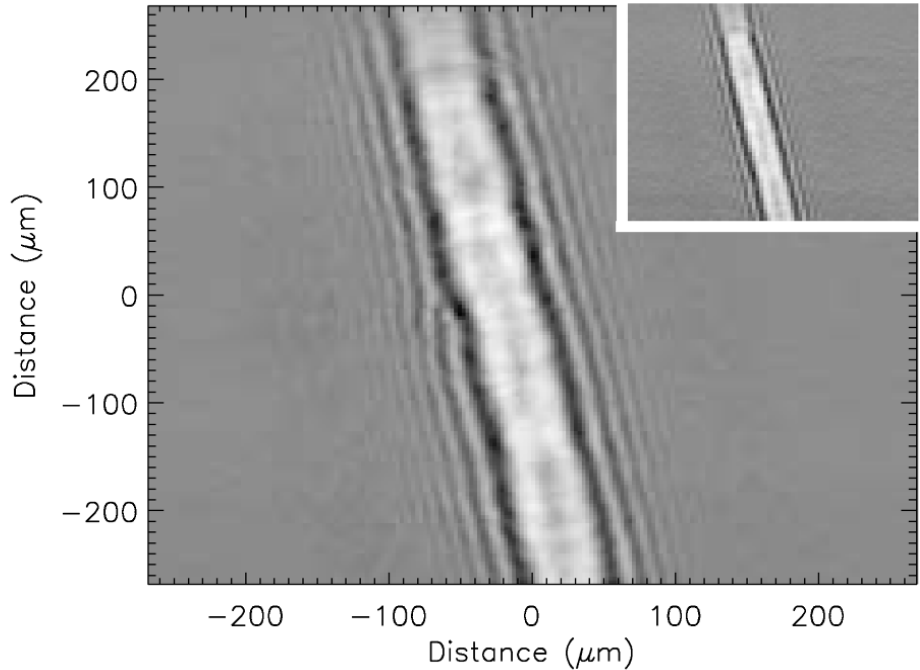

*Figure 17: Interferogram showing the forked fringe characteristic of the presence of an optical phase vortex. Inset show the data with the vortex absent. Data acquired with 9keV X-rays at the 2-ID-D beamline at the Advanced Photon Source. Reproduced from ref [190].*

The literature on the study and applications of phase dislocations – singular optics – is now very extensive and these optical structures have found applications in a number of areas in optical physics and optical trapping studies. X-ray optical vortices were first observed by Peele and collaborators [190] (Figure 17) and the propagation properties were reported in more detail in a subsequent paper [191]. Higher charge vortices ($m>1$) were also reported in this latter paper as well as in work by Cojoc et al [192].

The applications of optical vortices in the field of X-ray physics have not yet been fully developed but a couple have emerged. McNulty and colleagues have proposed that X-ray photons carrying angular momentum may open the way to new forms of spectroscopy [193] through new forms of dichroism. As a possible probe, this group has also shown that it might be possible to directly create high intensity X-ray optical vortices using an undulator source [194, 195]. Another proposal has suggested that highly energetic photons carrying angular momentum might be used to probe nuclear properties [196]. However none of these proposals have yet been experimentally demonstrated.

A less speculative proposal is to illuminate an object in a scanning microscope with a focussed optical vortex [197] created by a spiral zone plate. The phase gradient around the vortex core creates a degree of interference between adjacent points, somewhat analogous to the method on differential interference microscopy, resulting in a phase contrast image.

However, while optical vortices are not particularly familiar objects in conventional X-ray optics, it needs to be recognised that they emerge almost inevitably in any diffracted field. Indeed, vortices can be created through the simple interference of only three plane waves [198] and so it is almost inevitable that vortices will appear in most diffracted fields. The analysis of phase in the previous section showed that, when light propagation is viewed as a flow of energy, then the energy flow can naturally be viewed as containing both scalar and vector components, with the latter carrying any vorticity in the energy flow. The energy flow associated with the vector phase component has the property of being divergenceless, $\nabla \bullet \nabla \times \boldsymbol{\Phi}_V(\boldsymbol{r}) = 0$, and so cannot be observed via a technique that simply depends on variations of the intensity on propagation. Optical vortices are highly symmetric with the result, for example, that optical vortices with opposite topological charges produce identical three-dimensional intensity distributions [199] and so cannot be distinguished on the basis of intensity alone. It has been shown, however, that symmetry-breaking operations, such as the introduction of cylindrical lenses [53, 200], can enable vortex structures to be correctly recovered [201].

In the field of phase-imaging, then, it is essential to recognise that the presence of vortices is possible, even likely, and they have the potential to compromise the accuracy of the phase recovery. This is an aspect of the field to which this review will return. Note that the text by Paganin [202] contains an extensive discussion on phase dislocations and vortices in the context of X-ray science and is an excellent resource on this topic and other matters covered in the present review.

**5.4 Quantitative Methods in Phase Contrast Imaging**

*5.4.1 Propagation-based phase contrast*

*5.4.1.1 General Principles*

In optical imaging many objects of biological interest have negligible absorption but a large impact on the optical phase. This observation underlines the fact that the phase and the absorption can, at least in principle, vary independently of one another. A coherent field is determined throughout space by its complex amplitude over a surface and so a field with, say, N resolution elements is determined by N complex numbers, or 2N real numbers. As an intensity measurement yields a real number, in general the measurement of a complex wavefield with N resolution elements will require 2N independent intensity measurements.

A field containing vorticity has an additional degree of freedom and so two planes of data may not be enough. The simple geometric picture offered by the Wigner function can be instructive. If a light is travelling through space in a straight line, then two points define the trajectory; if the light ray describes a spiral path due to the presence of vortices then an additional data point is needed to define its path – a number of curves will be able to pass through any two points, but is not through any three. This generic argument is supported by simulations performed for electron imaging [203].

This requirement can only be relaxed if one has access to additional measurements or *a-priori* information about the object. As will be seen, the *a-priori* information can come in many different forms.

There has been considerable work on the development of methods for the non-interferometric recovery of phase and a number of algorithms have been developed. In the next section, these approaches are briefly reviewed. The reader is also referred to a paper by Langer et al [204] that compares many of the methods in this section and in which useful guidance is given on the regimes in which some of the methods here are best implemented for tomographic imaging.

*5.4.1.2 The Transport of Intensity Equation*

The transport of intensity equation, eq82, uses a measured intensity distribution, $I(r)$, and its spatial derivative along the optic axis, $\partial I(r)/\partial z$, to solve for the phase. The intensity derivative may be formed in a number of ways, including acquiring the intensity measurement at two closely spaced planes and forming the difference, or by observing the change in the intensity distribution in a microscope image due to a small defocus.

While the transport-of-intensity equation pertains to the field and so may be legitimately applied at any plane, typically, a measurement of the phase distribution of an object is desired and so the measurement plane is located as close as possible to the object, or the plane of an image of the object.

The mathematical foundations of the transport of intensity equation have been studied in some detail, primarily by Gureyev and colleagues. In 1995, in the context primarily of adaptive optics for visible and X-ray applications, Gureyev et al [27] showed that the transport of intensity equation has a unique solution in the absence of phase singularities and therefore was a potentially useful approach to the non-interferometric recovery of phase. This group went on to seek approaches to the solution of the equation using orthonormal polynomials such as Zernike polynomials [205] for uniform illumination and more general polynomials for non-uniform illumination [206]. This latter work enabled the fast Fourier transform to be employed for uniform objects [207] but was seen to be extremely slow for cases in which the amplitude of the light displays spatial variation. In the case of an in-focus image, or where the planes are close to the object to be imaged, the use of planes separated by a small distance to form an estimate of the longitudinal intensity derivative results in a high degree of tolerance to low coherence, particularly longitudinal coherence, as noted by Paganin and Nugent [187], due to the very small optical path length differences involved. The small separation of the detection planes also results in a relatively noisy estimate of the longitudinal derivative, but the solution of the second-order transport-of-intensity equation amounts essentially to a double integration more than countering the noise amplification through the construction of the derivative. The result is a method that is rather more tolerant to noise than might at first seem likely.

Gureyev et al [208] looked at the development of an eikonal description of partially coherent propagation. A number of linear approaches have been developed for the Fresnel diffraction regime for coherent [209] and partially coherent [210, 211] illumination and also for coherent imaging systems [212]. Gureyev [213] has also looked at the possibility of combining deterministic and iterative image recovery algorithms for improving the results of a transport of intensity based phase recovery.

*5.4.1.3  Fourier-based solution*

Through the employment of some simplifying assumptions, Paganin and Nugent [187] proposed a Fourier-transform based approach that solved many of these problems and opened up the way to rapid quantitative recovery of phase information. In order to remove the non-uniqueness issues implied by the presence of vortices, Paganin & Nugent introduced an auxillary function, $\Psi(r)$, with the properties that

$$\langle S(r) \rangle = \nabla \Psi(r) \tag{84}$$

$$\nabla^2 \Psi(r) = 0 \tag{85}$$

and showed that this leads to a simple direct expression for the recovery of the phase

$$\phi(r) = -\hat{k}\nabla^{-2}\left\{\nabla \bullet \left[\frac{1}{I(r)}\nabla\nabla^{-2}\frac{\partial I(r)}{\partial z}\right]\right\} \tag{86}$$

where the Laplacian and inverse Laplacian operators may be conveniently implemented using Fourier transforms. This algorithm has seen considerable application in the applications of non-interferometric phase recovery ideas to microscopy, as described in sections 5.6.2 and 5.6.3.

*5.4.1.4 Phase-Only Object*

As discussed in section 5.4.1.1, either two data-sets or *a-priori* information is required. A particularly powerful piece of *a-priori* information is that the object influence only the phase of the incident wave.

The phase-only approach [185, 207] uses intensity data obtained over a plane a short distance from the object (Figure 16). It is assumed that the object variation is sufficiently slow that the measurement differs only slightly from the object free measurement, and that object absorption is negligible. The data is taken only at one plane at a distance *z* from the object and the transport of intensity equation is used where it is assumed known that the intensity distribution at the object plane is uniform with intensity value given by $\langle I(r) \rangle$, which is the spatial average of the measured intensity over the object. The *a-priori* information therefore permits an estimate of the intensity derivative to be formed from

$$\frac{\partial I(r)}{\partial z} \simeq \frac{I(r) - \langle I(r) \rangle}{z}, \qquad (87)$$

where $\langle \ \rangle$ here denotes and average over the measured intensity distribution. Given the assumed constant intensity, the transport of intensity equation itself assumes the form

$$\frac{\partial I(r)}{\partial z} \simeq \frac{1}{k} \langle I(r) \rangle \nabla^2 \phi(r), \qquad (88)$$

The phase can then be recovered from the expression,

$$\phi(r) = k \mathfrak{I}^{-1} \left\{ \frac{1}{|q|^2} \mathfrak{I} \left\{ \frac{1}{\langle I(r) \rangle} \frac{I(r) - \langle I(r) \rangle}{z} \right\} \right\}, \qquad (89)$$

where $\mathfrak{I}$ denotes the Fourier transform operation and $q$ is the variable conjugate to $r$. Eq89 is a quantity that is easily evaluated using Fourier techniques [185, 207]. Note that the phase is undetermined at $|q| = 0$, providing the physically reasonable conclusion that the phase is undetermined to within a constant. It can also be seen that the method is unlikely to be stable for small values of $|q|$, leading to amplification of low spatial frequency components of any noise in the data.

*5.4.1.5 Homogeneous Object*

The homogeneous object approximation assumes that the object consists of a single material with known complex refractive index [214]. That is to say, provided the object obeys the projection approximation (section 2.3), the phase and amplitude of a wave leaving it will have a known relationship. In principle, then, a measurement of the amplitude distribution of the wave leaving the object will contain precisely the same information as a phase measurement, but the phase measurement will provide much improved contrast.

With *a-priori* information about the constitution of the object, only one intensity data set is required. If the material is known, then the ratio of the real and imaginary parts of the refractive index is known, and the complex value for a given pixel in the reconstruction can be determined with a single intensity measurement. Paganin et al [214] first considered this problem and found that one can recover the thickness distribution from the expression

$$t(\mathbf{r}) = -\frac{1}{2k\beta}\ln\mathfrak{I}^{-1}\left[\frac{\beta}{\beta+\frac{2\pi}{k}z\delta|\mathbf{q}|^2}\frac{\hat{I}_z(\mathbf{q})}{I_0}\right], \qquad (90)$$

where $\beta$ and $\delta$ are the optical constants discussed in section 4.1. Note that a very similar idea, based on the phase-attenuation duality of the interactions of high energy X-rays with matter, has also been proposed for medical phase contrast imaging [26, 215].

Unlike the approach in sections 5.4.1.3 and 5.4.1.4, the denominator in the brackets of eq90 does not vanish as $|\mathbf{q}| \to 0$ and so this method is rather more stable to noise than pure transport of intensity equation methods.

*5.4.1.6 Method based on Guigay equations*

This method proposed by Guigay [30, 77] uses the coherent limits in equations 60 & 61. Simple inspection will quickly reveal that an intensity observation taken at two separated planes will produce two simultaneous equations for each pixel that can be solved for the complex value at that pixel. These ideas are adapted from methods in electron microscopy and have also been adapted for the examination of laboratory-source X-ray phase radiography [76]. Turner et al have broadened the formal regime of applicability for the method [216] and demonstrated some interesting cases where the various assumptions fail.

*5.4.1.7 Multi-plane intensity measurements*

Iterative methods become important when it is not possible to write an analytic and invertible relationship between the wavefield and the measurement. In this case the phase recovery is performed using approaches, typically iterative, in which a solution is found that is consistent with the measured intensity measurements. This approach allows the incorporation of additional *a-priori* information and the inclusion of consideration of experimental uncertainties, such as noise.

In the multi-plane approach, intensity measurements are obtained over planes at a range of distances along the optical axis. This approach has its origins in the through-focal series methods developed for transmission electron microscopy [217] and has been applied to develop quantitative three-dimensional X-ray images [186].

The use of multiple planes allows a range of spatial frequencies to be probed at different sensitivities and so a large amount of data is used leading to greater stability with respect to the effects of noise.

Some of the more recent developments in iterative phase recovery using multiple measurement planes are covered in section 5.6.3, concerned with electron imaging. An approach that mixes the method based on multiple-plane methods and the transport-of-intensity equation approach has been published by Guigay and colleagues [218].

*5.4.1.8   Multiple wavelength approach*

This approach, due to Gureyev et al [219] recognises that wavelength, $\lambda$, and propagation distance, $z$, always appear in the diffraction integral in the combination $\lambda z$. Thus, assuming the optical properties of the object are wavelength independent over the range used, a small change in wavelength can have the same effect on the diffraction equation as does a small change in propagation distance.

This method does not appear to have been put into practice with experimental data.

*5.4.1.9   Tomographic imaging*

Quantitative tomographic imaging was first demonstrated by Cloetens et al using their holotomography approach [186]. This approach was based on modifications of the algorithms originally developed for electron microscopy using through-focal series (see section 5.4.1.7) and performing reconstructions using a series of data in which the object is rotated about its axis allowing images of a range of projections. Related methods have been used to recover three-dimensional images of relatively thin objects known to have a laminated structure [220] in a method termed laminography.

McMahon and colleagues have obtained a three-dimensional phase reconstruction of an atomic force microscope tip [98] in a zone-plate microscope arrangement, and have used harder X-ray to quantitatively extract the complex refractive index to enable the extraction of information about the composition of an object [99].

Bronnikov [221, 222] has developed a theory of phase-contrast tomography that unifies the tomographic and phase reconstruction algorithms into a single procedure, and has demonstrated the approach using simulated data.

*5.4.1.10 The intensity-functional algorithm*

There have recently been some further approaches to quantitative phase recovery that use other forms of information. One interesting approach borrows ideas from density functional theory [223]. As emphasised at the beginning of section 5.4, the recovery of quantitative phase and amplitude information from intensity measurements requires either two sets of data or some *a-priori* information. An approach proposed by Quiney et al [224] assumes that the incident field is well known, and can in any case be measured using the iterative techniques to be considered later (see section 6.2), and that the functional form (a form termed here the intensity-functional description) describing the propagation of the intensity of the incident beam can be applied, to an adequate approximation, to the diffracted field – an assumption that amounts to the requirement that the object interacts only weakly with the incident field. In this way, an expression can be obtained for the longitudinal derivative of the intensity from a single measurement plane and the phase then recovered via the transport of intensity equation.

The case discussed in detail in ref[223] assumed that the incident field is coherent and Gaussian in shape, $\psi(r) = \exp\left[-\mu|r|^2\right]$, where $\mu$ specifies the appropriate width of the incident field, and so has a derivative of intensity along the optical axis described by the functional form

$$\frac{\partial I(r)}{\partial z} = -\frac{\mu}{z} \int |q|^2 \, g(q) \exp\left[i\frac{k}{z} q \bullet r\right] dq. \tag{91}$$

where z is the distance from the object to the plane of measurement and $g(q)$ is the autocorrelation of the measured intensity. One uses the measurement of intensity to form $g(q)$, eq91 to form $\partial I(r)/\partial z$, and then the transport of intensity equation, eq82, to recover the phase.

This method has yet to be demonstrated on experimental data, but has yielded promising results with simulated data and added noise.

*5.4.2 Wigner Phase-Space Deconvolution*

Section 6 will consider the problem of imaging from coherent diffraction, a method that uses the diffraction pattern from an object to form an image of it. This section considers a method that arguably falls between phase imaging and coherent diffraction methods. The first experimental work in this broad area considered the role of Wigner phase space deconvolution (WPSD) [225] and was published by Chapman [226].

Wigner phase space deconvolution, as its name suggests, considers the imaging problem in terms of the Wigner function. In its simplest form, an object is illuminated by a known wavefield and the two-dimensional diffraction pattern recorded as the illuminating field is scanned across the object, resulting in a full four-dimensional data set. In the context of this review, the method of WPSD is rather straightforward to understand. Let us first assume the projection approximation – the object is effectively perfectly thin – and illuminated by a field $\psi(r)$ with finite extent. If the illumination is displaced from the origin by a vector $X$ and the far-field intensity measured for all values of $X$ then the four-dimensional data set is described by (see eq30)

$$I_{ff}(s,X) = \int \psi(r_1 - X)\psi^*(r_2 - X)T(r_1)T^*(r_2)\exp\left[-iks\bullet(r_2 - r_1)\right]dr_1 dr_2. \quad (92)$$

The coordinate system used to form the Wigner function is introduced, so eq92 can be written

$$I_{ff}(s,X) = \int \psi\left(r+\frac{y}{2}-X\right)\psi^*\left(r-\frac{y}{2}-X\right)T\left(r+\frac{y}{2}\right)T^*\left(r-\frac{y}{2}\right)dr\exp[-iks\bullet y]dy. \quad (93)$$

The Wigner function for the illuminating field is defined to be

$$B_\psi(r,u) = \int \psi\left(r+\frac{y}{2}\right)\psi^*\left(r-\frac{y}{2}\right)\exp[-iku\bullet y]dy, \quad (94)$$

so that

$$\psi\left(r+\frac{y}{2}\right)\psi^*\left(r-\frac{y}{2}\right) = \left(\frac{k}{2\pi}\right)^2 \int B_\psi(r,u)\exp[iky\bullet u]du. \quad (95)$$

A similar result obviously also holds for the Wigner function of the scattering function, so that, after substitution of eq95, and its equivalent for the transmission function, into eq94 and some elementary re-arrangements and simplifications,

$$I_{ff}(s,X) = \left(\frac{k}{2\pi}\right)^2 \int B_\psi(r-X,u)B_V(r,s-u)dudr \quad (96)$$

is obtained. With appropriate variable changes, this can be easily to seen to have the form of a four-dimensional convolution of the Wigner function of the illuminating beam with the Wigner function of the scattering function.

The argument leading to eq96 applies equally well if the illumination is partially coherent, where $B_\psi(r,u)$ describes the Wigner function of the partially coherent field (see section 3). It is therefore possible to use this method to obtain an image using partially-coherent illumination. Conversely, a known test object could be used to recover the complete partially coherent properties of the illuminating field, an approach that has not yet been put into practice.

The function on the left hand side of eq96 is a four-dimensional measured data set. If it is assumed that the illuminating field is known, or can be measured, then the scattering function can be recovered using four-dimensional deconvolution methods. It is this idea that was experimentally realised by Chapman [226]. The work of Chapman is important, because it, with the more or less simultaneous work of Nugent et al [185], is the first deterministic and non-interferometric recovery of phase information in X-ray imaging. However the data volume to recover an image using WPSD is huge, requiring full four-dimensional data sets, and so this method has not subsequently been used. Chapman [226] also pointed out that it is also possible to recover images with a resolution beyond the diffraction limit implied by the numerical aperture of the illumination, though he did not demonstrate this. This is also an idea that emerges in the method of ptychography.

### 5.4.3 *Ptychography*

Ptychography is an approach that has its roots in electron microscopy [227] but is emerging as an important approach for high resolution X-ray phase imaging. It has a similar data acquisition methodology as Wigner phase-space deconvolution, insofar as one scans the object with an illuminating beam and records the diffraction pattern at each location of the illumination. WPSD is an entirely general imaging approach that is suited to fields with very low spatial coherence. On the other hand, in order to perform the required four-dimensional deconvolution, a complete and precise measurement of the four-dimensional function defined in eq96 is needed and this requires that the illuminating field be scanned on a very fine grid, requiring extremely large quantities of data. The method of ptychography appears to be relatively sensitive to issues of partial coherence but can be performed using a much coarser grid and the recovery can be performed using an iterative approach [228-230].

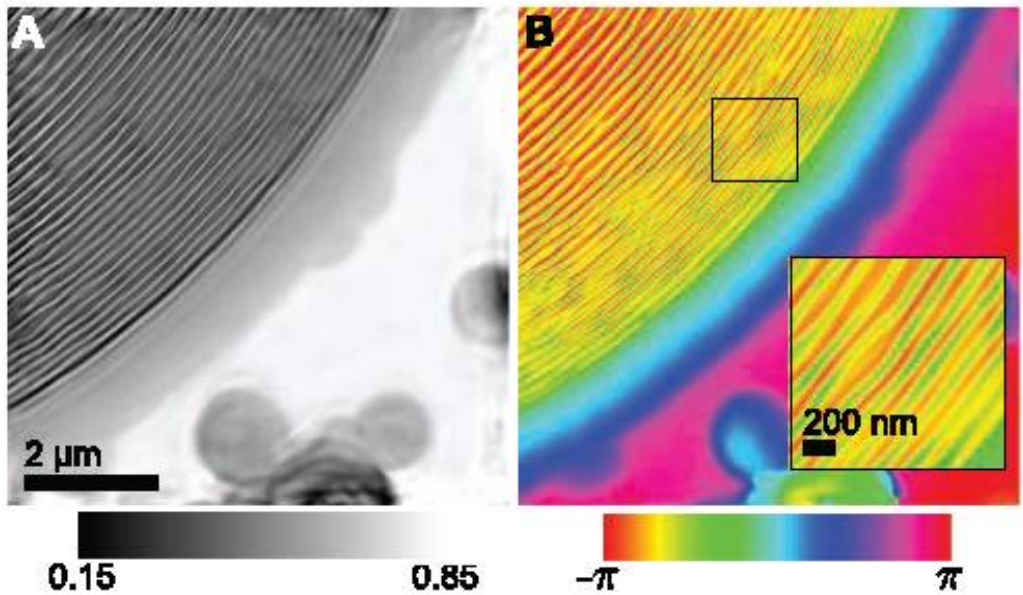

*Figure 18: The amplitude (A) and phase (B) of a buried zone plate structure recovered with high-spatial resolution using an extension of the method of ptychography. Reproduced from ref[233].*

In the implementation of ptychography for X-ray problems [231], the object is illuminated by a field that is limited in extent and the diffraction pattern is recorded. The illumination is then translated so that there is some overlap with the previous field [232] and the diffraction pattern again recorded. This process is repeated until the area of interest has been completely scanned. The data set is intrinsically four-dimensional, as with WPSD, however the number of diffraction patterns required for the iteration to converge is greatly reduced, allowing the method to be rather easier to use than is WPSD. A recent study has explored the degree of overlap that is required [232]. The algorithms have been shown to be able to fit the entire diffraction pattern, to recover the probe through the iterative method, and to allow the high resolution imaging of structures below the surface [233] and able to simultaneously recover the wavefront of the illuminating beam [234], allowing very high-spatial resolutions to be achieved. A high-resolution image of a buried structure is shown in Figure 18.

*5.4.4 Fourier transform holography*

Imaging from coherent diffraction clearly has its conceptual roots with the development of holographic imaging and there were a considerable number of attempts to develop methods in X-ray holographic imaging [235-240]. It is fair to say that these efforts, while interesting, did not ultimately yield a valuable approach to high-resolution X-ray imaging until the work of Eisebitt et al [241].

Eisebitt [241-243] showed that Fourier transform holography, first demonstrated for X-rays by McNulty et al [237], is able to yield quite striking images of magnetic domains. These authors have extended the scope of Fourier transform holography to encompass multiple reference beams [244] and applied it to a number of investigations of magnetic structures [245-247]. See Figure 19.

A major driver for the development of these methods is the possibility of imaging with a resolution well beyond that which will be possible using diffractive structures such as zone plate lenses. The essential feature of Fourier transform holography is that the object is illuminated over a finite area but with a small reference source, typically simply a pinhole, located nearby. To describe this, a transmission function of the form

$$T(r) = A(r) + p(r - d) \qquad (97)$$

is employed where $A(r)$ is the transmission of the object of interest and $p(r-d)$ describes a pinhole displaced from the object by a vector *d*, then the observed intensity distribution is

$$I_{ff}(s) = \int \left\{ \begin{array}{l} \left[ A\left(r + \frac{y}{2}\right) + p\left(r + \frac{y}{2} - d\right) \right] V\left(r + \frac{y}{2}\right) \\ \left[ A\left(r - \frac{y}{2}\right) + p\left(r - \frac{y}{2} - d\right) \right] V^*\left(r - \frac{y}{2}\right) \end{array} \right\} \exp[-iks \bullet y] dr dy. \qquad (98)$$

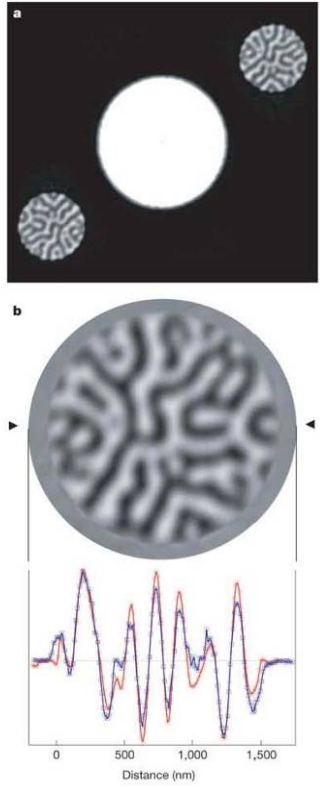

*Figure 19: Reconstructed images of magnetic structures of a Co/Pt multilayer film obtained using soft X-ray Fourier transform holography. The data was acquired using 778 eV X-rays at the BESSY-II storage ring. Reproduced from ref [241].*

The Fourier transform of this intensity measurement is

$$I_{ff}(q) = \int \left\{ \begin{array}{l} \left[ A\left(r + \frac{\lambda q}{2}\right) + p\left(r + \frac{\lambda q}{2} - d\right) \right] \times \\ V\left(r + \frac{\lambda q}{2}\right) \left[ A\left(r - \frac{\lambda q}{2}\right) + p\left(r - \frac{\lambda q}{2} - d\right) \right] V^*\left(r - \frac{\lambda q}{2}\right) \end{array} \right\} dr. \quad (99)$$

This autocorrelation function includes terms describing the cross-correlation of the pinhole with the object displaced by a vector $d$ from the optical axis, which are images of the object formed with a resolution determined by the size of the pinhole. Eisebitt et al [241] have speculated that it might be possible to use the higher-angle diffraction by the object to enhance the resolution beyond the limit determined by the size of the reference pinhole.

This group also used a number of pinhole references to permit the reconstruction of multiple holograms and thereby improve the signal to noise ratio in the image [244] and more redundancy in the reconstruction. Reference [248] reports an experiment in which the pinhole is replaced with a uniformly redundant array (section 3). The result is a structured reference wave that contains many more photons than one produced using a small pinhole. An improved image signal to noise ratio without loss of spatial resolution is obtained. Schlotter at al [244] report the use of a complex reference wave but one that would be termed, in the language of coded aperture imaging, as a non-overlapping array [249].

*5.4.5   The method of Podorov et al*

A method has recently been proposed by Podorov et al [250] that has a close relationship to the method of Fourier transform holography. In this approach the diffraction by the object is subject to interference with the radiation scattered by the edge of the aperture in which the object is contained. The methodology has been expanded and examined by Fienup and colleagues [251-253] as an interesting new paradigm, however it is apparent that the spatial resolution of the method is determined by the sharpness and straightness of the aperture edge [251] and so the path to very high resolution is by no means clear; this method seems unlikely to find widespread application to high-resolution imaging problems. Martin and Allen [254] have also considered a non-iterative phase recovery technique that uses features of the geometry of the illumination and of the aperture surrounding the object.

*5.4.6   X-ray Imaging and Phase Recovery*

The above methods have in turn enabled the development of phase sensitive imaging for zone-plate based full-field X-ray microscopy. Although quantitative phase imaging has not been extensively applied in this area, Nugent et al [255] have developed a consistent transfer function theory of the imaging process and tested its predictions using an X-ray imaging microscope [256]. Zone plate imaging has also been used to perform quantitative phase imaging [257] and this approach has been extended to enable tomographic imaging using relatively soft X-rays [98] of the tip of an atomic force microscope probe. A quantitative analysis of phase imaging using full-field zone plate microscopy has also been published [258] and has found the same tolerance to partial coherence that is observed in the case of optical microscopy.

It is also worth mentioning that other forms of phase-sensitive X-ray microscopy have been developed, including scanning X-ray microscopy [259] using a structured detector [260]. Methods have also been developed for differential phase contrast [261] and Zernike phase contrast [262-264] and spiral zone plates [197], though these methods do not yield quantitative results.

## 5.5 Detectors and X-ray phase contrast

Many of the applications of X-ray phase contrast imaging rely on the observation of the X-ray intensity after it has propagated some distance from the sample, and many applications use relatively hard X-rays. The key requirements of the detector are that the X-rays are able to be detected with high efficiency and with relatively high spatial resolution. Diffraction-enhanced imaging (section 4.4.3) uses a scanned method of data acquisition and so is able to use linear detection system that can offer a high detective quantum efficiency but typically rather limited spatial resolution [265]. CCD cameras offer better spatial resolution and are able to capture the two-dimensional frame needed for many of the other imaging approaches and so are most commonly used. In particular, a phosphor screen can convert the incident X-rays into visible light which can then be matched to a cooled CCD camera and so permit flexibility in the available resolution.

Energy discrimination is not required for the imaging systems discussed here and dynamic range, though important, is not as critical a limitation as for the coherent methods discussed in the next section. Most experiments are performed, therefore, using high quality X-ray CCD systems [265, 266]

## 5.6 Developments using non-interferometric phase measurement ideas

### 5.6.1 *Neutron Imaging*

There are many commonalities between the fields of X-ray and neutron optics; many of the techniques developed in one area may be transferred to the other. The relatively low coherence of X-ray sources in comparison to optical laser sources has seen the development of X-ray methods that are tolerant to low coherence and are therefore applicable to areas that do not have access to high-coherence sources, such as neutron science. While interferometry has been extensively applied to neutron science, it was not until the work of Allman et al [267] that it was realised that the methods of X-ray phase contrast imaging and phase measurement can also be applied to neutron radiography.

Allman et al [267] showed both that phase-contrast could be observed and that it could be quantified using transport of intensity equation methods. The same group subsequently developed a theoretical formalism for the phase quantification built around the ideas outlined in section 5.2 and showed how the values of phase could be assigned a physically meaningful interpretation, confirming this with quantitative neutron phase measurements on a silicon block sample [268]. Interestingly, it was also found that contrast could also be obtained via small angle scattering from the sample so that moving the neutron detector well downstream from the sample enabled small angle scattering to be an effective image contrast mechanism [269].

Subsequently applications for neutron phase imaging have begun to emerge and, in parallel, other techniques have emerged, with the work of Pfeiffer and colleagues being of particular note [156, 270].

*5.6.2 Optical Microscopy*

Propagation-based phase recovery, though primarily developed in the context of X-ray imaging methods that do not use image forming oprics, has also enabled new approaches to imaging using optical lens-based systems. For completeness, this section and the next briefly touches on these, though they do not conform to the strict scope defined for this review. They are important nonetheless as they represent an example of developments in X-ray science influencing methods in more mature areas such as optical and electron microscopy.

The transport of intensity equation was first proposed for use in optical microscopy by Striebl [184]. While suggestive, this paper stopped well short of a solution as it was not at that time known how to efficiently solve for the phase. The role of partial coherence in phase imaging has been discussed by Gureyev et al. [27]. This paper used the Wigner function to examine the propagation of partially coherent wavefields and showed that, for sufficiently small propagation distances, the field behaves as if it is coherent. Following the success in applying these methods to X-ray imaging, Barty et al [271] proposed and demonstrated that the method could be of practical benefit for optical microscopy. The longitudinal intensity derivative was obtained via a very small defocus of the optical system with a symmetric defocus producing a particularly good estimate. These authors showed that, for small optical defocuses, the method was very tolerant to relatively poor spatial coherence and they demonstrated quantitative optical phase microscopy using an optical fibre and a human cheek cell as test objects. This work was subsequently extended to optical phase tomography [272]

Barone-Nugent et al [273] analysed this form of optical microscopy in some more detail, formulating a description of the imaging process based on the three-dimensional optical transfer function theory of Striebl [274]. The predictions of this theoretical analysis were confirmed experimentally and the tolerance to low levels of spatial coherence was confirmed. Similar theoretical results were obtained by Sheppard [275] using a different approach.

The application to three-dimensional objects is important for tomographic imaging and it has been noted that the lack of knowledge of the additive phase component may complicate the interpretation for some objects [276], though this seems unlikely for isolated objects. Paganin et al explored the effects of noise is a quantitative manner [277] and demonstrated that the use of multiple defocus planes significantly assists with noise tolerance. Bellair et al [278] explored the phase measurement obtained from thick objects and found that the experimental results were in good agreement with the optical transfer function analysis [273, 275] and that it was possible to interpret the phase measurement as a projection measurement through the object in quite a wide range of circumstances.

Quantitative phase microscopy of this form has now been used to address a number of scientific problems. Ross et al [279] have used the method to improve the visualization of cells for experiments with microbeam irradiation. Curl and colleagues have used the approach to observe cell culture growth patterns [280]. Differential interference contrast imaging yields an image of the phase gradient in a particular direction, a form of contrast that can make edge detection quite difficult. Phase microscopy allows a contrast that is linear with phase and so can improve automated edge detection. Curl et al used this feature to measure the rate of growth of cell cultures [281] and cell morphology [281, 282] as well as to combine the method with confocal microscopy to enable quantitative measurements of refractive index in live cells [283]. Dragomir and colleagues have observed the birefringence in live cardiac cells [284] and obtained good agreement with independent methods.

Barone-Nugent et al [273, 285] used quantitative phase microscopy to obtain excellent contrast on palaeobotanical images of samples from the Triassic era. In this work, the method was used to improve the contrast of the images rather than to obtain quantitative phase information. The method was used in an analysis of the fossil record from Leigh Creek in Australia [285].

The method has also found applications in the measurement of optical properties of samples. Following the initial demonstrations of measurements with optical fibres [271, 272], Roberts et al used phase measurements for orthogonal polarisations to obtain the strain-induced birefringence in an optical fibre [286]. Ampen-Lassen and colleagues have developed an approach to obtaining the detailed refractive index distribution in an optical fibre [287, 288] and Dragomir et al [289] have been able to look in detail *in-situ* at the splicing of optical fibres. Aruldoss, Roberts and colleagues [290, 291] have brought intensity transport ideas to the analysis of imaging through turbid media. The transport of intensity equation has also been identified as a simple and accurate approach to the testing of optical surfaces [292]. Barbero and Thibos [293] have performed a study of the consistency and accuracy of the method for wavefront reconstruction.

*5.6.3 Electron Microscopy*

The idea that defocused images can yield phase information has a long and venerable history, going right back to the optical star test in which a defocused image of a point of light is used to assess aberrations (i.e. phase errors) in an optical system. Early workers in electron microscopy also noted this effect, and defocused images came to be routinely used to create phase contrast. It was also the electron microscope phase imaging problem that gave the initial impetus to the algorithm proposed by Gerchberg and Saxton [294]. This algorithm, which has been adapted to many different applications, was proposed to solve for the phase of an electron field given a measurement of its in-focus intensity distribution and its far-field diffraction pattern. The scheme then iteratively finds a solution for the phase that is consistent with both measurements. However for the purposes of electron microscopy it is not always convenient to obtain an image of both the in-focus and far-field diffraction patterns.

An intermediate approach was proposed by Coene et al [181, 295, 296] who explored an iterative approach to finding a complex exit wave field that is consistent with many images taken at different levels of defocus. Such an approach assumes that the field can be regarded as completely coherent so as to enable standard diffraction physics to recover a field consistent with all of the measurements. It has been found that the scheme which iterates between all the measurements imposing the measured magnitude at each plane and allowing the phase to find a consistent value, indeed converges on the correct value rapidly and consistently.

Bajt et al [297] demonstrated quantitative transport of intensity-based electron phase microscopy, obtaining images of the electron phase distribution induced by the magnetic field surrounding a cobalt sample. Prior to this work, quantitative electron phase imaging of such objects was obtained using the more complex technique of electron holography. Bajt et al [297] compared their results with electron holography and found very good agreement. McMahon et al [298] used the method to obtain phase images of biological objects. Using ideas put forward by Bajt et al [297], McMahon et al [298] used the complex wave information to simulate other forms of phase contrast, such as differential interference contrast, an idea later extended to X-ray phase imaging [299].

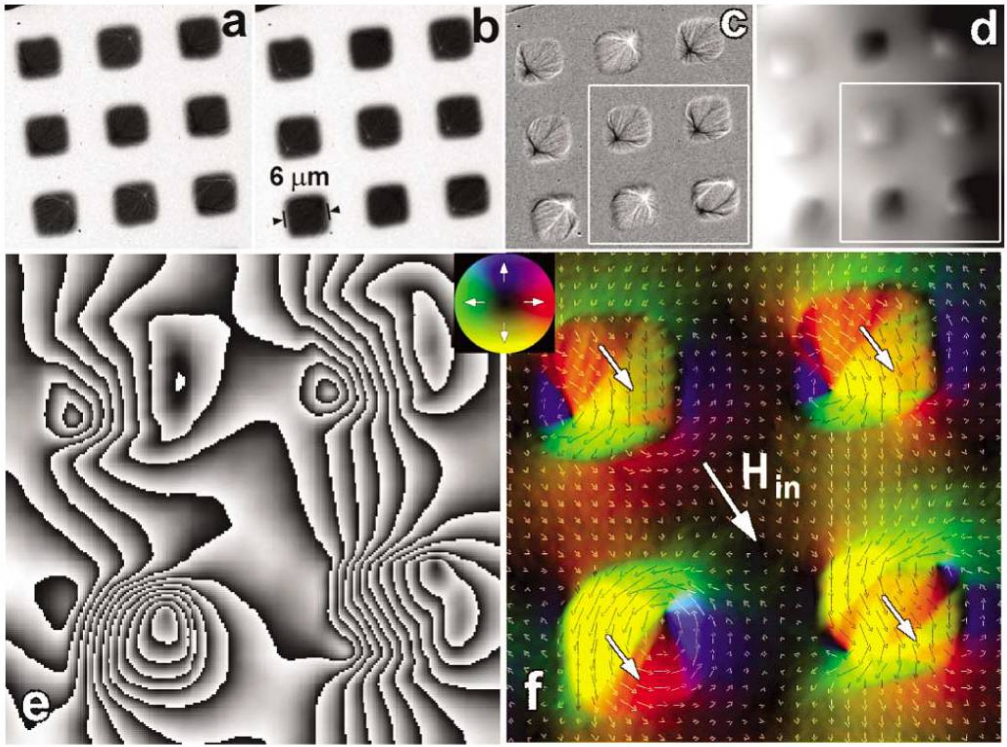

*Figure 20*: *Experimental underfocused (a) and overfocused electron images (b) of 30nm thick Co islands. A reconstructed phase-contrast image (c) and an image of the recovered phase (d); the enlarged images of phase contours, showing the magnetic flux distribution (e) and computed projected induction map (f) for the boxed areas in (c) and (d), shown both by color-code and arrow-vector maps. The inset illustrates the vector amplitude and direction. Large arrows in (f) show the predominant magnetization of Co islands. Reproduced from ref[309]*

The ability to obtain quantitative phase information in this way raised concerns in the conference literature that complete coherence is required for the precise measurement of phase. This argument was examined by Beleggia et al [300] who showed that the quantitative phase method was essentially classical in nature and so was not subject to this limitation. This issue was also explored by Nugent et al [255] who created a transfer function theory for quantitative phase imaging for weak, thin objects. In this work, the transport of intensity limit was explored and it was shown that, for a differential defocus, the image formation was effectively indistinguishable from coherent image formation when the coherence length is only 25 resolution elements in length, where a resolution element is the diffraction-limited resolution of the optical system. This is far below the need for the coherence length to encompass the entire object, as is required for electron holography and has its physical origins in the fact that all of the path length differences in the image formation process are very short. This theoretical work was experimentally tested for X-ray zone-plate based microscopy [256].

Allen and Oxley [301] have looked at phase recovery methods using the transport-of-intensity equation using the Paganin-Nugent approach [187] and compared the results with the multigrid approach to solving the differential equation and to an iterative phase recovery algorithm. They note that phase vortices (section 5.3) will often be present in practice and find that the iterative approaches are tolerant to the presence of phase vortices where the more direct solution methods are not, as evidenced, for example, in the limitation implicit in eq84 & 85. The examination of the effect of phase vortices was extended by Allen et al [203] who found using numerical experiments that iterative approaches can be tolerant to vortices but that at least three planes of intensity data are required in order to achieve a reliable solution, consistent with the dimensional arguments in section 5.4. In a subsequent paper, Allen et al [302] considered the possibility of using phase recovery to correct aberrations, including in the presence or vorticity in the energy flow, and found that the iterative approach offers an interesting possible way forward. Allen et al [303] explored the application of a global approach to iterative wavefront retrieval, reconstructing the wave-function simultaneously in all experimental planes. This approach is robust in the presence of experimental noise. Their approach was compared with the maximum likelihood method [181, 296] and they found that the two methods give good agreement when applied to high resolution electron microscopy images and note that the iterative approach is straightforward to implement and is reasonably tolerant to the effects of partially coherent illumination. Martin et al [304] further numerically investigated the impact of partial spatial coherence on iterative and transport-of-intensity approaches to phase recovery and identify regions of applicability for the two approaches. Volkov and Zhu [305] proposed a symmetrisation method to estimate the boundary conditions needed to solve the transport of intensity equation in electron-microscopy applications. Beleggia et al [300] considered the phase sensitivity of the transport of intensity method, showing that it is quite sensitive to small phase shifts. Volkov and Zhu [306] proposed a related phase recovery method that they identify as having better tolerance to noise, high phase gradients and the presence of phase vortices. Zhu et al [307] have used non-interferometric transport of intensity methods to investigate the field- and orientation-dependence of magnetic domains in permanent magnets and Volkov et al [308] have used these methods to look at the evolution of magnetic structure in magnetic arrays. Volkov and Zhu [309] derive a modified transport of intensity equation for magnetic structures and use it to map the magnetic flux and projected induction in magnetic and superconducting materials (see figure 20).

The ideas of direct phase recovery have now been applied to a number of scientific problems investigated using transmission electron microscopy. Martin et al [310] have looked at the evolution of a gold-vacuum interface in which the images are recovered using an iterative approach. Petersen et al [311] have used transport of intensity methods to look at the dopants in a p-n junction. This paper also reports the use of multiple image planes to obtain a better estimate of the longitudinal derivative. The same group [312] also used the transport of intensity equation to image the structure of MgO nano-cubes and also measure the mean inner potential of the cubes, and went on to use the transport of intensity equation to image phase shifts [313] that they attributed to surface plasmon excitation. This paper suggested that the phase retrieval techniques offer a new approach to the observation of such structures.

# 6  Coherent Diffractive Imaging

## 6.1  Overview

The application of an understanding of the physics of diffraction to the study of materials has been one of the great success stories of modern science. The development of the methods of crystallography has been built on the periodic nature of the crystalline structure. The extraordinary success of this technique led one of the pioneers of X-ray crystallography, David Sayre, to ask whether one might apply a related technique to non-periodic objects [314]. The success of such a proposal would see the imaging of structures with a resolution limited only by the angle of diffraction, not by the resolution available from a physical lens. The argument presented by Sayre was an elementary one. Crystallographers regard the sampling of a diffraction pattern as implying that one measures the intensities of each of the diffraction spots in the pattern. Sayre posited that a measurement of the diffraction pattern at twice this sampling rate ("oversampling" from a crystallographer's perspective) would potentially provide sufficient data for the complete reconstruction of the object. This is the "implication" in the title of the short paper, and issues of the independence of the measurements were not addressed; it was a tantalising but underdeveloped idea at that stage.

The second motivator to this question was the program to develop high-resolution soft X-ray microscopy in the so-called water window [315, 316], a wavelength region in which there is natural absorption contrast between carbon and oxygen, allowing imaging of biological objects in their natural state.

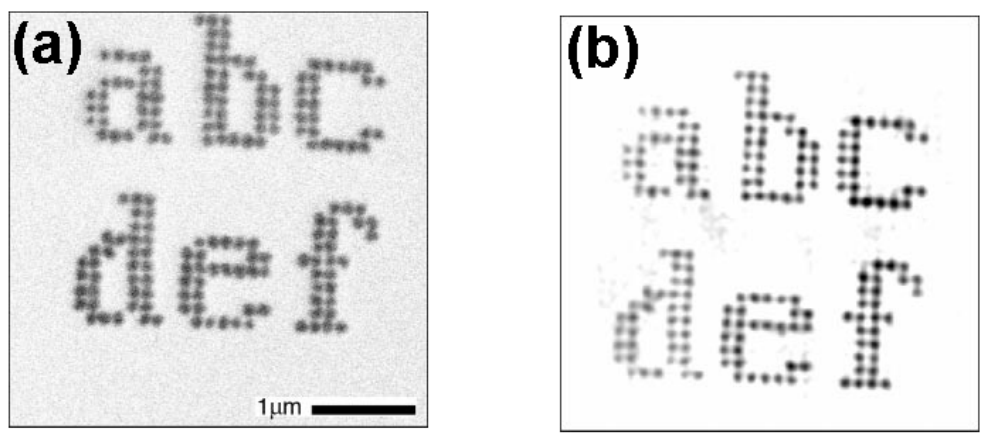

*Figure 21: The first coherent diffractive image. (a) An electron microscope image of the gold test sampl) is compared with (b) the image reconstructed using coherent X-ray diffraction (b). The data was acquired at using 740 eV X-rays at the X1A beamline at the National Synchrotron Light Source. Reprinted from ref [321]*

The method by which it was proposed to reconstruct the image has its origins in ideas in electron microscopy. The so-called Gerchberg-Saxton algorithm [294] employs an iterative method in which the phase distribution within an object is found that is consistent with both an intensity image of it and with its far-field diffraction pattern. The proposed algorithm has been found to consistently converge on the correct solution, and the underlying ideas have been applied to a wide variety of problems. Clearly, however, if an object is so small that a direct image cannot be acquired then reconstruction via the Gerchberg-Saxton algorithm is not possible. Bates considered this problem [317] and argued, though did not prove, that an object is uniquely determined by its autocorrelation function (or, equivalently, the intensity of its far-field diffraction pattern) if its spatial extent – its support – is known. Bates also identified those aspects of the reconstruction that cannot be recovered. For the most part, these are entirely unimportant, such as translation, phase conjugation, inversion and the value of the absolute phase of the wave, the latter being a quantity with no physical content anyway. Bates showed that it was extremely unlikely that an object with a given support could produce identical diffraction patterns, though examples have been found [318].

Fienup [318, 319] demonstrated that a modification of the Gerchberg-Saxton algorithm could solve the problem posed by Sayre where only support information is used at the object plane.

The application of these ideas to X-ray imaging has been the subject of considerable analysis [320]. A principal issue was that the scattering from the object is so weak that the unscattered synchrotron beam will swamp the signal of interest. This problem is mitigated by the introduction of a beam stop to prevent the undiffracted X-ray beam from destroying the detector; however this also eliminated substantial low spatial frequency information from the acquired data. Miao et al [321] circumvented this problem by using an electron microscope image of the object to re-create the missing low spatial frequency data, a trick that in a sense returns the method to the original Gerchberg-Saxton approach, and experimentally demonstrated that the method could be made to work with X-ray data (Figure 21). It is the paper of Miao et al [321] that has largely been responsible for the rapid recent growth of interest in coherent diffraction for imaging.

The second major stimulus has been the proposed development of X-ray free electron lasers. A potential use of these sources is contained in the proposal that the coherent

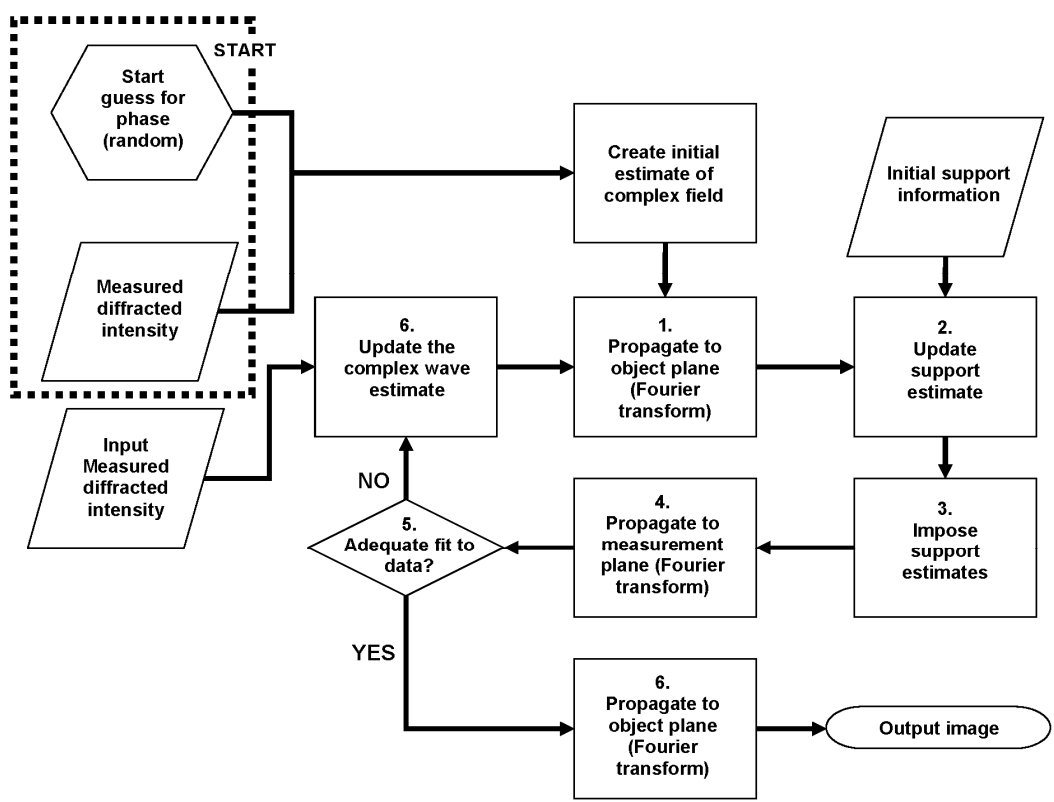

*Figure 22: Outline of the fundamental underlying algorithm that permits the reconstruction of a sample distribution from its diffraction pattern. The variations in the implementation of the algorithm primarily lie in the manner in which steps 2 & 3 are implemented.*

X-ray pulses might be sufficiently bright that it will be possible to observe diffraction from a single molecule and thereby recover the structure of that molecule with atomic resolution [322, 323] using iterative reconstruction processes. In parallel, it is emerging that coherent diffraction may well be a very useful form of high-resolution imaging independent of applications to structural biology using X-ray free electron lasers.

The key idea for the development of the imaging is to find an object distribution that is consistent with the support, assumed known for the moment, and with the measured diffraction pattern. An iterative technique is used to find this consistent object distribution and the (almost) unique result of Bates is invoked to declare that the solution is the correct one. Before proceeding with the development of the imaging ideas, the variations on the image recovery algorithms are briefly discussed.

## 6.2   Iterative image recovery algorithms

The basic structure of the iterative algorithm for recovering images from coherent X-

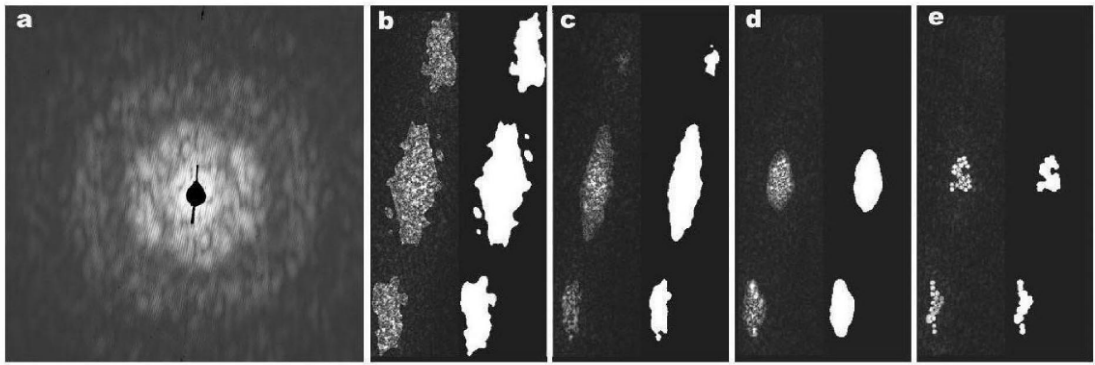

*Figure 23: Evolution of the "shrink-wrap" algorithm. (a) Coherent X-ray diffraction pattern colloidal gold particles. (b)-(e) shows convergence of the algorithm in which the support for the reconstructed object is updated dynamically. Each panel shows the latest estimate of the sample distribution (left) and the estimate of the support (right). The support is dynamically updated as the algorithm progresses. The data was acquired using 620eV X-rays at the Advanced Light Source. Reproduced from ref [331]*

ray diffraction patterns is outlined in figure 22. The known (ie. physical extent of the object, or support) or measured (ie. diffracted intensity) information is applied during the iteration as constraints in the object plane and the detector plane, respectively. The key differences in the various implementations lie in the manner in which the support constraint (steps 2&3, Figure 22) are implemented. The resulting implementations are quite different and as a result of these differences, they will have rather different convergence properties in practice. The simple imposition of the known support is known as the Error Reduction (ER) algorithm [318]. In practice, this algorithm has a tendency to stagnate before finding a suitable solution.

Fienup introduced a relaxation parameter into the object update stage of the algorithm producing an approach – the hybrid input-output algorithm [318, 324] – that tends to be less susceptible to stagnation. This algorithm has led to a number of analyses and improvements, primarily in the manner in which the object estimate is updated. These include the difference map [325], saddle-point optimisation [326], hybrid reflection projection [327], relaxed averaged alternating reflections [328] and charge-flipping [329]. A unified analysis of the methods has been published by Marchesini [330]. It must be observed, however, that none of these approaches have yielded an algorithm that has been found to converge in all cases.

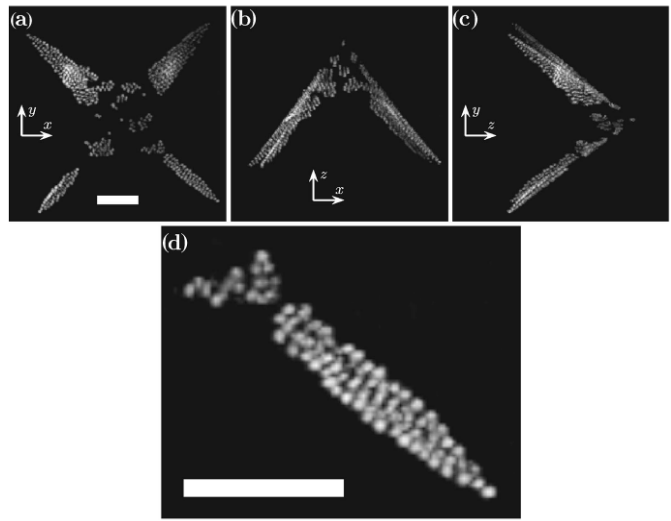

*Figure 24*: *(a)-(c) Projections along the three-dimensional image of a pyramidal structure obtained using coherent diffractive imaging; (d) shows an enlarged region with a 500nm scalebar. Data were acquired using 750 eV X-rays at beamline 9.0.1 at the Advanced Light Source. Reprinted from Chapman et al [342]*

An important innovation was proposed by Marchesini et al. [331] in which the support for the object (step 2, Figure 22) is obtained dynamically, using the so-called shrink-wrap algorithm (Figure 23). This method eliminates weak parts of the reconstructed diffracting structure from the iterative process and so, with care, can independently find the support for the structure. Another related modification to step 2 is that of "charge-flipping" [329, 332-334] in which the sign of low amplitude parts of the reconstruction are changed with each iteration. This method has been found to be quite successful [335] on experimental data.

All of the above algorithms are deceptively simple to implement but their success in application to experimental data is not universal and requires considerable skill and experience. As will be argued in section 6.4, this observation is due, at least in part, to the less than perfect coherence of X-ray sources.

The regions of applicability of one of the more popular algorithms, the difference map approach [325] has been explored by Allen et al [336] who found, using a limited range of test images, that all reconstructions converged if one is able to identify the correct degree of oversampling and the right update parameters in the iteration. Martin and Allen [337] explored the role of vorticity in the iterative scheme and showed that the overall angular momentum is not conserved in the iterative process and concluded that there is no strict constraint on the vorticity in the initial starting point.

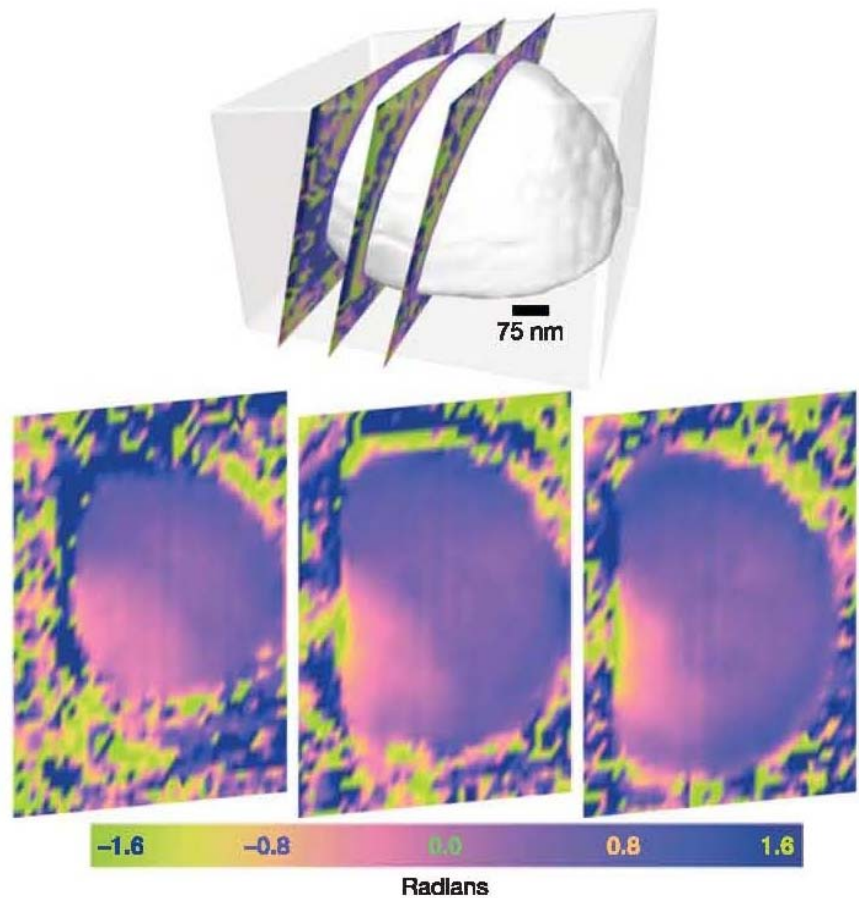

*Figure 25: Tomographic reconstruction of the strain distribution in a lead nanocrystal. Phase maps cutting through the crystal at three parallel planed separated by 138nm are shown. Data were acquires at using 9keV X-rays at the 34-ID-C beamline at the Advanced Photon Source. Reproduced from ref [353].*

## 6.3 Experimental demonstrations

The first demonstration of coherent diffractive imaging was published by Miao et al [321], a paper that stimulated a great deal of interest in the methodology. A significant number of demonstrations and developments have flowed from this original work.

The motivation for the development of the method is the desire to develop a flexible high-resolution imaging technique, ultimately for the imaging of isolated biological objects, though applications to other systems are rapidly emerging, as will seen here.

Much of the development has concentrated on the development of algorithmic approaches but there have been a number of important demonstration experiments, and some applications. The methods demonstrated by Miao et al have been extended to include some preliminary experimental and simulation work on three-dimensional imaging of biomolecular samples [338], and materials science samples [339-341] However the most spectacular and convincing demonstration of the three-dimensional imaging is that of Chapman et al [342] who showed true three-dimensional tomographic imaging of a test three-dimensional object (Figure 24).

The other important area of extension has been demonstrated by Williams et al [343] who showed that it is possible to create diffraction using curved incident beams, termed Fresnel coherent diffractive imaging. This was first argued theoretically, based on ideas developed for the transport of intensity equation [200], but it was subsequently realised that the simple use of a curved incident beam has significant benefits in terms of the reliability and convergence of the image recovery [344, 345],

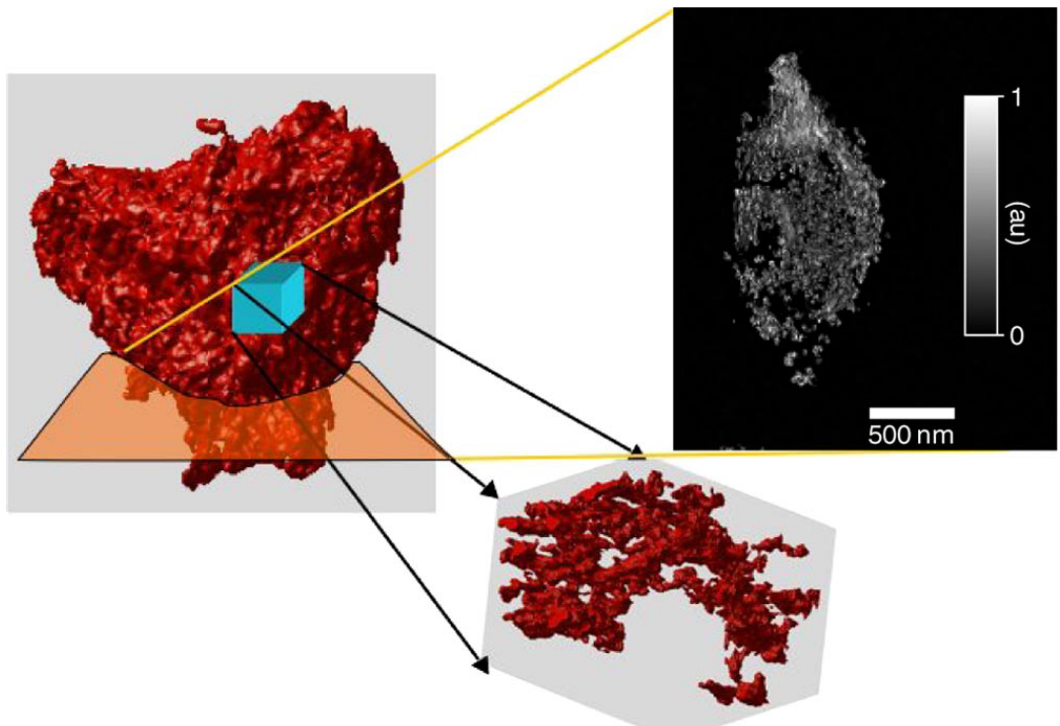

*Figure 26: Section and isosurface rendering of a 500nm cube within three-dimensional image of a ceramic nanofoam. A 500 micron scalebar indicates the scale of the image. Data were acquired using 750 eV X-rays at the Advanced Light Source. Reproduced from ref [358]*

and that the relationship between a finite object and its diffraction pattern is in fact unique [346]. This approach was demonstrated experimentally using a gold test object [343]. This method requires that the incident beam be very well characterised and diffractive imaging was also used for this purpose, where the wave diffracted by a zone plate was very precisely characterised using the known pupil of the zone plate as the support [224]. Importantly, in a demonstration that brings coherent imaging and ptychography into some alignment, it was further shown that the incident curved beam can itself be used to define a support within an extended object [347] thereby opening up coherent imaging for application to extended objects (ie. infinite support).

X-ray waveguides have also been used to produce the curved beams needed for Fresnel coherent diffraction [348].

A further important development is the application of coherent imaging to nanocrystals. Elementary analysis of diffraction physics will reveal that the diffraction pattern from a finite crystal will be described as the convolution of the diffraction pattern from an infinite crystal with the Fourier transform of the complex shape function of the crystal itself. That is, the structure of each of the Bragg peaks contains information about the shape of the crystal. This idea was first put into practice by Robinson et al [349] and a theoretical analysis of the effects of partial coherence on the technique was also presented around the same time [350] and which explained some of the features in the image. A simple shape function implies that the Bragg peaks should be spatially symmetric; an asymmetric Bragg peak implies a complex shape function, where the phase of the complex distribution reflects strain in the crystal [351]. The extension of these ideas to the three-dimensional imaging of the structure in nanocrystals was later reported [352]. In this latter work, the full three-dimensional diffraction pattern around the diffraction peak could be recovered by rotating the crystal through a relatively small angle and the three-dimensional structure then recovered. A closely related method has more recently been used to produce a detailed three-dimensional map of a deformation field within a lead nanocrystal [353], an example of which is shown in Figure 25. Schroer and colleagues [354] have recently reported measurements of the shape of a nanocrystal with 5nm spatial resolution and using a focussed X-ray beam and indicating that a focussed beam enables high-resolution images to be obtained in a relatively short period of time provided that the object is radiation hard. CDI has been extended to provide a degree of elemental resolution [355] and to view structures within biological objects [340]. A couple of studies have also been devoted to the problem of imaging the structure of quantum dots [356, 357]. The mechanics of a ceramic nanofoam have also been investigated using these methods (see Figure 26) [358] and the structural

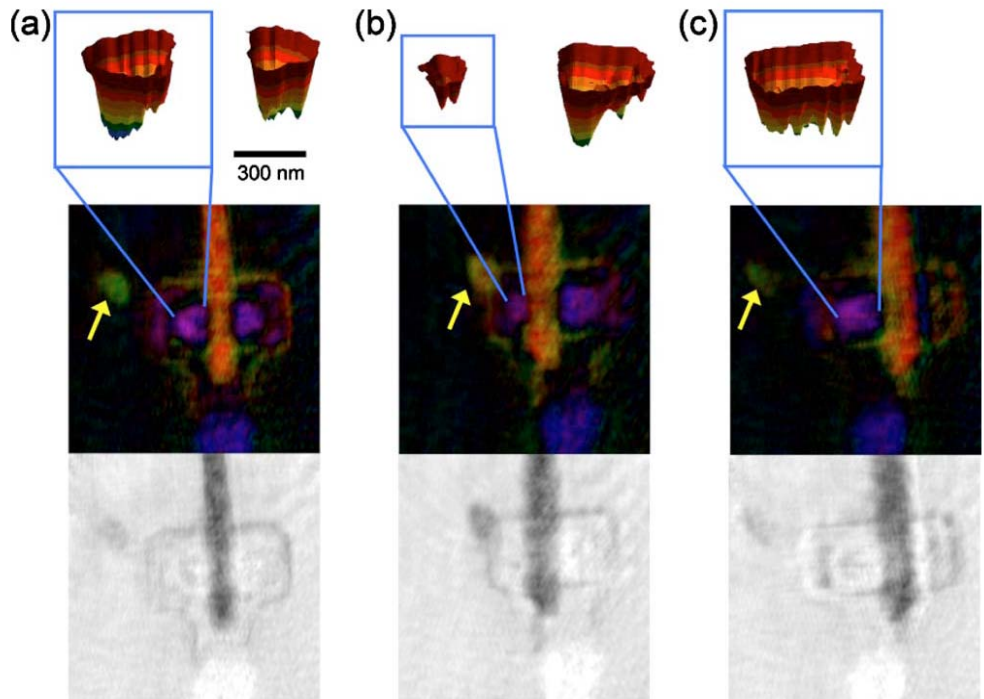

*Figure 27: Reconstruction of void defects in fusebay sample viewed at a range of angles. The top row shows 3D images inferred from the phase difference through voids compared to the surrounding material. The middle row shows the product of phase and amplitude of reconstructed X-ray wave leaving the sample. The yellow arrow indicates the position of an aluminum contamination identified via the measured phase shifts.  Notice how the relative phase shift through the contaminant changes as it overlapsother structures in the projection (b).The bottom row shows the reconstructed amplitude of the transmitted wave. The data were acquired at using 1.8keV X-rays at the 2-ID-B beamline at the Advanced Photon Source. Reproduced from ref [359]*

information obtained was found to be consistent with independent small-angle X-ray scattering experiments.

The Keyhole diffraction imaging method [347], which is applicable to the imaging of extended objects, has been applied to the analysis of integrated circuit samples [359] (see Figure 27) and allowed the analysis of defects and voids in the structure. A similar application of ptychography, to buried zone plate structures, has also been reported by Thibault and colleagues [233] (see Figure 18 and section 5.4.3)..

The motivation for much of this work has been in the direction of bioimaging and some reports are now emerging in that area. Initial work by Miao and colleagues showed some images of *E coli* bacteria [360]. These images were of demonstrative value only and facilities at the Advanced Light Source were later used to image a yeast cell and compare the results to images from transmission X-ray microscopy [361]. Williams et al [362] have used Fresnel coherent diffractive imaging to look at the structure of the malaria parasite in an infected red blood cell and compared the results with other forms of microscopy.

The issue of resolution was discussed by Thibault et al [361] who proposed that the resolution be determined by the highest spatial frequency at which the diffraction pattern is reliably reconstructed. This is an approach that seems to have gained reasonably widespread acceptance.

Other forms of soft X-ray sources are in the process of being developed, including soft X-ray lasers [363, 364] and high-harmonic generation sources [10, 365], and both classes of source have been shown to produce a high degree of spatial coherence [10, 365-367]. The applications of these very interesting sources are emerging, and one possibility is the use of coherent diffraction to enable high spatial resolution imaging [368, 369]. The results of these experiments still require very long exposure times, use very simple objects and long wavelengths, and have relatively poor spatial resolution. It is safe to say that there is a lot of further development required before these sources can enable high-resolution imaging that will yield scientifically valuable results. However other laser-based X-ray sources are emerging as a potentially important technology for X-ray microscopy [370] but will probably not have sufficient coherence for CDI.

Coherent diffraction has considerable promise as a high-resolution imaging modality using synchrotron sources, particularly for objects that are reasonably resistant to the effects of very high radiation doses, such as may be available in materials science and condensed matter physics. It should be noted, however, that the ability to yield biologically interesting results using this method has yet to be convincingly demonstrated.

A primary underlying motivation, however, is the program of performing single molecule imaging using X-ray free electron lasers. The hope here is that the pulse will be so short that the diffraction process will be completed before the molecule disintegrates [322, 323]. Current modelling indicates that the time scale for the disintegration may be of the order of a few femtoseconds, and considerable work has been performed on the analysis of this problem [371-374], with the conclusion being that, without some efforts to control the sample expansion [375] the pulse from the free electron laser will be too long. These studies are largely concerned with structural integrity and do not pay a great deal of attention to the properties of the electrons, the particles responsible for the scattering of the X-rays. Recent work with longer wavelength (~30nm) free electron lasers [376] suggests that the physics of the interaction of atoms and molecules with these short wavelength coherent fields is fundamentally different and so there is much theoretical and fundamental experimental work required before one can confidently predict the diffraction patterns of molecules from intense free electron laser pulses. Nonetheless, some very interesting experimental work has been performed, with Chapman and colleagues showing [377] that a nanofabricated sample can be imaged with a coherent free-electron laser pulse before the sample disintegrates; an interesting and influential demonstration but one that is also still many orders of magnitude away from conditions relevant to biomolecular imaging. The same group has gone on to develop novel pulsed holographic imaging methods [378] and found that the dynamics of the disintegration is consistent with theoretical modelling of the expansion of the ions produced by the pulse [379].

In keeping with the theme of this review, it should be noted that the developments in coherent diffractive X-ray imaging, with its origins in iterative techniques developed for electron imaging, are now feeding back into the electron community through the application of the ideas to the development of coherent electron imaging for the observation of structure in carbon nanotubes [380] and the atomic distribution within nanocrystals [381].

### 6.4 The role of coherence

In spite of the state of refinement of the algorithms it was observed that even high-quality X-ray data could not guarantee that a solution could be found, though it was noted that where experiments had been performed with laser light the reconstruction algorithms found a solution reliably and rapidly [382]. All of the image recovery algorithms relate the scattered field at the object to the scattered field at the detector via a Fourier transform, a relationship that is only valid for an optical field that is completely coherent. This is an assumption that does not hold for X-rays produced by a third-generation synchrotron, but is expected to be a good model for light produced by an X-ray free-electron laser. Accordingly, Spence et al [383] developed an analysis that predicted that coherent diffracting image requires that the coherent patch of the light have twice the width of the object, a requirement that has its origins in the over-sampling argument of Sayre [314, 384]. However this argument is incomplete as the size of the coherent patch is subject to the precise definition of coherence length used to decide whether the object does or does not meet the criterion. While it is a useful picture to regard the coherent area of a light field as containing completely coherent light within it completely incoherent light elsewhere, it is not a good analytical concept. Almost inevitably, and as discussed in detail in section 3, the correlation distribution in a light field has a Gaussian distribution and the coherence length is a measure of the width of this Gaussian. The degree of correlation therefore falls off monotonically as a function of the separation of the points between which the correlation is measured. A more rigorous model for CDI was developed by Williams et al [385] who found, when combining their results with computer simulations, that the ability to obtain a reliable reconstruction was critically dependent on a very high degree of spatial coherence, an observation in line with the experience of workers in the field.

The effects of coherence may be considered by once again applying eq30 in the projection approximation

$$I_f(s) = I_0 \int g(r_1' - r_2') T(r_1') T^*(r_2') \exp\left[-iks \bullet (r_1' - r_2')\right] dr_1' dr_2'. \tag{100}$$

Following Williams et al [385], and the guidance of section 3, a Gaussian statistically stationary [58] coherence function is again introduced (See Table 2) and the coordinate system in section 2.5 is again used to obtain

$$I_f(s) = I_0 \iint T\left(r - \frac{x}{2}\right) T^*\left(r + \frac{x}{2}\right) dr \, exp\left[-ik_0 s \bullet x - \frac{|x|^2}{2\ell_c^2}\right] dx, \tag{101}$$

Obviously, in the limit of $\ell_c \rightarrow \infty$ this reverts to the coherent limit in which the intensity of the diffraction pattern is the Fourier transform of the autocorrelation function of the scattering distribution, and the results follow for the uniqueness of the relationship between that diffraction pattern and the scattering function. However for a finite coherence length, eq101 describes a convolution over intensity and the effect of such a convolution can fundamentally change the nature of the function being convolved.

It is instructive to recall that an analytic function is fully defined by the locations of its complex zeroes [386, 387], and the diffraction pattern from a finite object is analytic. The blurring effect of the convolution is to immediately remove any zeroes from the measurement and so eliminate the possibility that any reconstruction can be completely consistent with the data. The conclusion has been confirmed via a series of simulations [385] in which it was found that even a small degree of partial coherence, for which the coherence length is still far greater than the dimension of the object, would remove the possibility that the algorithm can converge on any solution, let alone the correct one. These conclusions have been confirmed experimentally [388]. It is possible that the unreliability of the convergence of reconstructions of X-ray data, and the reliable convergence with optical laser data [382] has its origins in the less than perfect coherence from most X-ray sources.

This observation may also perhaps be viewed from a more general perspective. Scanning microscopy employs a zone plate for the production of a very tight focal distribution, and the rule-of-thumb is that the spatial coherence length of the incident radiation be larger than the zone plate itself, so that it is "fully coherently illuminated", and the temporal coherence length exceed any path differences, requiring that the monochromaticity of the incident light, $\lambda/\Delta\lambda$ be greater than the number of zones in the zone plate. Let us here concentrate on the spatial coherence requirement. The convolution description of coherence [38] implies that the focal distribution of the zone plate is the convolution of the diffraction limited focal distribution with the Fourier transform of $g(r_1 - r_2)$. By the van Cittert-Zernike theorem [389], and ignoring the effects on any intervening beamline optics, $g(r_1 - r_2)$ corresponds to an appropriately scaled Fourier transform of the source distribution in the undulator [38]. The requirement of "coherent illumination" is that the convolution does not significantly broaden the diffraction limited spot, and requiring that the original source be either sufficiently small or sufficiently distant. In most modern synchrotron sources, the source places a resolution limit of a few tens of nanometres. That is, to significantly improve imaging beyond this limit is to run into the effects of partial coherence, a result that is *not* dependent on the size of the object. The aim of CDI is inevitably to improve spatial resolution to the ultimate limit and this argument would lead to the conclusion that the assumption of perfect coherence is going to break down and that the value of arguments about the relative size of the object to the coherence length of the illumination is limited, and will lead to unrealistic expectations of the ability to achieve very high resolution. This conclusion is entirely consistent with the results of Williams et al [385].

It was noted, and an explanation was given [388], that CDI using curved beams is more tolerant to reductions in spatial coherence, partly because of the reduced ambiguity in the data – a Fresnel diffraction pattern often has a more direct relationship to the diffracting object, where this is generally not the case for a Fraunhofer diffraction pattern.

More recently, Chen at al [390] have considered the role of longitudinal coherence - optical bandwidth – in diffractive imaging using high-harmonic generation sources. A method was demonstrated in which spectral information was incorporated into the image reconstruction approach. It remains to be seen whether this work has meaningful implications for imaging using X-ray free electron lasers, which themselves have a significant optical bandwidth and, recent measurements suggest, possibly less than perfect spatial coherence [45].

## 6.5 Detectors for coherent imaging

The detector demands for coherent imaging are very stringent. In particular, the diffracted data has a very large dynamic range (around 6 orders of magnitude) and so the dynamic range of the detection system should also be correspondingly large. Currently, as CCD systems do not offer this, the required dynamic range is typically achieved by summing a large number of frames of data using cooled CCD camera systems.

The ability to reconstruct an image relies on the ability to properly record the diffraction pattern. The Fourier transform relationship between the field and the detector implies that the larger the object the finer is the sampling required at the detector plane – the oversampling condition discussed in section 6.1. In practice, this means that there is a premium for the detector to have adequate angular resolution, a requirement that is met by CCD systems. The emerging forms of hybrid pixel detectors offer enormous potential in this application but so far only offer a pixel size that is too large for most applications [391]. The emerging X-ray free electron laser sources are driving a considerable amount of effort in this direction and small X-ray active pixel sensors are now beginning to emerge [392]. An active pixel detector system with a pixel size of $50 \mu m$ or less would be ideal, and one can anticipate that such a detector will be available in the next few years.

# 7 Coherent Scattering as a Probe of Material Structure and Dynamics

The development of coherent X-ray techniques for probing the dynamics and detailed structure of matter is briefly reviewed. In large part, the methods draw very heavily from methods that have been developed in either the electron or visible optics

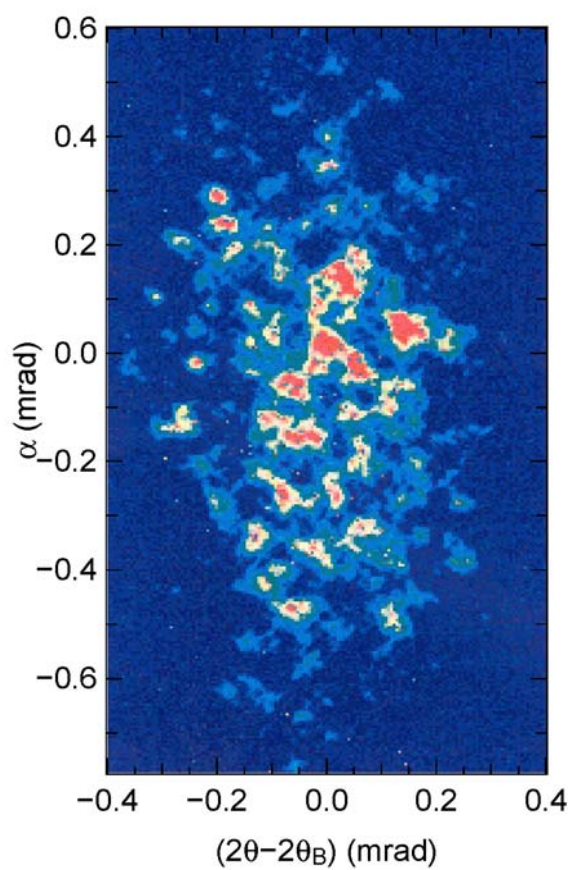

*Figure 28: First observation of X-ray speckle from a synchrotron source. The data was acquired around the Bragg peak of $Cu_3Au$ using 8.3keV photons at the X25 beamline at the National Synchrotron Light Source. Adapted from ref [351].*

communities.

## 7.1 X-ray photon correlation spectroscopy

An X-ray photon correlation spectroscopy (XPCS) seeks to measure the intensity correlation function of scattered of light for the study of the statistical physics of systems [393]. The applications using visible light have a long and rich history but have only recently been developed using coherent X-ray sources. The origins of the use of coherent X-ray speckle lies with the first observation of X-ray speckle scattered by an object [394], and the first observation of X-ray speckle from light scattered by a surface [395] (Figure 28). The field has been subject to a number of recent reviews [396-398] and so the discussion here will be limited to the broad methodological developments.

It is well established that diffraction by a static system does not alter the degree coherence of the radiation – to permit the possibility of inducing partial coherence into a closed static system would violate Liouville's theorem. In XPCS, the change in the properties of the system are probed, in essence, by observing the changes in the properties of the field over time, and so measuring the fluctuation time in the coherence properties of the diffracted wave.

Let us again consider the far-field diffraction of a partially coherent field by a thin scattering object

$$I_{ff}(\mathbf{s},t) = \int g(\mathbf{x}) T\left(\mathbf{r}+\frac{\mathbf{x}}{2},t\right) T^*\left(\mathbf{r}-\frac{\mathbf{x}}{2},t\right) \exp[-ik\mathbf{s}\bullet\mathbf{x}] d\mathbf{x} d\mathbf{r} \qquad (102)$$

where the scattering object is allowed to fluctuate in time and the effects of the time delay between the light from the two scattering points to reach the observation point are ignored.

The field of X-ray photon correlation spectroscopy (XPCS) requires that one form the intensity correlation function

$$\gamma^{(2)}(\mathbf{s},\tau) = \frac{\langle I(\mathbf{s},t)I(\mathbf{s},t+\tau)\rangle}{\langle I(\mathbf{s},t)\rangle^2}, \qquad (103)$$

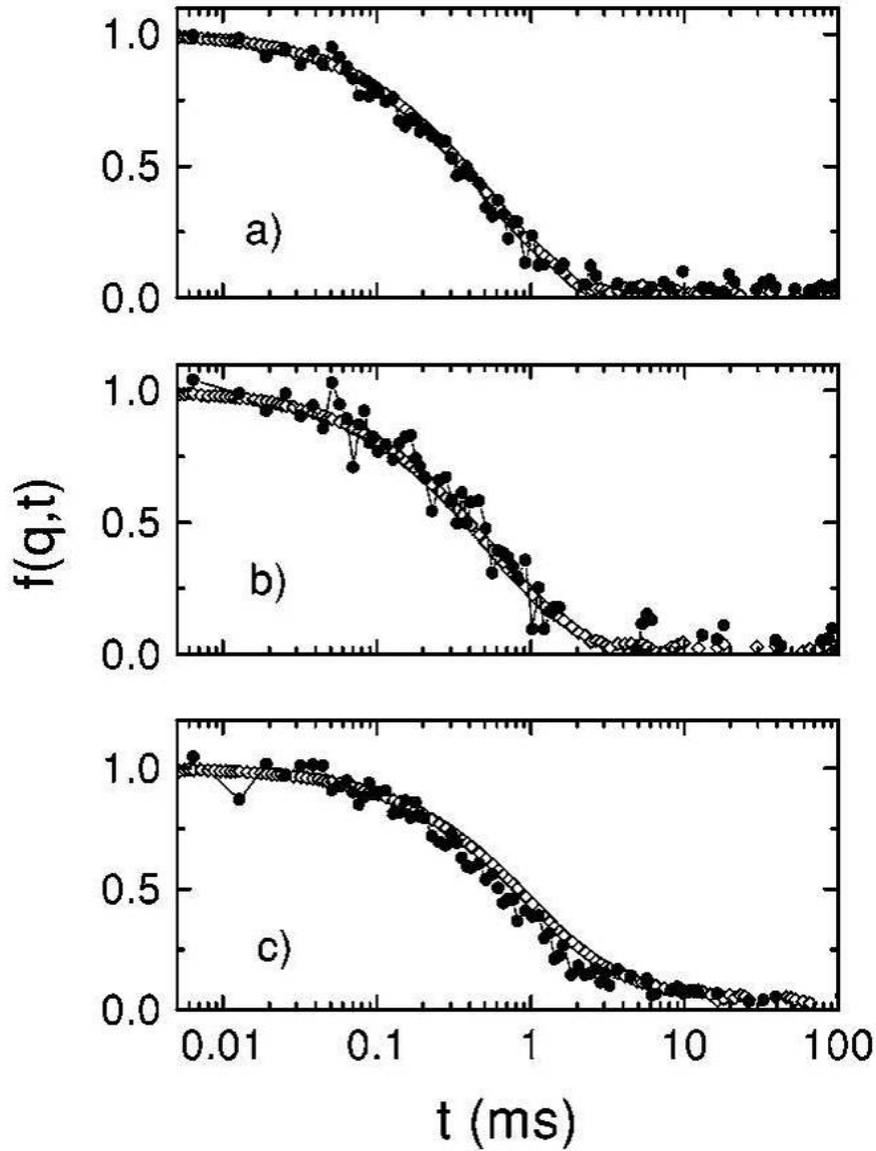

*Figure 29: Comparison of intermediate scattering functions from X-rays (dots) and visible light (circles) from a dense colloidal speciment. The X-ray data were acquired using 8.2keV X-rays at the ID10 beamline at the European Synchrotron Radiation Facility. Reproduced from ref [400]*

which is simply eq4 written in its far-field form and in terms of a single scattering vector. This is related to the scattering function of the system via the so-called Siegert relationship, eq5,

$$\frac{\langle I(s,t)I(s,t+\tau)\rangle}{\langle I(s,t)\rangle^2} = 1 + \left|\gamma^{(2)}(s,\tau)\right|^2. \qquad (104)$$

In the case of XPCS it is assumed that the incident field is coherent. Liouville's theorem assures us that the resulting scattered field is coherent, however one can form an ensemble average of the field realisations, as is required for the formation of a correlation function in eq103, by obtaining repeated measurements over time. This data may then be analysed, via eq104, to recover information about the structure of the scattering medium and, in particular, its dynamics. These ideas have been applied extensively using laser light and have only relatively recently been possible using X-rays.

XPCS experiments may be performed in two broad configurations: homodyne and heterodyne. The former relies on a measurement of the intensity diffraction pattern described by eq102. Heterodyne methods allow the scattered field to interfere with a static reference field that enables some information to be obtained about the phase of the correlation function. In particular [399] one obtains for the heterodyne method

$$\gamma^{(2)}(s,\tau) \approx 1 + 2\left(\frac{I_s}{I_{ref}}\right)\text{Re}\left[\gamma_1(s,\tau)\right], \qquad (105)$$

where $I_s$ and $I_{ref}$ refer to the intensities of the scattered and reference waves respectively. The reference signal may be either introduced externally or may arise from the object itself.

The use of X-rays in this application became available with the development of third-generation synchrotrons. X-rays have the advantage of possibly providing greater spatial resolution, better penetration and significantly reduced multiple scattering, the last point enabling a more straightforward interpretation of the data. A comparison of X-ray and visible light methods [400] has shown that the methods yield consistent results and also confirmed that X-rays are able to provide a valuable complementary source of structural information (see Figure 28).

The limitation imposed on the use of X-rays arises from the fact that the sources are not yet fully coherent and this can limit the ability to quantitatively interpret the measurements. The effect of partial coherence is, essentially, to reduce the contrast in the observed speckle, as described by the convolution in eq102. For small departures from spatial coherence, the visible light scattering community models the effect of partial coherence via the relationship

$$\gamma^{(2)}(\tau) = 1 + \beta \left| \gamma^{(1)}(\tau) \right|^2, \tag{106}$$

where $\beta$ is simply the fringe visibility that would be produced by the light for fringes with the appropriate spatial frequency. This model is correct for Gaussian spatial statistics, and a more detailed analysis has been provided [401] that illustrates that the physics can be rather more complex than indicated by eq106 for surface scattering. Sikhasrulidze et al have looked at the effects of coherence and detector resolution in reflection geometry [402]. A corollary of the less than perfect coherence in the illumination is that the sources are rather less bright than the laser sources available in the visible regime and so one must take care to optimise the experimental configuration. An analysis of this optimisation problem has been published by Mochrie and colleagues [403, 404]. Thurn-Albrecht et al [405] demonstrated the possibility of using XPCS for non-transparent (for visible light) media and that the method can be applied over a wide range of time scales.

The methods of XPCs are best suited to the exploration of the internal dynamics of systems and have been used for the exploration of the diffusion processes of particles in solution [405-407], to the dynamics of polymer blends [408-410], the study of clays [411], the structure of surface height fluctuations in liquids [412-416], the properties of liquid crystal membranes [417-419], magnetic speckles [420], the non-equilibrium dynamics of binary allows [421, 422] and metal polymer composites [416, 423].

A study of the dynamics of critical fluctuations in a binary alloy ($Fe_3Al$) at equilibrium [424] found many features in agreement with theoretical expectations but there was some quantitative disagreement that the authors attribute to less than perfect coherence and the effects of the averaging used to counter the effects of relatively low signal levels, in turn arising from imperfect spatial coherence in the illuminating beam. Mochrie and colleagues [408], in their application of the method to the study of the dynamics on block copolymer micelles, found good agreement with expectations and demonstrating the ability to probe wave vectors well beyond those accessible with visible light.

Price and colleagues [417] noted that, as the coherent flux from a synchrotron source varies as $\sim \lambda^2$, the use of longer wavelength X-rays will lead to greater brightness (coherence) and therefore to the possibility of probing shorter timescales. Using the Advanced Light Source and an X-ray wavelength of 4.4nm they were able to probe dynamics at a timescale as short as 2 microseconds. Subsequent experiments have been able to probe timescales down to 50ns [419]. These soft X-ray methods have been used to obtain direct measurements of magnetic devices [425].

Other variants on the method are emerging. For example, Banyopadhyay [426] has proposed a new approach based on the idea of measuring the dependence of the variance in the speckle as a function of the exposure time. Cipelletti [427] has proposed a method termed time-resolved correlation in which it is possible to look at dynamics with unusual properties such as those involving relatively sudden changes. The idea of looking at intermittent dynamics, with a non-Gaussian statistical distribution, is also leading to an examination of higher-order correlation functions and this has been explored by Duri and colleagues [428, 429]. This is an area that will no doubt be the subject of further development.

Cerbino et al [430] have looked at X-ray speckle without the observation necessarily taking place in the far-field, designating it "near-field". In this regime, they find greater tolerance to lower levels of spatial coherence. The authors note that other workers have been using related methods, and they consider the work of Kim and Lee [128], work that we have already discussed in this review in the context of phase-contrast imaging. Other related work [88-90, 431] might also be cited. This connection underlines that the methods for uses of coherent X-rays are all strongly related, but also that the boundaries are never clear; a phase-contrast imaging might equally well be considered a "near-field" speckle pattern; however it is not clear whether there is much to be gained from this perspective. The observed tolerance to relatively poor spatial coherence is well-understood through studies of the effects of partial spatial and temporal coherence in X-ray phase contrast imaging [187].

## 7.2  Detectors for XPCS

XPCS methods seek to probe time-dependent phenomena and an ideal detector should be highly efficient and collect over a large angular range, while having adequate resolution to properly resolve the speckle. The requirement on pixel size and detector efficiency is therefore comparable with those for CDI, so that direct detection CCD systems are very often used. However XPCS does not have as stringent a requirement for dynamic range as for CDI, but this is replaced by the need for speed of read-out. As such, XPCS is largely performed with direct-detection CCD systems that have been optimized for speed of data acquisition and read-out [432].

## 7.3 Fluctuation microscopy and medium range order

It might be argued that a recurring them of this review has been the development of techniques that will enable the probing of non-crystalline materials. Crystalline materials have, of course, been the subject of a great deal of attention over the last decades. The motivation behind coherent diffractive imaging and X-ray photon correlation spectroscopy is the ability to probe order and structure at short ranges. The probing of matter at medium ranges has been a more challenging problem.

The electron imaging community has developed methods for the probing of medium range order using the method of fluctuation microscopy [433, 434]. The essential idea behind fluctuation microscopy is to image a object using hollow-cone illumination and to observe the diffraction of the electrons using a dark-field imaging arrangement. The angle of the hollow-cone illumination is changed and the manner in which the variance of the speckle distribution changes as a function of illumination angle is used to deduce properties of the medium range order in the object.

Electron based fluctuation microscopy is able to probe order at the $\geq 0.5 nm$ scale [434], longer wavelength soft X-rays may be used to probe ordering at the scale of self-assembled nanoscale materials, $\geq 5nm$. For experimental reasons, it is more convenient to implement X-ray fluctuation microscopy in a different form. Instead of illuminating with a hollow-cone illumination, pinholes of varying dimensions are scanned across the object and the diffraction observed.

Define the function

$$\Pi(x) = \begin{cases} 1 & x \leq 1/2 \\ 0 & x > 1/2 \end{cases}, \qquad (107)$$

So that, using eq30, the diffraction pattern produced by a pinhole of radius $R_0$ located at $r_n$ can be described by

$$I_{ff}(s, r_n, R_0) = \int \Pi\left(\frac{|r_1 - r_n|}{2R_0}\right) \Pi\left(\frac{|r_2 - r_n|}{2R_0}\right) T(r_1) T^*(r_2) \exp\left[-iks \bullet (r_2 - r_1)\right] dr_1 dr_2. \quad (108)$$

One can form the variance of the intensity patterns through an average over the pinhole position, $r_n$, via

$$\sigma^2(R_0, s) = \frac{\langle I^2(s, r_n, R_0)\rangle}{\langle I(s, r_n, R_0)\rangle} - 1. \quad (109)$$

Fan et al [435] showed that the intensity measurements in eq108 can be directly related to the theory of electron fluctuation microscopy and that the corresponding results may be readily adopted. As shown by Gibson et al [436], one can predict that plots of $R^2 / \sigma^2(R_0, s)$, where $R = 1/2R_0$, versus $R^2$ should be linear and that the correlation length in the material can be deduced from the slope and intercept of the line. The validity of this deduction and its first application was reported by Fan et al [437].

In summary, X-ray fluctuation microscopy is a potentially important emerging coherent X-ray technique for probing medium range order, but has yet to be fully developed and exploited.

No consensus has yet emerged, but fluctuation microscopy does not have the speed requirements needed for XPCS nor the dynamic range requirements for CDI so CCD systems are likely to be adequate for the foreseeable future.

## 7.4 X-ray free electron lasers

The limitation in all of the above work is the relatively low brightness of the available X-ray sources. X-ray lasers will produce beams with a much greater level of brightness and this will lead to a number of important new scientific opportunities. There is also little doubt that the very high brightness and peak power of these sources will lead to a need to re-think many of the experimental methods discussed in this review, and the invention of new approaches.

Of course, a major driving forced for coherent imaging is the desire to image single molecules using diffractive imaging methods such as outlined in this review. A major obstacle to be overcome is the effects of damage on the target molecule [323]. A major issue concerns the timescale of the explosion process and a number of simulations studies of this process have been published [438-440]. A consensus is emerging that the movement of the nuclei within an exploding molecule can be ignored for up to around 5fsec, but that the electrons responsible for scattering the electrons will move on a far shorter timescale. Importantly, the electrons are responsible for the X-ray scattering and so it is expected that the impact on the X-ray diffraction pattern will, for all realistic pulse lengths, be immediate [441]. The essential message behind these simulations is that the molecule will inevitably be strongly influenced by the probe beam and that progress on this front will need to account for the interaction with the beam or, as with the development of laser science, to use the interaction as a probe of the properties.

The use of a laser beam as the probe has a long history but consideration of a coherent X-ray beam as a probe is in its infancy. Of considerable interest is the series of papers by Mukamel and colleagues on the use a very short (less than 1 fsec) coherent X-ray pulse as a precise probe of molecular dynamics [442-446]. This is an exciting possibility but will await the development of X-ray pulses in the attosecond regime.

X-ray photon correlation spectroscopy is essentially a technique that looks at systems that are in some sort of equilibrium, allowing the fluctuations to be probed over time. So, while XPCS does benefit from a very high degree of coherence, the short pulse and its propensity to disturb the object will require consideration of how the method may best be adapted. As with visible dynamic light scattering, there is a considerable advantage in having fully coherent sources and these are now becoming available through the development of X-ray free electron lasers. It is anticipated that the much increased coherent flux of the X-ray laser sources will allow XPCS to find an important home at these facilities and will display the ability to probe materials at very short time- and length-scales [447] using a form of pump-probe method that can allow time-series data to be acquired.

## 8  Summary and conclusions

Coherent X-ray science is an area that is at the meeting point of a number of fields and the necessarily broad scope of this review, ranging from astronomical imaging through X-ray to optical microscopy, underlines the degree to which these seemingly disparate fields are beginning to interact and benefit from each other.

The field of coherence measurement has had a long tradition in optical physics, but the requirement for a full four-dimensional characterisation of the field is an area that has seen new impetus with the development of coherent X-ray optical techniques. This review has described the state of development to date in this area and it has been seen that the detail with which the coherence state of X-ray fields has been characterised now exceeds that in the visible optics regime. The interest in the coherence properties of the X-rays has led some authors to introduce the concept of X-ray decoherence, though it is now suggested that this is a phenomenon arising from unresolved X-ray speckle in the wave-field.

| Technique | Sources | Area of application | Coherence requirements | Spatial resolution | Detector requirements | Preferred detector types | Relevant sections of review | Comments |
|---|---|---|---|---|---|---|---|---|
| Propagation-based | Laboratory-based; synchrotrons | Materials; medical imaging | Low | 10 μm | Small pixel size; high efficiency; large energy range | CCD | 4.1; 4.2; 4.3 | Resolution limited by detectors. Has already produced much good science at 3rd generation synchrotron sources. Can be made quantitative. |
| | Electron, neutrons and light sources | Materials science and biological microscopy | Low | 1 nm (electrons) | High efficiency | CCD | 5.5.1; 5.5.2; 5.5.3 | These methods are seeing applications in various forms of microscopy and radiography. |
| Interferometry | Laboratory-based; synchrotron | Medical imaging, Non-destructive testing | Low | 10 μm | Small pixel size; high efficiency. | CCD | 4.4.1 | Effective using laboratory sources. Can produce valuable medical images. Can be quantitative. |
| Moire/Talbot methods | Laboratory sources | Materials; medical imaging | Low | 10 μm | Small pixel size; high efficiency; large energy range | CCD | 4.4.2 | Effective using laboratory sources. Can produce valuable medical images. Can be quantitative. |
| Diffraction-enhanced | Laboratory-based; synchrotron | Medical imaging | Low | 10 μm | Small pixel size; high efficiency; large energy range | CCD or linear arrays | 4.4.3 | Effective using laboratory sources. Can produce valuable medical images. Hard to quantitate. |
| Holography | Synchrotron sources; XFEL sources | Magnetic materials | High | 10 nm | Small pixel size; high efficiency. | CCD | 5.4.4 | Resolution limited by reference pinhole, but may be enhanced using CDI style processing |
| Wigner Phase Space Deconvolution | Synchrotron sources | Limited practical value | Low | 10 nm | Small pixel size; high efficiency. | CCD | 5.4.2 | An interesting demonstration but probably will not see a lot of practical application. |
| Ptychography | Synchrotron sources | Imaging extended objects | High | 10 nm | Small pixel; high efficiency; high dynamic range | Pixel array detectors | 5.4.3 | Requires serial data acquisition so not suitable for single-shot experiments. High dynamic range detector needed for extension to very high resolution. |
| Plane-wave CDI | Synchrotron sources; XFEL sources | Imaging nanocrystals and small isolated objects | High | 10 nm Possibly 0.1 nm using XFEL sources. | Small pixel; high efficiency; high dynamic range | Pixel array detectors | 6 | May achieve much higher resolution using X-ray laser sources. Adequate pixel array detectors not yet available so CCDs still detector of choice |
| Fresnel CDI | Synchrotron sources; XFEL sources | Imaging regions of extended objects | High | 10 nm Possibly 1 nm using XFEL sources. | Small pixel; high efficiency; high dynamic range | Pixel detectors | 6 | May achieve much higher resolution using X-ray laser sources. Adequate pixel detectors not yet available so CCDs still detector of choice |
| XPCS | Synchrotron sources; XFEL sources | Materials dynamics; soft matter | Very high | 10nm | Small pixel size; high efficiency; rapid read-out | Rapid read-out CCD | 7.1 | Requires very high phase-space density of photons to probe high-resolution dynamics. A serial method so not suited to single-shot experiments |
| Fluctuation microscopy | Synchrotron sources | Materials; soft matter | High | 10 nm | Small pixel size; high efficiency. | CCD | 7.2 | This was developed in electron imaging and probes medium range order but has yet to see wide application in X-rays. |

*Table 3*: A brief overview summary of the techniques that have been considered in this review. The spatial resolution is a rough indication of the resolution that may be possible and will be highly dependent on the implementation and the source used.

X-ray phase contrast imaging has rapidly emerged as an influential methodology, with the prime impact being in the ability to visualise structure through the refraction of the X-rays. This method has found many applications in the detailed imaging and analysis of a range of objects, from fuel sprays through medical imaging to palaeontology. The ability to quantify the phase through the phase-contrast mechanism has drawn from ideas developed in optical astronomy and electron microscopy, and has in turn led to new approaches in optical microscopy and electron microscopy. Moreover, the relatively low coherence requirements of X-ray phase contrast imaging have led to the demonstration that phase-contrast techniques are also possible for neutron sources.

Coherent diffraction is becoming increasingly important and these methods have drawn heavily from ideas developed for high-resolution electron imaging, including iterative phase-retrieval algorithms. This area has yet to deliver on its full scientific potential, but this is likely to be achieved with the advent of the X-ray free-electron laser. Nonetheless, coherent diffractive imaging is certainly emerging as a powerful high-resolution X-ray imaging technique. Interestingly, methods proposed but not successfully implemented for electron imaging, such as ptychography, are now working well with coherent X-rays, a result that is likely attributable to greater coherence being available with X-ray sources than from electron sources. The high-resolution diffraction imaging methods demonstrated with X-rays are now being used for very high-resolution electron imaging of nano-tubes [380] and nano-crystals [381], further emphasising that the fields are increasingly interactive.

X-ray photon correlation spectroscopy is drawing from the techniques developed over many years in visible light and using laser sources. It is yielding higher spatial resolutions and greater penetration into objects. However it is an area that will also significantly benefit from the true spatial coherence promised by the X-ray free electron laser sources.

A very broad summary of the techniques covered in this review is presented in Table 3.

The trends identified here will continue to accelerate and it is fair to anticipate that the extended availability of X-ray free electron lasers will see huge extensions of the fields of interaction to draw even more strongly from laser physics, quantum optics and non-linear science. These latter areas have not been covered to any great extent in this review and will be the excellent topic of a further review in the coming years.

## 9  Acknowledgements

The author acknowledges helpful critical feedback on the manuscript from Les Allen and Garth Williams of the University of Melbourne and from Mark Sutton of McGill University. The author also acknowledges useful discussions with Oleg Shpyrko of the University of California San Diego. He would like to acknowledge the support of the Australian Research Council through it Fellowship and Centres programs.